\documentclass[letterpaper,titlepage,11pt]{article}
\pdfoutput=1
\usepackage[colorlinks=true,urlcolor=blue,citecolor=magenta]{hyperref}
\usepackage{amssymb,amsmath,amsfonts}
\usepackage{epsfig}
\usepackage{epsfig}
\usepackage{graphicx}
\usepackage{epstopdf}
\usepackage{caption}
\usepackage{subcaption}
\usepackage{amscd}
\usepackage{amsthm}
\usepackage{latexsym}
\usepackage{amsbsy}
\usepackage{bbm}
\usepackage[english]{babel}
\usepackage{psfrag}
\usepackage{tabularx}

\allowdisplaybreaks

\def\clock{{\count0=\time
           \divide\count0 60
           \ifnum\count0<10 0\fi\the\count0
           \multiply\count0 -60 \advance\count0 \time
           :\ifnum\count0<10 0\fi \the\count0
         }}
\newcommand{\timestamp}{{\small\vbox{\hbox{\tt\jobname.tex}
\hbox{\the\day/\the\month/\the\year, \clock}}}}

\numberwithin{equation}{section}

\begin{document}

\begin{titlepage}
\rightline{\vbox{   \phantom{ghost} }}

 \vskip 1.8 cm

\centerline{\huge \bf
Boundary Stress-Energy Tensor and Newton--}
\vskip .3cm

\centerline{\huge \bf  Cartan Geometry in Lifshitz Holography}

\vskip 1.5cm

\centerline{\large {{\bf Morten H. Christensen$^1$, Jelle Hartong$^1$, Niels A. Obers$^1$, Blaise Rollier$^2$}}}

\vskip 1.0cm

\begin{center}

\sl $^1$ The Niels Bohr Institute, Copenhagen University  \\
\sl  Blegdamsvej 17, DK-2100 Copenhagen \O , Denmark
\vskip 0.4cm
\sl $^2$  Institute for Theoretical Physics,
University of Amsterdam,\\
Science Park 904, Postbus 94485, 1090 GL Amsterdam, The Netherlands

\end{center}
\vskip 0.6cm

\centerline{\small\tt mchriste@fys.ku.dk,  hartong@nbi.dk, obers@nbi.dk, B.R.Rollier@uva.nl}

\vskip 1.3cm \centerline{\bf Abstract} \vskip 0.2cm \noindent

For a specific action supporting $z=2$ Lifshitz geometries we identify the Lifshitz UV completion by solving for the most general solution near the Lifshitz boundary. We identify all the sources as leading components of bulk fields which requires a vielbein formalism. This includes two linear combinations of the bulk gauge field and timelike vielbein where one asymptotes to the boundary timelike vielbein and the other to the boundary gauge field. The geometry induced from the bulk onto the boundary is a novel extension of Newton--Cartan geometry that we call torsional Newton--Cartan (TNC) geometry. There is a constraint on the sources but its pairing with a Ward identity allows one to reduce the variation of the on-shell action to unconstrained sources. We compute all the vevs along with their Ward identities and derive conditions for the boundary theory to admit conserved currents obtained by contracting the boundary stress-energy tensor with a TNC analogue of a conformal Killing vector. We also obtain the anisotropic Weyl anomaly that takes the form of a Ho\v{r}ava--Lifshitz action defined on a TNC geometry. The Fefferman--Graham expansion contains a free function that does not appear in the variation of the on-shell action. We show that this is related to an irrelevant deformation that selects between two different UV completions.

\end{titlepage}


\tableofcontents

\section{Introduction}


\subsection{Scope and Motivation}

Ever since the birth of the AdS/CFT correspondence \cite{Maldacena:1997re} there has been a continuous effort to find more examples of holographic correspondences. In recent years this has received an extra boost due to the growing interest in using holography as a tool to study strong coupling physics of potential relevance to phenomenology such as holographic QCD, quark-gluon plasma physics, far from equilibrium dynamics, thermalization, high-$T_c$ superconductors and many other condensed matter systems (see \cite{Gubser:2009md} for reviews). Apart from the obvious phenomenological interest, this development provides a field theory inspired guidance to look for interesting problems on the gravitational side of the duality. In fact, very often the dual field theories are not known explicitly and one (quasi) defines them in the appropriate regime of the coupling constant at large $N$ via the holographic duality. These developments have led to a tremendous activity in the field of applied holography leading e.g. to many new interesting asymptotically AdS black holes solutions and the construction of new types of holographic dualities involving non-asymptotically AdS space-times such as Schr\"odinger, Lifshitz and hyperscaling violating geometries.

The motivation of the present paper  lies in understanding the basic ingredients of
holographic dualities for scale invariant field theories with dynamical exponent
$z>1$. Such theories are of relevance to condensed matter theory (CMT) where one
frequently finds effective field theory descriptions of a system near some quantum critical
point that is invariant under the Lifshitz symmetry group ($z$-dependent scale
transformations $t\rightarrow\lambda^z t$, $\vec x\rightarrow\lambda\vec x$,
space-time translations and spatial rotations). Lifshitz geometries \cite{Koroteev:2007yp,Kachru:2008yh,Taylor:2008tg} together with
hyperscaling violating geometries \cite{Charmousis:2010zz,Huijse:2011ef,Dong:2012se} as well as more general Bianchi type space-times 
\cite{Iizuka:2012iv} have occurred as effective IR geometries that could furnish as
a groundstate geometry of some CM system. Depending on one's interest one can then
consider either an AdS (with or without hyperscaling violation) or a Lifshitz UV
completion (again with or without hyperscaling violation). In this work we will see an
example where a $z=2$ Lifshitz IR geometry becomes either $z=2$ Lifshitz in the UV
or AdS with hyperscaling exponent $\theta=-1$.

The study of scale invariant (or covariant as in the case of hyperscaling violating) geometries provides
furthermore a great opportunity to extend our understanding of holography beyond the
familiar AdS/CFT context. There are many interesting open problems regarding the
precise holographic nature (already at the level where a gravitational approximation
applies) of Schr\"odinger, Lifshitz and hyperscaling violating geometries. Many such
questions are in one way or another related to the precise properties of the analogue
Fefferman--Graham (FG) theorem which so far is only known when contact with an AdS space-time can be made (e.g. by dualities such as 
TsT in the case of $z=2$ Schr\"odinger holography \cite{Balasubramanian:2008dm,Hartong:2010ec,Guica:2011ia,Hartong:2013cba} or by dimensional reduction as in the case of $z=2$ Lifshitz holography which will be elaborated on further below). 
As a result, many properties of the holographic dictionary such as
the boundary geometry, holographic renormalization, and one-point functions including
the boundary stress-energy tensor are currently ill understood.

Extending the holographic paradigm to space-times that go beyond
the original AdS-setting has thus received a great impetus in recent years. As remarked above, this has been motivated in part by
applying holographic ideas to the study of strongly coupled condensed matter
systems, which often exhibit non-relativistic scaling, and thus necessitate the
consideration of bulk space-times
with asymptotics different from AdS. Moreover, beyond the success of holography to
study different types of strongly coupled quantum field theories, it is interesting
to examine more generally to what extent holography is applicable in spaces with
different asymptotics. This may shed further light on the nature of quantum gravity
and elucidate puzzles in black hole physics.

From now on we will focus our attention on holography for Lifshitz space-times. By
far the majority of work on Lifshitz holography has been within the context of the
massive vector model \cite{Kachru:2008yh,Taylor:2008tg} because it is simple in matter content and because it can account for all values of
$z$ by suitably choosing the cosmological constant and mass parameter. For a holographic study of this model see \cite{Ross:2009ar,Ross:2011gu,Baggio:2011cp,Mann:2011hg,Griffin:2011xs,Baggio:2011ha}. From an analysis of the linearized perturbations \cite{Danielsson:2009gi,Ross:2009ar,Ross:2011gu,Baggio:2011cp} it is known that one must separately study the following three cases: i). $1<z<2$, ii). $z=2$ (see also \cite{Cheng:2009df,Baggio,Holsheimer:2013ula}) and iii). $z>2$. For a
perturbative approach to values of $z$ close to one see the recent works
\cite{Korovin:2013bua}. Black hole solutions of the massive vector model have been studied in
\cite{Danielsson:2009gi,Bertoldi:2009vn,Brynjolfsson:2009ct}. For related work on solutions such as Lifshitz black holes in other models for Lifshitz holography see e.g. \cite{Taylor:2008tg,Gregory:2010gx,Tarrio:2011de,Braviner:2011kz}. Probe fields and correlation functions have been studied in \cite{Kachru:2008yh,Taylor:2008tg,Edalati:2012tc,Keeler:2013msa,Zingg:2013xla}.

As an alternative approach one could give up generality and study a specific model where $z$ is fixed but with the advantage that one controls the asymptotic expansions to the equations of motion. Such a scenario is known to be possible when $z=2$ for the following reason. A Lifshitz space-time with $z=2$ can be uplifted to an asymptotically
AdS space-time in one dimension higher \cite{Balasubramanian:2010uk,Costa:2010cn}.
This observation has motivated the search for Lagrangians that in 4 dimensions
admit $z=2$ Lifshitz solutions and that can be uplifted to 5 dimensions where they
admit asymptotically AdS solutions
\cite{Donos:2010tu,Cassani:2011sv,Chemissany:2011mb}. The central idea is then to
construct the FG expansions of the solutions in 5 dimensions and to reduce this to
4 dimensions. A first step in this direction was taken in \cite{Chemissany:2012du}
where the focus was on deriving the counterterms in 4 dimensions. Using this philosophy, we have presented in \cite{Christensen:2013lma} the basic ingredients of an explicit holographic dictionary for Lifshitz holography at the level of the supergravity approximation. The aim of this paper is to give the details of the calculations that underlie the results of \cite{Christensen:2013lma} and to present a number of other important properties of this holographic correspondence. In particular we explicitly address the holographic dictionary, including the
corresponding boundary geometry, the identification of sources+vevs and the computation of Ward identities as well as many other important properties of the boundary stress-energy tensor such as the anisotropic Weyl anomaly and conserved boundary currents, as will be detailed below.

\subsection{Summary and Outline} 

As an aid to the reader, we present here an outline of the paper, along with a summary of the main results. 

{\bf Section 2}: We begin in Section \ref{sec:background} with a brief summary of the model that we use. The starting point is the 5-dimensional (renormalized) action \eqref{eq:actionmodel} of Einstein gravity with a negative cosmological constant coupled to an axion-dilaton system. Note that throughout the paper we use the following notation: 5-dimensional quantities/indices are hatted while 4-dimensional quantities do not have a hat. Further, $a,b$-type indices refer to the boundary space-time and underlined indices denote tangent space. Our input will be the FG expansion of the solution to the equations of motion of the 5-dimensional action near the boundary (see \cite{Papadimitriou:2011qb,Chemissany:2012du}), along with the identification of the sources and vevs in this model and the Ward identities satisfied by the vevs. The reduction from 5 to 4 dimensions is a Scherk-Schwarz reduction which we choose to perform such that the 5-dimensional axion has the form $\hat\chi=ku+\chi$ with $\chi$ a 4-dimensional axion and where $u\sim u+2\pi L$ parametrizes the reduction circle. The reduction ansatz for the remaining fields is of the standard Kaluza--Klein (KK) form. This Scherk--Schwarz reduction gauges the axion shift symmetry with the KK vector leading to a massive vector in 4 dimensions via the 
St\"uckelberg mechanism. This is a consistent reduction meaning that all solutions of the 4-dimensional theory can be uplifted to solutions of the 5-dimensional theory. When $k\neq 0$, which we will always assume, the 4-dimensional solutions split up into two classes depending on the asymptotic behavior of the KK dilaton. We will mostly focus on solutions belonging to the class for which there exists a Lifshitz UV completion. We show that this requires a certain constraint on the sources. 

The reduction is spacelike everywhere in the bulk of the 5-dimensional space-time but must remain null on the boundary (which is the origin of the constraint just mentioned). There are two scales involved, $k$ and $L$ and we will be working in the regime where $kL\ll 1$. We show that this is compatible with the usual requirements of weak curvature and type IIB string coupling. In this situation we can ignore the KK tower of massive states and we obtain a 3-dimensional boundary theory. The weak coupling description of this theory has not been worked out in detail. Some general comments can be made about it. The axion in the bulk sources a theta angle in $\mathcal{N}=4$ SYM and its reduction will give rise to a Chern--Simons term in 3-dimensions. Further since the reduction is along a null circle (from a boundary perspective) the theory will have a $z=2$ dynamical exponent. It has therefore been dubbed a Lifshitz--Chern--Simons gauge theory \cite{Mulligan:2010wj}. Hence a specific subset of the 5-dimensional asymptotically AdS solutions can be reduced to 4-dimensional asymptotically Lifshitz geometries  \cite{Chemissany:2012du} while maintaining a well-defined low energy approximation of type IIB string theory. The 4-dimensional action is given in \eqref{eq:4Daction}. 

Section \ref{subsec:AlLif} gives a preview of our results for the sources that are obtained upon dimensional reduction. We discuss a sequence of boundary conditions: asymptotically Lifshitz, asymptotically locally Lifshitz and UV Lifshitz which is such that the next item is weaker than the previous one. For each of these boundary conditions we list the corresponding boundary geometry in table \ref{table}. The reduction suggests (see also \cite{Kanitscheider:2008kd}) that the 4-dimensional asymptotic expansions are naturally formulated in a non-radial gauge. The section concludes with a discussion of the issues one faces when trying to transform to radial gauge. 

{\bf Section 3}:
We obtain the most general boundary conditions that determine the $z=2$ Lifshitz UV completion in section \ref{sec:UVcompletion},
which we denote by Lif UV.  Due to the constraint mentioned above, namely that the reduction circle must remain null on the boundary, the Lif UV boundary conditions are most conveniently implemented using a vielbein formalism. This is the subject of section \ref{subsec:bdryconditions}. In fact if we demand that the sources appear as the leading components of bulk fields then it is mandatory to use vielbeins. By relating the 4-dimensional bulk frame fields to those in 5 dimensions \eqref{eq:framechoice1}-\eqref{eq:framechoice5} we obtain a simple relation between the 4- and 5-dimensional sources as given in \eqref{eq:Phi04d5d}--\eqref{eq:e0ia}. We notice the appearance of two special combinations \eqref{eq:framechoice4} and \eqref{eq:framechoice5} of the bulk gauge field with the bulk timelike frame field. The leading component of \eqref{eq:framechoice4} will be the boundary timelike vielbein $\tau_{(0)a}$ while the leading component of \eqref{eq:framechoice5} will be the boundary gauge field $A_{(0)a}$. With these ingredients at hand, we can then compute the boundary geometry along with the variation of the on-shell action and obtain the Ward identities. 

 {\bf Section 4}:  In section \ref{sec:bdrygeom} we obtain and study in detail the boundary geometry of the $z=2$ Lifshitz space-times of our model,
and the resulting torsional Newton-Cartan structure is one of our central results.  The appearance of Newton-Cartan structures
is expected, since in our case the boundary geometry is obtained by null-reduction of the 5-dimensional asymptotically locally AdS boundary geometry. To study the metric structure of the boundary we start by obtaining the transformations induced on the boundary vielbeins by bulk local Lorentz transformations that leave $dr/r$ invariant.  
The result is that the boundary vielbeins transform under the contracted Lorentz group consisting of local $SO(2)$ rotations and Galilean boosts (see 
\eqref{eq:invariancetau0_a}-\eqref{eq:rotatione0^a}). The construction of covariant derivatives containing two types of connections, one for local tangent space transformations and for coordinate transformations together with the corresponding vielbein postulates is discussed in subsection \ref{subsec:vielbeinpostulate}. Here we will see for the first time the important role played by torsion. The vielbein postulates relate the two types of connections. 

In section \ref{subsec:Gamma0} we choose our connection $\Gamma_{(0)ab}^c$ for covariant derivatives of tensors that are inert under local tangent space transformations. The choice is naturally suggested by the null reduction of the 5-dimensional asymptotically locally AdS boundary geometry and amounts to taking $\Gamma_{(0)ab}^c$ to be of the same functional form as in Newton--Cartan geometry but with the important difference that we do not set to zero a specific torsion tensor. Hence the name torsional Newton--Cartan (TNC). Even though we obtained the boundary geometry for a specific model admitting a $z=2$ Lifshitz UV completion, nothing depends essentially on $z=2$ but rather on the fact that the local tangent space group is a contraction of the Lorentz group. Since this will be the case for any $z>1$ we expect the TNC geometry to be generic for Lifshitz holography. 

In section \ref{subsec:NewtonCartan} we discuss two important special cases namely twistless torsional Newton--Cartan (TTNC) obtained by taking $\tau_{(0)a}$ to be hypersurface orthogonal and Newton--Cartan (NC) obtained by taking $\tau_{(0)a}$ to be closed. For the case of TTNC (and thus automatically also for NC) we work out the geometry induced on the hypersurfaces to which $\tau_{(0)a}$ is orthogonal in section \ref{subsec:curvature}. It turns out that this is described by Riemannian geometry (there is no torsion in directions tangential to these hypersurfaces). The geometric notions defined in section \ref{sec:bdrygeom} enable us to write the Ward identities of our model in a covariant form, and furthermore play an important role in the study of the anisotropic Weyl anomaly.

{\bf Section 5}: The vevs are calculated in section \ref{sec:sourcesvevs}. The variation of the on-shell action is given in \eqref{eq:4dRen_Action}. The Ward identities for the vevs can be readily obtained by reducing the 5-dimensional PBH transformations. This is done in section \ref{subsec:Wardidentities4D} and gives rise to a set of algebraic and differential relations for the vevs and sources. The differential expressions will be written in a covariant form using the TNC geometry in section \ref{subsec:covWard}. The vevs appearing in the variation of the on-shell action \eqref{eq:4dRen_Action} can be related to the 5-dimensional vevs by dimensional reduction. This is done in section \ref{subsec:dimredvevs}. The relation between the 5- and 4-dimensional vevs shows a number of interesting features: i) it implies various additional Ward identities in 4 dimensions, and ii) the relation given in  \eqref{eq:tau}--\eqref{eq:tuu} is not invertible, i.e. we cannot express all 5-dimensional vevs in terms of 4-dimensional ones. The implications of this are discussed in section \ref{subsec:holographicreconstruction} which we will summarize shortly below. 

As regards to point i), namely the extra Ward identities, some of these are to be expected and are simply related to the fact that we work with a vielbein formalism so that there will be Ward identities related to the local tangent space transformations, i.e. the Galilean boosts and the $SO(2)$ rotations and this is indeed what we find. However we find one more Ward identity \eqref{eq:vevrelation2} whose origin we explain in section \ref{subsec:localtangentspacetrafos} and we show that it is intimately connected with the constraint that the reduction circle on the boundary is null. The relation \eqref{eq:vevrelation2} allows us to remove the term proportional to $\delta\Phi_{(0)}$ in the variation of the on-shell action \eqref{eq:4dRen_Action} leaving us with only unconstrained sources. Further sections \ref{subsec:localtangentspacetrafos} and \ref{subsec:gaugeandscalevevs} discuss the various transformation properties of the sources and vevs under the various local symmetries. This leads us to define the unique gauge and local tangent space invariant boundary stress-energy tensor $\mathcal{T}_{(0)b}^a$ as given in \eqref{Tshifted}. We derive the scale dimensions of its tangent space components which in \cite{Ross:2009ar,Ross:2011gu} have been referred to as the energy density, the momentum density, the energy flux and the stress. Interestingly the energy flux appears to be a dimension 5 operator. Nevertheless we are able to compute it because it should really be viewed as a contraction of a dimension 3 vev with a dimension 2 source.

{\bf Section 6}: The last section \ref{sec:properties} is a collection of various physical properties of the boundary stress-energy tensor. The Ward identity for the boundary stress-energy tensor is not of the form of a conservation equation. This is common for stress-energy tensors defined by variation with respect to vielbeins whenever there are vectors in the theory \cite{Hollands:2005ya}. The main difference between $\mathcal{T}_{(0)b}^a$ and the HIM \cite{Hollands:2005ya} boundary stress-energy tensor is due to the TNC boundary geometry, i.e. the fact that we cannot raise/lower indices and the presence of torsion. In section \ref{subsec:bdrycurrents} we derive necessary and sufficient conditions for there to be conserved currents obtained by contracting the boundary stress-energy tensor with some vector $K_{(0)}^a$. This leads to a set of conditions on $K_{(0)}^a$ that can be thought of as the analogue of the conformal Killing equation in the context of TNC. In section \ref{subsec:anomaly} we evaluate the anisotropic Weyl anomaly density $\mathcal{A}_{(0)}$ and show that it takes the form of a Ho\v{r}ava--Lifshitz (HL) action but with the important difference that it is defined on a TNC geometry as opposed to a Lorentzian geometry as is the case for HL theories. An important role here is played by the boundary gauge field which is necessary to make kinetic terms appearing in $\mathcal{A}_{(0)}$ invariant under local Galilean boosts. The expression for $\mathcal{A}_{(0)}$ contains three different types of terms: those that are zeroth order in derivatives, second order kinetic terms and fourth order spatial derivative terms. In section \ref{subsec:holographicreconstruction} we come back to the issue that not all components of the 5-dimensional vevs can be rewritten in terms of 4-dimensional vevs. In fact we show that there is a specific component of the 5-dimensional boundary stress-energy tensor that decouples from all the Ward identities upon reduction and that appears in the 4-dimensional FG expansion without a dual sources. This is due to the fact that we have turned its source off. This is once again related to the constraint coming from the fact that the reduction circle must remain null on the boundary in order to have a Lifshitz UV completion. From the point of view of perturbations around the $z=2$ Lifshitz space-time this corresponds to turning off an irrelevant deformation. If we were to allow for this irrelevant deformation we would obtain a UV completion that is of the hyperscaling violating type with $\theta=-1$ and $z=1$ as is shown in appendix \ref{subsec:h0uuneq0}.

{\bf Discussion and Appendices}: In the discussion section \ref{sec:discussion} we further elaborate on some of our findings and suggest future directions for research. Appendix \ref{sec:holoren5D} contains background material on the details of the 5-dimensional theory where section \ref{subsec:boundaryfoliations} can be used to convert our results to any 4-dimensional boundary ADM gauge. Appendices \ref{app:radialgauge} and \ref{app:reducedanomaly} contain calculational details that have been omitted in the main text regarding the transformation to radial gauge and the reduction of the 5-dimensional Weyl anomaly, respectively. Appendix \ref{app:reconstruction} collects the 4-dimensional Fefferman--Graham expansions written in metric (i.e. non-vielbein) language. Finally as already mentioned appendix \ref{subsec:h0uuneq0} deals with the second UV completion obtained by taking the reduction circle spacelike on the boundary.

\section{Background}\label{sec:background}

As this work is a continuation of earlier work \cite{Chemissany:2011mb,Chemissany:2012du} we briefly summarize the model used there. This also allows us to introduce notation and to motivate more precisely our interest in the structure of the sources and vevs of the theory, the corresponding Ward identities, and the boundary geometry.

\subsection{The model}\label{subsec:model}

The model that we use can be obtained by dimensional reduction of the following 5-dimensional action \cite{Donos:2010tu,Cassani:2011sv,Papadimitriou:2011qb}
\begin{equation}\label{eq:actionmodel}
S_{\text{ren}} =\frac{1}{2\kappa_5^2}\int_{\mathcal{M}}d^5x\sqrt{-\hat g}\left(\hat R+12-\frac{1}{2}\partial_{\hat\mu}\hat\phi\partial^{\hat\mu}\hat\phi-\frac{1}{2}e^{2\hat\phi}\partial_{\hat\mu}\hat\chi\partial^{\hat\mu}\hat\chi\right)+\frac{1}{\kappa_5^2}\int_{\partial\mathcal{M}}d^4x\sqrt{-\hat h}\hat K+S_{\text{ct}}\,,
\end{equation}
where $\kappa_5^2=8\pi G_5$ with $G_5$ the 5-dimensional Newton's constant and where $\hat h$ denotes the determinant of the metric on $\partial\mathcal{M}$. The action \eqref{eq:actionmodel} can be obtained by a Freund--Rubin compactification of type IIB supergravity. The AdS$_5$ length has been set equal to one. Throughout this paper we will denote 5-dimensional quantities/indices by putting a hat on them. The action $S_{\text{ct}}$ contains all the counterterms (see \eqref{eq:Sct1}--\eqref{eq:anomalycounterterm} for their explicit expressions). The equations of motion are
\begin{eqnarray}
\hat{\mathcal{E}}_{\hat\mu\hat\nu} & = & \hat G_{\hat\mu\hat\nu}-6\hat g_{\hat\mu\hat\nu}-\hat T^{\text{bulk}}_{\hat\mu\hat\nu}=0\,,\label{eq:Einsteineqs}\\
\hat{\mathcal{E}}_{\hat\phi} & = & \hat\square\hat\phi-e^{2\hat\phi}(\partial\hat\chi)^2=0\,,\label{eq:phieom}\\
\hat{\mathcal{E}}_{\hat\chi} & = & \hat\nabla_{\hat\mu}\left(e^{2\hat\phi}\partial^{\hat\mu}\hat\chi\right)=0\,,\label{eq:chieom}
\end{eqnarray}
where the bulk energy-momentum tensor is
\begin{equation}
\hat T^{\text{bulk}}_{\hat\mu\hat\nu} =
\frac{1}{2}\partial_{\hat\mu}\hat\phi\partial_{\hat\nu}\hat\phi+\frac{1}{2}e^{2\hat\phi}\partial_{\hat\mu}\hat\chi\partial_{\hat\nu}\hat\chi
-\frac{1}{4}\hat g_{\hat\mu\hat\nu}\left((\partial\hat\phi)^2+e^{2\hat\phi}(\partial\hat\chi)^2\right)\,.
\end{equation}

Dimensional reduction of the action \eqref{eq:actionmodel} can be performed
using the ansatz
\begin{eqnarray}
        d\hat{s}^{2} &=& \hat{g}_{\hat{\mu} \hat{\nu}} dx^{\hat{\mu}}dx^{\hat{\nu}} =
e^{-\Phi}g_{\mu \nu}dx^{\mu}dx^{\nu} + e^{2\Phi}\left(du + A_{\mu}dx^{\mu}
\right)^2
        \,,\label{eq:KKansatzmetric} \\
        \hat{\chi} &=& \chi + ku\,, \\
        \hat{\phi} &=& \phi\,,
\end{eqnarray}
where the four dimensional unhatted fields are independent of the fifth coordinate
$u$ which is periodically identified $u \sim u + 2\pi L$, giving
\begin{eqnarray}
S_{\text{ren}} & = & \frac{2\pi L}{2\kappa_{5}^{2}} \int d^{4}x \sqrt{-g}\left(R -
\frac{3}{2}\partial_\mu\Phi\partial^\mu\Phi -
\frac{1}{4}e^{3\Phi}F_{\mu\nu}F^{\mu\nu} -
\frac{1}{2}\partial_\mu\phi\partial^\mu\phi - \frac{1}{2}e^{2\phi}D_{\mu}\chi
D^{\mu}\chi - V \right) \nonumber \\
        && + \frac{2\pi L}{\kappa_{5}^{2}}\int d^{3}x\sqrt{-h} K + S_{\text{ct}}\,,\label{eq:4Daction}
\end{eqnarray}
where
\begin{eqnarray}
        D_{\mu}\chi & = &  \partial_{\mu}\chi-kA_{\mu}\,,\label{eq:covderchi} \\
        F_{\mu \nu} & = & \partial_{\mu}A_{\nu} - \partial_{\nu}A_{\mu}\,, \\
        V & = & \frac{k^2}{2}e^{-3\Phi+2\phi} - 12e^{-\Phi}\,,\label{eq:potential}
\end{eqnarray}
and we take $k \neq 0$. The reduced counterterm action was obtained in \cite{Chemissany:2012du} and is given in \eqref{eq:reducedcountertermaction}. The corresponding equations of motion of the reduced theory are then given by
\begin{eqnarray}
 \mathcal{E}_{\mu \nu} & = &G_{\mu \nu}  + \frac{1}{8}e^{3\Phi}g_{\mu
\nu}F_{\rho\sigma}F^{\rho\sigma}  - \frac{1}{2}e^{3\Phi}F_{\mu
\rho}F_{\nu}{}^{\rho}  + \frac{1}{4}e^{2\phi}g_{\mu \nu}D_\rho\chi D^\rho\chi -
\frac{1}{2}e^{2\phi}D_{\mu}\chi D_{\nu}\chi\nonumber \\
 & &+ \frac{3}{4}g_{\mu \nu}\partial_\rho\Phi\partial^\rho\Phi-
\frac{3}{2}\partial_{\mu}\Phi \partial_{\nu}\Phi  + \frac{1}{4}g_{\mu
\nu}\partial_\rho\phi\partial^\rho\phi- \frac{1}{2}\partial_{\mu} \phi
\partial_{\nu} \phi + \frac{1}{2}g_{\mu \nu} V\,,\label{eq:Einsteineqs4D} \\
 \mathcal{E}^{\nu} & = & \nabla_{\mu}\left(e^{3\Phi}F^{\mu \nu} \right) +
ke^{2\phi}D^{\nu}\chi\,,\label{eq:eomA}\\
 \mathcal{E}_{\Phi} & = & 3\square \Phi - \frac{3}{4}e^{3\Phi}F_{\mu\nu}F^{\mu\nu} +
\frac{3}{2}k^{2}e^{-3\Phi + 2\phi} - 12e^{-\Phi}\,, \label{eq:eomPhi}\\
 \mathcal{E}_{\phi} & = & \square \phi - e^{2\phi}D_\mu\chi D^\mu\chi -
k^{2}e^{-3\Phi+2\phi}\,, \label{eq:eomphi}\\
\mathcal{E}_\chi & = & \nabla_{\mu}\left(e^{2\phi}D^\mu\chi\right)\,.\label{eq:eomchi}
\end{eqnarray}

\subsection{Lifshitz space-times}\label{subsec:Lifshitzspace}

The equations \eqref{eq:Einsteineqs4D} to \eqref{eq:eomchi} admit the pure $z=2$ Lifshitz space-time as a solution,
\begin{eqnarray}
ds^2=g_{\mu\nu}dx^\mu dx^\nu & = & e^{\Phi_{(0)}}\left(\frac{dr^2}{r^2}-e^{-2\Phi_{(0)}}\frac{dt^2}{r^4}+\frac{1}{r^2}\left(dx^2+dy^2\right)\right)\,,\label{eq:Lifshitzmetric}\\
A=A_\mu dx^\mu & = & e^{-2\Phi_{(0)}}\frac{dt}{r^2}\,,\\
\Phi & = & \Phi_{(0)}=\phi_{(0)}+\log\frac{k}{2}\,,\label{eq:PhipureLifshitz}\\
\phi & = & \phi_{(0)}=\text{cst}\,.
\end{eqnarray}
From a 5-dimensional perspective this solution is a $z=0$ Schr\"odinger space-time \cite{Balasubramanian:2010uk,Donos:2010tu,Costa:2010cn} and reads
\begin{eqnarray}
d\hat s^2 & = & \frac{dr^2}{r^2}+\frac{1}{r^2}\left(2dtdu+dx^2+dy^2\right)+\frac{k^2}{4}g_s^2du^2\,,\label{eq:Schz=0}\\
\hat\phi & = & \hat\phi_{(0)}=\phi_{(0)}=\log g_s=\text{cst}\,,\\
\hat\chi & = & ku+\text{cst}\,.
\end{eqnarray}

For the remainder of this subsection we find it convenient to reintroduce the AdS$_5$ length parameter $l$. The supergravity approximation, i.e. small curvature and weak string coupling, requires
\begin{equation}\label{eq:AdS5conditions}
\frac{l}{l_s}\gg 1\,,\qquad g_s\ll1\,.
\end{equation}
The first condition is the usual limit of large 't Hooft coupling $l/l_s=\lambda^{1/4}$.
The radius of the circle over which we compactify from 5 to 4 dimensions is given by (in units of string length)
\begin{equation}
\frac{2\pi L_{\text{phys}}}{l_s}=\frac{1}{l_s}\int_0^{2\pi L}du \sqrt{g_{uu}}=\frac{1}{l_s}(2\pi L)\frac{lkg_s}{2}\,,
\end{equation}
where
\begin{equation}
e^{\Phi_{(0)}}=\frac{lkg_s}{2}\,.
\end{equation}
In order not to have any light string winding modes we demand that
\begin{equation}\label{eq:Lphysoverls}
\frac{L_{\text{phys}}}{l_s}=\frac{l}{l_s}\frac{Lkg_s}{2}\gg 1\,.
\end{equation}
Hence in order for the 5-dimensional supergravity approximation to hold (on a background with a circle) we need \eqref{eq:AdS5conditions} and \eqref{eq:Lphysoverls}. We will always assume these conditions to be satisfied. We conclude that the circle is spacelike everywhere in the bulk and can be taken large in units of $l_s$. Despite this, the circle on the boundary metric obtained by rescaling \eqref{eq:Schz=0} by $r^2$ and setting $r=0$ is null. This will have important consequences that will be discussed below.

In type IIB string theory the axion shift symmetry is broken to a symmetry under integer shifts (which here follows from single-valuedness of the axion wavefunction along $u$) so that
\begin{equation}
2\pi Lk\in\mathbb{Z}\,.
\end{equation}
It follows from \eqref{eq:Lphysoverls} that the physical size $L_{\text{phys}}$ of the compactification radius in units of the AdS length is given by
\begin{equation}
\frac{L_{\text{phys}}}{l}=L\frac{kg_s}{2}\,.
\end{equation}
so that the decompactification limit corresponds to
\begin{equation}
\frac{L_{\text{phys}}}{l}\gg 1\,.
\end{equation}

Next we consider the opposite regime with $\frac{L_{\text{phys}}}{l}\ll 1$ (with $l/l_s$ sufficiently large such that \eqref{eq:Lphysoverls} remains satisfied) and argue that this is the range in which the boundary theory becomes 3-dimensional. To this end we look at a probe scalar $\hat\varphi$ on the 5-dimensional background \eqref{eq:Schz=0} described by the equation
\begin{equation}
\left(\hat\square-m^2\right)\hat\varphi=0\,.
\end{equation}
Decomposing
\begin{equation}
\hat\varphi=\sum_{n}e^{inu/L}\varphi_n\,,
\end{equation}
where the $\varphi_n$ are complex valued we obtain for each $\varphi_n$ the equation\footnote{We incidentally note that a good probe equation of motion for a charged scalar field on Lifshitz involves a minimal coupling term to the background gauge field.}
\begin{equation}
\left(D_\mu D^\mu-m^2_{\text{Lif}}\right)\varphi_n=0\,,
\end{equation}
where
\begin{equation}
D_\mu=\partial_\mu-i\frac{n}{L}A_\mu\,,
\end{equation}
and
\begin{equation}
m^2_{\text{Lif}}=e^{-\Phi_{(0)}}\left(m^2+e^{-2\Phi_{(0)}}\frac{n^2}{L^2}\right)\,.
\end{equation}
Using that the Lifshitz radius $l_{\text{Lif}}$ is given by (see equation \eqref{eq:Lifshitzmetric})
\begin{equation}
l^2_{\text{Lif}}=l^2e^{\Phi_{(0)}}\,,
\end{equation}
we have
\begin{equation}
m^2_{\text{Lif}}l^2_{\text{Lif}}=m^2l^2+\frac{4n^2}{k^2g_s^2L^2}=m^2l^2+\frac{l^2n^2}{L^2_{\text{phys}}}\,.
\end{equation}
In order to stay well below the KK mass scale we thus need
\begin{equation}
\frac{L_{\text{phys}}}{l}\ll 1\,.
\end{equation}

Above the decompactification scale $kL\gg g_s^{-1}$ the theory is 4-dimensional $\mathcal{N}=4$ SYM in the background of a nontrivial theta angle (sourced by the axion). For $kL\sim g_{s}^{-1}$ we cannot ignore the KK modes and the theory is a DLCQ of $\mathcal{N}=4$ SYM but where the DLCQ is deformed by the axion flux. When $kL\ll g_s^{-1}$ the boundary theory is a 3-dimensional Lifshitz--Chern--Simons non-Abelian gauge theory \cite{Mulligan:2010wj}. Of course throughout we need $l/l_s$ sufficiently large and $g_s$ small.

\subsection{AlLif space-times and beyond}\label{subsec:AlLif}

It is instructive to look at \eqref{eq:Schz=0} from the point of view of a 5-dimensional Fefferman--Graham expansion. Since this was already done in \cite{Chemissany:2012du} we shall be brief.
To this end we write the 5-dimensional metric as
\begin{equation}
\label{5Dhhat}
        d\hat{s}^{2} =
\frac{dr^{2}}{r^2} + \hat{h}_{\hat{a}\hat{b}}dx^{\hat{a}}dx^{\hat{b}}   \  ,
\end{equation}
and using the general metric expansion \eqref{eq: sol metric gauge} we conclude by comparing with \eqref{eq:Schz=0} that we have
\begin{eqnarray}
\hat h_{(0)uu} & = & 0\,,\label{eq:hath00}\\
\hat h_{(2)uu} & = & \frac{k^2g_s^2}{4}\,.
\end{eqnarray}
The latter condition implies via \eqref{eq:h2ab} that for the $z=0$ Schr\"odinger space-time
\begin{equation}\label{eq:HSO}
\hat R_{(0)uu}=0\,.
\end{equation}
Using that $\partial_u$ is a null Killing vector and thus tangent to a null geodesic congruence it has been shown in \cite{Chemissany:2012du} that provided \eqref{eq:HSO} holds $\partial_u$ is hypersurface orthogonal.

In the pure Lifshitz solution of the previous subsection the two dilatons $\Phi$ and $\phi$ were constant and related via $\Phi-\phi=\log\tfrac{k}{2}$. When considering more general space-times a natural generalization of this would be to consider $\Phi$ and $\phi$ such that they asymptote to something of order $r^0$. We would thus demand that we have
\begin{eqnarray}
\Phi & = & \Phi_{(0)}+\ldots\,,\\
\phi & = & \phi_{(0)}+\ldots\,,
\end{eqnarray}
where the boundary values $\Phi_{(0)}$ and $\phi_{(0)}$ are arbitrary functions of the boundary coordinates. The reduction ansatz \eqref{eq:KKansatzmetric} tells us that
\begin{equation}
e^{2\Phi} = \hat h_{uu}\,.\label{eq:Phi}
\end{equation}
In order that $\Phi$ asymptotes in general to something of order $r^0$ we need that \eqref{eq:hath00} holds.

The case where \eqref{eq:hath00} is dropped is discussed in appendix \ref{subsec:h0uuneq0}. For a large part of the paper we do not consider this case because it does not contain the Lifshitz space-time of the previous subsection\footnote{In appendix \ref{subsec:h0uuneq0} we show that our 4-dimensional model does contain geometries that asymptote to a hyperscaling violating geometry with $\theta=-1$ and $z=1$. However in 4-dimensions one cannot continuously deform (while staying close to the UV at $r=0$) the class of solutions containing asymptotic $\theta=-1$ and $z=1$ space-times to the class of solutions containing asymptotic $\theta=0$ and $z=2$ space-times. We discuss the role of these two different UV theories further in section \ref{subsec:holographicreconstruction}.}. From the lowest order in the expansion of \eqref{eq:Phi} we obtain
\begin{equation}\label{eq:constraint4D}
e^{2\Phi_{(0)}}=-\frac{1}{2}\hat R_{(0)uu}+\frac{k^2}{4}e^{2\phi_{(0)}}\,,
\end{equation}
where we used \eqref{eq:h2ab}. It is therefore not possible for both the boundary values $\Phi_{(0)}$ and $\phi_{(0)}$ to fluctuate arbitrarily. This constraint can be viewed as a 4-dimensional analogue of the condition that the reduction circle, which is spacelike everywhere in the bulk, is null on the boundary, i.e. \eqref{eq:hath00}. As discussed in the previous subsection this is not equivalent to a standard DLCQ reduction of the boundary theory. We showed that there is a well-defined parameter regime in which the reduction is well-defined (consistent and within suitable parameter ranges in order for the low energy approximation to apply) and we can truncate the KK tower of massive particles with the boundary theory being described by a Lifshitz--Chern--Simons gauge theory. 

Since from the point of view of the boundary of the 5-dimensional AlAdS space-time the reduction is along a null circle we expect that the boundary structure of the 4-dimensional space-time shows non-relativistic structures. Indeed we will see later that we obtain Newton--Cartan boundary geometries as well as generalizations thereof. To this end it is very useful to introduce frame fields.

From the reduction ansatz \eqref{eq:KKansatzmetric} and \eqref{5Dhhat} we learn that the 4-dimensional metric
 can be written  as
\begin{equation}\label{eq:4Dmetric}
ds^2=e^\Phi\frac{dr^2}{r^2}+h_{ab}dx^a dx^b\,,
\end{equation}
where $r$ is the 5-dimensional radial gauge coordinate and where
\begin{equation}
h_{ab} = \left(\hat h_{uu}\right)^{1/2}\left(\hat{h}_{ab}-\frac{\hat{h}_{au}\hat{h}_{bu}}{\hat{h}_{uu}}\right)\,. \label{eq:red5dmetric} \\
\end{equation}
We write the 4D metric $h_{ab}$ in a frame field basis as follows
\begin{equation}\label{eq:4Dmetricinframefields}
h_{ab}=-e^{\underline{t}}_{a}e^{\underline{t}}_{b}+\delta_{\underline{i}\underline{j}}e^{\underline{i}}_{a}e^{\underline{j}}_{b}\,.
\end{equation}
Here and in the following we use the notation that underlined indices are (flat) tangent space indices and $\underline{a}= \{ \underline{t},
\underline{i}\}$ with $i=1,2$.
Since we impose \eqref{eq:hath00} it follows that the second term in \eqref{eq:red5dmetric} starts at order $r^{-4}$. Hence we have for the frame fields
\begin{eqnarray}
e^{\underline{t}} & = & r^{-2}e^{-\Phi_{(0)}/2}\tau_{(0)a}dx^a+\ldots\,,  \label{etexp} \\
e^{\underline{i}} & = & r^{-1}e^{\Phi_{(0)}/2}e_{(0)a}^{\underline{i}}dx^a+\ldots\,. \label{eiexp}
\end{eqnarray}
We have included specific powers of $e^{\Phi_{(0)}}$ that will prove very convenient\footnote{The reason for this is explained
below eq.~\eqref{eq:omega5}
under the heading ``local dilatations''. This choice is therefore justified a posteriori. }
in the analysis of the vevs in section \ref{sec:sourcesvevs}.

It can be shown that (see section \ref{subsec:4Dsources}),
\begin{equation}
\label{Rhat00}
\hat R_{(0)uu}=\frac{1}{2}\left(\varepsilon_{(0)}^{abc}\tau_{(0)a}\partial_b \tau_{(0)c}\right)^2\,,
\end{equation}
where
\begin{equation}
\varepsilon_{(0)}^{abc}=\epsilon^{\underline{a}\underline{b}\underline{c}}e^a_{(0)\underline{a}}e^b_{(0)\underline{b}}e^c_{(0)\underline{c}}=e_{(0)}^{-1}\epsilon^{abc}\,,
\end{equation}
with $\epsilon^{\underline{a}\underline{b}\underline{c}}$ and $\epsilon^{abc}$ the Levi-Civit\`a symbol in flat and curved indices, respectively and where $e_{(0)}$ is the determinant of $e_{(0)a}^{\underline{a}}$.
We take $\epsilon^{\underline{t}\underline{i}\underline{j}}=-\epsilon^{\underline{i}\underline{j}}$.
It follows that hypersurface orthogonality of $\tau_{(0)a}$, i.e. the vanishing of $\varepsilon_{(0)}^{abc}\tau_{(0)a}\partial_b \tau_{(0)c}$ is equivalent to hypersurface orthogonality of the null Killing vector $\partial_u$ with respect to the AlAdS boundary metric\footnote{We note that boundary metrics of 5-dimensional AAdS space-times that admit a hypersurface orthogonal null Killing vector also played a key role in the construction of 5-dimensional $z=2$ asymptotically Schr\"odinger space-times by using TsT transformations \cite{Hartong:2010ec}. This could be easily generalized to the case of AlAdS$_5$ space-times whose boundary metric admits a hypersurface orthogonal null Killing vector.}, i.e. the vanishing of $\hat R_{(0)uu}$.
Using the expression \eqref{Rhat00} we can rewrite the constraint \eqref{eq:constraint4D} as
\begin{equation}\label{eq:constraint4D2}
e^{2\Phi_{(0)}}=-\frac{1}{4}\left(\varepsilon_{(0)}^{abc}\tau_{(0)a}\partial_b \tau_{(0)c}\right)^2+\frac{k^2}{4}e^{2\phi_{(0)}}\,.
\end{equation}

Next we turn our attention to the KK vector of the 4D theory, which takes the form
\begin{eqnarray}
A_{r} & = & 0\,, \\
A_{a} & = & \frac{\hat{h}_{au}}{\hat{h}_{uu}}\,, \label{eq:Aa}
\end{eqnarray}
using the reduction ansatz \eqref{eq:KKansatzmetric}. We thus get
the expansion
\begin{equation}
A_a=r^{-2}e^{-2\Phi_{(0)}}\tau_{(0)a}+\ldots\,.
\end{equation}
after using the relations \eqref{eq:Phi} and \eqref{eq:red5dmetric}. We see that asymptotically the bulk gauge field is proportional to the timelike frame field $e^{\underline{t}}_a$.

We have so far discussed the sources that are the leading components of the frame fields and scalar fields and we observed that the leading term in the expansion of the KK vector is given in terms of those. There are two more sources that will play a role later on, and that are given here for completeness
\begin{eqnarray}\label{eq:bdrygaugefield}
A_a-e^{-3\Phi/2}e_{a}^{\underline{t}} & = & A_{(0)a}+\ldots\,,\\
\chi & = & \chi_{(0)}+\ldots\,,
\end{eqnarray}
where $A_{(0)a}$ is the boundary gauge field and $\chi_{(0)}$ the boundary axion. We will define AlLif and their deformations independently of what happens with $A_{(0)a}$ and $\chi_{(0)}$. The notion of boundary gauge field as defined in \eqref{eq:bdrygaugefield} has to the best of our knowledge been overlooked in the Lifshitz literature. It forms an essential part of the boundary geometry. This boundary geometry, as will be explained in section \ref{sec:bdrygeom}, will turn out to be Newton--Cartan geometry extended with a certain torsion tensor. The geometrical role of the boundary gauge field will be further discussed in section \ref{subsec:localtangentspacetrafos}.

Consider the following two types of asymptotic structures:
\begin{enumerate}
\item AlLif: $\Phi_{(0)}-\phi_{(0)}=\log\tfrac{k}{2}$.
\item Lifshitz UV: no extra conditions other than \eqref{eq:constraint4D}.
\end{enumerate}
The condition that $\Phi_{(0)}-\phi_{(0)}=\log\tfrac{k}{2}$ is via \eqref{eq:constraint4D2} equivalent to the hypersurface orthogonality condition
\begin{equation}\label{eq:HSOtau0}
\tau_{(0)[a}\partial_{b}\tau_{(0)c]}=0
\end{equation}
for $\tau_{(0)a}$. In section \ref{subsec:NewtonCartan} we will see that the boundary geometry is torsional Newton--Cartan geometry with torsion proportional to $\partial_a\tau_{(0)b}-\partial_b\tau_{(0)a}$.

A special subclass of AlLif boundary conditions is obtained by setting
\begin{equation}\label{eq:NC}
\partial_a\tau_{(0)b}-\partial_b\tau_{(0)a}=0\,.
\end{equation}
As shown in section \ref{subsec:NewtonCartan} this gives rise to Newton--Cartan boundary geometry (i.e. without torsion). We can then without loss of generality choose coordinates such that $\tau_{(0)a}=\partial_a t$. It also corresponds to a class of space-times for which the Lifshitz scale transformation is still an asymptotic symmetry. We will refer to this subset as asymptotically Lifshitz space-times (ALif).

The proper time between two events connected by some path $\gamma$ is given by $\int_\gamma\tau_{(0)}$. When $\tau_{(0)a}$ is hypersurface orthogonal we can choose a coordinate system in which $\tau_{(0)i}=0$ so that $\tau_{(0)a}=\tau_{(0)t}\partial_a t$. This is an ADM decomposition in which surfaces of constant $t$ describe absolute simultaneity. If furthermore \eqref{eq:NC} is satisfied $t$ becomes absolute time.

We summarize the various asymptotic structures in the table \ref{table}.
\begin{table}[h!]
      \centering
      \begin{tabular}{|c|c|c|c|}
      \hline
Asymptotics & $\tau_{(0)}\wedge d\tau_{(0)}$  & $d\tau_{(0)}$ & Boundary Geometry\\
             \hline
             ALif & $0$  & $0$ & NC\\
             \hline
             AlLif & $0$  & $\neq 0$ & TTNC\\
            \hline
            Lif UV & $\neq 0$  & $\neq 0$ & TNC\\
             \hline
      \end{tabular}
      \caption{Indicated are the 3 different boundary conditions discussed in the text depending on the behavior of $\tau_{(0)}$. The last column indicates the type of boundary geometry.}\label{table}
\end{table}
In the last column we have indicated the type of boundary geometry that corresponds to the boundary conditions where NC denotes Newton--Cartan, TNC torsional Newton--Cartan and TTNC twistless torsional Newton--Cartan. These concepts will be defined in section \ref{sec:bdrygeom}.

In the remainder of this paper we will use the most general boundary conditions, i.e. the ones we call the Lifshitz UV. These include all deformations that take one away from AlLif boundary conditions. The goal will be to study the boundary geometry and compute the vevs and their Ward identities for this most general case. We will in particular focus our attention on computing the boundary stress energy tensor as defined by \cite{Hollands:2005ya,Ross:2009ar} and its Ward identities.

The definition of an AlLif space-time as given in \cite{Ross:2011gu} uses a radial gauge. Here we find it more convenient to work in the gauge \eqref{eq:4Dmetric}. In the next section we will consider the problem of transforming to radial gauge.

\subsection{Radial gauge}\label{subsec:radialgaugeEframe}

It is common practice to study solutions to the equations of motion admitting Lifshitz solutions in Einstein frame in radial gauge. To this end we will study the problem of rewriting our non-radial gauge Einstein frame metric \eqref{eq:4Dmetric} in radial gauge, i.e. we wish to perform the following coordinate transformation
\begin{equation}\label{eq:radialgaugecoordinatetrafo}
ds^2=e^\Phi\frac{dr^2}{r^2}+h_{ab}dx^adx^b=l^2_{\text{Lif}}\left(\frac{dr'^2}{r'^2}+h'_{ab}dx'^adx'^b\right)\,,
\end{equation}
where $l^2_{\text{Lif}}$ is the Lifshitz radius. To do this in full generality is prohibitively difficult so we will restrict ourselves to infinitesimal coordinate transformations. To this end we will assume that $\Phi$ can be approximated by
\begin{eqnarray}
\Phi & = & 2\log l_{\text{Lif}}+\delta\Phi\,,\\
\delta\Phi & = & \epsilon\delta_{[1]}\Phi+\frac{1}{2}\epsilon^2\delta_{[2]}\Phi+O(\epsilon^3)\,,
\end{eqnarray}
where $\epsilon$ is some small expansion parameter similar to the expansion parameter that would be used when studying perturbations around a Lifshitz background. To achieve the desired coordinate transformation we transform the left hand side of \eqref{eq:radialgaugecoordinatetrafo} using
\begin{equation}
r=r'-\xi^r(r',x')+\frac{1}{2}\xi^\nu\partial'_\nu\xi^r+O(\epsilon^3)\,,\qquad x^a=x'^a-\xi^a(r',x')+\frac{1}{2}\xi^\nu\partial'_\nu\xi^a+O(\epsilon^3)\,,
\end{equation}
where we expand $\xi^\mu$ as
\begin{equation}
\xi^\mu=\epsilon\xi^\mu_{[1]}+\frac{1}{2}\epsilon^2\xi^\mu_{[2]}+O(\epsilon^3)\,.
\end{equation}
For further details we refer the reader to appendix \ref{app:radialgauge} and we proceed by stating the end result of the calculation presented in that appendix.
Expressing the radial gauge metric $h'_{ab}$ in terms of $h_{ab}$ and the functions appearing in the expansion of $\xi^\mu$ we obtain
\begin{eqnarray}
h'_{ab} & = & l^{-2}_{\text{Lif}}\left(h_{ab}-\epsilon\left(\xi^r_{[1]}\partial_r h_{ab}+L_{\xi_{[1]}}h_{ab}\right)-\frac{1}{2}\epsilon^2\left(\xi^r_{[2]}\partial_rh_{ab}+L_{\xi_{[2]}}h_{ab}-\xi^r_{[1]}\partial_r\left(\xi^r_{[1]}\partial_rh_{ab}\right)
\right.\right.\nonumber\\
&&\left.\left.-\xi_{[1]}^r\partial_r\left(L_{\xi_{[1]}}h_{ab}\right)-L_{\xi_{[1]}}\left(\xi^r_{[1]}\partial_r h_{ab}\right)-L_{\xi_{[1]}}L_{\xi_{[1]}}h_{ab}\right)+O(\epsilon^3)\right)\,,\label{eq:hprimeabbody}
\end{eqnarray}
where $L_{\xi_{[1]}}$ and $L_{\xi_{[2]}}$ denote the Lie derivative along $\xi_{[1]}^a$ and $\xi_{[2]}^a$, respectively. Here the generators of the infinitesimal coordinate transformation admit the following $r$ expansions
\begin{eqnarray}
\xi^r_{[1]} & = & r\left(\log r\xi^r_{[1](0,1)}+\xi^r_{[1](0)}\right)+O(r^3\log^2 r)\,,\\
\xi^a_{[1]} & = & \xi^a_{[1](0)}+O(r^2\log r)\,,\\
\xi^r_{[2]} & = & r\left(\log r\xi^r_{[2](0,1)}+\xi^r_{[2](0)}\right)+O(r^3\log^2 r)\,,\\
\xi^a_{[2]} & = & \xi^a_{[2](0)}+O(r^2\log r)\,,
\end{eqnarray}
where
\begin{eqnarray}
\frac{1}{2}\delta_{[1]}\Phi_{(0)} & = & \xi^r_{[1](0,1)}\,,\\
\frac{1}{2}\delta_{[2]}\Phi_{(0)} & = & \xi^r_{[2](0,1)}+\xi^a_{[1](0)}\partial_a\xi^r_{[1](0,1)}\,.
\end{eqnarray}

Constructing $h'_{ab}$ by using \eqref{eq:hprimeabbody} we find that at second order in $\epsilon$ there is a term of the form $r^{-4}\log^2 r$ coming from the $\xi^r_{[1]}\partial_r\left(\xi^r_{[1]}\partial_rh_{ab}\right)$ term. At first order in $\epsilon$ we find a term of the form $r^{-4}\log r$. More precisely at each order $\epsilon^n$ we find a coefficient which is a polynomial in $\log r$ of order $n$. This means that we cannot use the $r$-expansion of the metric in radial gauge as a near boundary expansion as long as we work perturbatively in $\epsilon$ since each higher order in $\epsilon$ leads to a more dominant near boundary term. Hence in order to know the radial gauge expansion of the metric of a AlLif or a Lif UV space-time with $\partial_a\Phi_{(0)}\neq 0$ we need to be able to sum to all orders in $\epsilon$ or alternatively be able to construct the expansion directly in radial gauge without reference to the expansion obtained by dimensional reduction from a 5-dimensional radial gauge.

Summing the $\epsilon$ expansion has been done for purely radial perturbations in \cite{Baggio,Holsheimer:2013ula} in the context of the massive vector model where it is shown that the resummation leads to negative powers of $\log r$ in agreement with what has been observed in \cite{Cheng:2009df}{}\footnote{We thank Kristian Holsheimer and Marco Baggio for useful discussions on this point.}.

It is important to stress that the variation $\delta\Phi$ in \eqref{eq:perturbationPhi} is non-constant. All constant terms at order $r^0$ have been absorbed in the Lifshitz radius. This means that $\delta\Phi$ is either a non-trivial function of the boundary coordinates and starts at order $r^0$ or it goes to zero as $r$ goes to zero and starts at some higher order in $r$. In the latter case the leading $r^{-4}$ terms in the $\epsilon$ expansion of $h'_{ab}$ do not receive logarithmic corrections. We therefore expect that the situation here is qualitatively different from
\cite{Baggio,Cheng:2009df} as we see no log deformations of the $r^{-4}$ term by going to radial gauge in the case of purely radial solutions.

\section{The Lifshitz UV Completion}\label{sec:UVcompletion}

In this section we will obtain the most general boundary conditions compatible with the constraint \eqref{eq:constraint4D}, which will determine the Lifshitz UV completion, denoted by Lif UV. This will be accomplished by working with frame fields and relating the 4-dimensional ones to those of the 5-dimensional theory.  In particular, we will see that it is only in terms of frame fields that the sources are always the leading components in the expansions. Later we will see that they are furthermore very useful in order to describe the boundary geometry and for the computation of the boundary stress-energy tensor.

\subsection{Frame fields}\label{subsec:framefields}

Consider the following frame field decomposition of the 5-dimensional metric
\begin{equation}
d\hat s^2=\frac{dr^2}{r^2}+\left(-\hat e^{+}_{\hat a}\hat e^{-}_{\hat b}-\hat e^{+}_{\hat b}\hat e^{-}_{\hat a}+\delta_{\underline{i}\underline{j}}\hat e^{\underline{i}}_{\hat a}\hat e^{\underline{j}}_{\hat b}\right)dx^{\hat a} dx^{\hat b}\,,
\end{equation}
where $\underline{i}=1,2$. Using the reduction ansatz \eqref{eq:Phi}, \eqref{eq:red5dmetric} and \eqref{eq:Aa} and the 4D frame field decomposition \eqref{eq:4Dmetricinframefields} we can relate the 5- and 4-dimensional frame fields via
\begin{eqnarray}
\hat e^{+}_u & = & -\hat e^{-}_u=\frac{1}{\sqrt{2}}e^\Phi\,,\label{eq:framechoice1}\\
\hat e^{\underline{i}}_u & = & 0\,,\label{eq:framechoice2}\\
\hat e^{\underline{i}}_a & = & e^{-\Phi/2}e^{\underline{i}}_a\,,\\
\hat e^{+}_a & = & \frac{1}{\sqrt{2}}e^\Phi\left(A_a+e^{-3\Phi/2}e^{\underline{t}}_a\right)\,,\label{eq:framechoice4}\\
\hat e^{-}_a & = & -\frac{1}{\sqrt{2}}e^\Phi\left(A_a-e^{-3\Phi/2}e^{\underline{t}}_a\right)\,.   \label{eq:framechoice5}
\end{eqnarray}
For the inverse frame fields we have
\begin{eqnarray}
\hat e^u_{+} & = & -\frac{1}{\sqrt{2}}e^{\Phi/2}\left(A_a-e^{-3\Phi/2}e_a^{\underline{t}}\right)e^a_{\underline{t}}\,,\label{eq:invframe1}\\
\hat e^u_{-} & = &-\frac{1}{\sqrt{2}}e^{\Phi/2}\left(A_a+e^{-3\Phi/2}e_a^{\underline{t}}\right)e^a_{\underline{t}}\,,\\
\hat e^u_{\underline{i}} & = & -e^{\Phi/2}A_a e^a_{\underline{i}}=-e^{\Phi/2}\left(A_a-e^{-3\Phi/2}e_a^{\underline{t}}\right)e^a_{\underline{i}}\,,\\
\hat e^a_{+} & = & \hat e^a_{-}=\frac{1}{\sqrt{2}}e^{\Phi/2}e^a_{\underline{t}}\,,\\
\hat e^a_{\underline{i}} & = & e^{\Phi/2}e^a_{\underline{i}}\,.\label{eq:invframe4}
\end{eqnarray}
Because of our choice of frame \eqref{eq:framechoice1} and \eqref{eq:framechoice2} we have
\begin{eqnarray}
\hat h_{ab} & = &  -\hat e^{+}_{a}\hat e^{-}_{b}-\hat e^{+}_{b}\hat e^{-}_{a}+\delta_{\underline{i}\underline{j}}\hat e^{\underline{i}}_{a}\hat e^{\underline{j}}_{b}\,,\label{eq:hat hab}\\
\hat h_{au} & = &  \hat e^{+}_{u}\left(\hat e^{+}_{a}-\hat e^{-}_{a}\right)\,,\label{eq:hat hau}\\
\hat h_{uu} & = & 2\hat e^{+}_{u}\hat e^{+}_{u}\, ,\label{eq:hat huu}
\end{eqnarray}
for the 5D metric expressed in terms of the 5D frame fields.

\subsection{Boundary conditions}\label{subsec:bdryconditions}

We now turn to the boundary conditions obeyed by the 5D frame fields.  It will be convenient to choose
\begin{equation}\label{eq:boundarycondition}
\hat e^{+}_a = \frac{1}{r^2}\hat e^{+}_{(0)a}+\ldots\,.
\end{equation}
Then we must take
\begin{equation}
\hat e^{-}_a = \hat e^{-}_{(0)a}+\ldots\,,
\end{equation}
in order that $\hat h_{ab}$ in \eqref{eq:hat hab} is $O(r^{-2})$. It also implies that we must take
\begin{equation}
\hat e^{+}_u = -\hat e^{-}_u=\hat e^{+}_{(0)u}+\ldots\,,
\label{epem}
\end{equation}
in order that $\hat h_{au}$ in \eqref{eq:hat hau} is $O(r^{-2})$. This implies using \eqref{eq:hat huu} that $\hat h_{uu}=O(1)$ so that
\begin{equation}\label{eq:uunull}
\hat h_{(0)uu}=0\,.
\end{equation}
We furthermore take
\begin{equation}
\hat e^{\underline{i}}_a = \frac{1}{r}\hat e^{\underline{i}}_{(0)a}+\ldots\,,
\end{equation}
to preserve manifest tangent space $SO(2)$ rotation invariance at leading order, where we
also used that  $\hat h_{ab}$ in \eqref{eq:hat hab} is $O(r^{-2})$.

We thus see that the boundary condition \eqref{eq:boundarycondition} is well suited for arbitrary boundary metrics obeying \eqref{eq:uunull}. From \eqref{eq:hat huu} and \eqref{epem} this in turn has the consequence that we get the following constraint on the sources
\begin{equation}\label{eq:h2uu}
2\hat e^{+}_{(0)u}\hat e^{+}_{(0)u}=\hat h_{(2)uu}=-\frac{1}{2}\hat R_{(0)uu}+\frac{k^2}{4}e^{2\hat\phi_{(0)}}\,,
\end{equation}
where we used \eqref{eq:h2ab}. We will assume that
\begin{equation}\label{eq:positivityh2uu}
\hat h_{(2)uu}>0\,,
\end{equation}
so that $\hat e^{+}_{(0)u}\neq 0$. We note that because $\hat R_{(0)uu} \geq 0$, as will be shown in the next subsection (equation \eqref{eq:5DexpressionRuu}), the condition \eqref{eq:positivityh2uu} is in general non-trivial.

Including subleading terms we thus have for the 5D frame fields the expansions
\begin{eqnarray}
\hat e^{+}_u & = & \hat e^{+}_{(0)u}+r^2\log r\hat e^{+}_{(2,1)u}+r^2\hat e^{+}_{(2)u}+O(r^4\log^2 r)\,,\label{eq:expeflatv-u}\\
\hat e^{+}_a & = & \frac{1}{r^2}\hat e^{+}_{(0)a}+\log r \hat e^{+}_{(2,1)a}+\hat e^{+}_{(2)a}+O(r^2\log^2 r)\,,\label{eq:expeflatv-a}\\
\hat e^{-}_a & = & \hat e^{-}_{(0)a}+r^2\log r \hat e^{-}_{(2,1)a}+r^2\hat e^{-}_{(2)a}+O(r^4\log^2 r)\,,\label{eq:expeflatu-a}\\
\hat e^{\underline{i}}_a & = & \frac{1}{r}\hat e^{\underline{i}}_{(0)a}+r\hat e^{\underline{i}}_{(2)a}+O(r^3\log r)\,,\label{eq:expeflati-u}
\end{eqnarray}
where the coefficients can be computed by using \eqref{eq:hat hab}--\eqref{eq:hat huu} and the expansions given in appendix \ref{ref:FGexpansions}.

The expansion of the inverse frame fields starts as
\begin{eqnarray}
\hat e^u_{+} & = & r^2\hat e^u_{(0)+}+\ldots\,,\label{eq:expinvframe1}\\
\hat e^a_{+} & = & \hat e^a_{-}=r^2\hat e^a_{(0)+}+\ldots\,,\\
\hat e^u_{-} & = & \hat e^u_{(0)-}+\ldots\,,\\
\hat e^u_{\underline{i}} & = & r\hat e^u_{(0)\underline{i}}+\ldots\,,\\
\hat e^a_{\underline{i}} & = & r\hat e^a_{(0)\underline{i}}+\ldots\,,\label{eq:expinvframe5}
\end{eqnarray}
with the relations
\begin{eqnarray}
\hat e^u_{(0)-} & = & -(\hat e^{+}_{(0)u})^{-1}\,,\label{eq:hat e u up underline{u}}\\
\hat e^u_{(0)+} & = & (\hat e^{+}_{(0)u})^{-1}\hat e^a_{(0)+}\hat e^{-}_{(0)a}\,,\\
\hat e^u_{(0)\underline{i}} & = & (\hat e^{+}_{(0)u})^{-1}\hat e^a_{(0)\underline{i}}\hat e^{-}_{(0)a}\,,\\
\hat e^a_{(0)+}\hat e^{+}_{(0)a} & = & 1\,,\\
\hat e^a_{(0)\underline{i}}\hat e^{+}_{(0)a} & = & 0\,,\\
\hat e^a_{(0)+}\hat e^{\underline{i}}_{(0)a} & = & 0\,,\\
\hat e^a_{(0)\underline{j}}\hat e^{\underline{i}}_{(0)a} & = & \delta^{\underline{i}}_{\underline{j}}\,.
\end{eqnarray}

\subsection{The 4-dimensional sources}\label{subsec:4Dsources}

To obtain the 4D sources, we now use the expansions for the 5D
vielbeins and their inverse
\eqref{eq:expeflatv-u}-\eqref{eq:expinvframe5} along with their
relations \eqref{eq:framechoice1}-\eqref{eq:invframe4} with the 4D
vielbeins, to write  leading components of the 5D frame fields in
terms of 4D quantities. For the vielbeins this gives
\begin{eqnarray}
\hat e_{(0)u}^{+} & = & \frac{1}{\sqrt{2}}e^{\Phi_{(0)}}\,,\label{eq:Phi04d5d}\\
\hat e_{(0)a}^{+} & = & \sqrt{2}e^{-\Phi_{(0)}}\tau_{(0)a}\,,\label{eq:eflatt-a}\\
\hat e_{(0)a}^{-} & = & -\frac{1}{\sqrt{2}}e^{\Phi_{(0)}}A_{(0)a}\,,\label{eq:A0a}\\
\hat e_{(0)a}^{\underline{i}} & = & e_{(0)a}^{\underline{i}}\,,\label{eq:e0ia}
 \end{eqnarray}
while for the inverse vielbeins one finds
 \begin{eqnarray}
\hat e^{u}_{(0)+} & = & -\frac{1}{\sqrt{2}}e^{\Phi_{(0)}}A_{(0)\underline{t}}  \,,\label{eq:varphi0-}\\
\hat e^{u}_{(0)-} & = & -\sqrt{2}e^{-\Phi_{(0)}}  \,,\label{eq:Phi0andpsi0}\\
\hat e^u_{(0)\underline{i}} & = & -A_{(0)\underline{i}}  \,,\label{eq:A0underlinei}\\
\hat e^a_{(0)+} & = & -\frac{1}{\sqrt{2}}e^{\Phi_{(0)}}v_{(0)}^a  \,,\label{eq:inve0t}\\
\hat e^a_{(0)\underline{i}} & = & e^a_{(0)\underline{i}}  \,,\label{eq:inve0i}
\end{eqnarray}
where
\begin{eqnarray}
\tau_{(0)a}v_{(0)}^a & = & -1\,,   \label{sourceinv1} \\
\tau_{(0)a}e_{(0)\underline{i}}^a & = & 0\,,\\
e_{(0)a}^{\underline{i}}v_{(0)}^a & = & 0\,,\\
e_{(0)a}^{\underline{i}}e_{(0)\underline{j}}^a & = & \delta^{\underline{i}}_{\underline{j}}\,,\\
A_{(0)a} & = & A_{(0)\underline{t}}\tau_{(0)a}+A_{(0)\underline{i}}e_{(0)a}^{\underline{i}} \label{sourceinv5} \,.
\end{eqnarray}
Focussing on the inverse vielbein relations, the equations above then define the corresponding 4-dimensional sources to be 
$v_{(0)}^a, e^a_{(0)\underline{i}}, \Phi_{(0)}, A_{(0)\underline{t}}, A_{(0)\underline{i}}$. For boundary vectors and frame field components we use the notation
$X_{(0)\underline{t}}=-X_{(0)a}v_{(0)}^a$, $ X_{(0)\underline{i}}=X_{(0)a}e_{(0)\underline{i}}^a$
and $X_{(0)a}=X_{(0)\underline{t}}\tau_{(0)a}+X_{(0)\underline{i}}e_{(0)a}^{\underline{i}}$.

Using these results we can then write the 5-dimensional boundary metric\footnote{We warn the reader that because of our vielbein boundary conditions the 5-dimensional sources $\hat e_{(0)u}^+$ and $\hat e_{(0)a}^+$ do not transform as components of a 5-dimensional vector.} $\hat h_{(0)\hat a\hat b}$ in terms of the 4-dimensional sources as
\begin{eqnarray}
\hat h_{(0)ab} & = & -\hat e^{+}_{(0)a}\hat e^-_{(0)b}-\hat e^{+}_{(0)b}\hat e^-_{(0)a}+\delta_{\underline{i}\underline{j}}\hat e^{\underline{i}}_{(0)a}\hat e^{\underline{j}}_{(0)b}=\tau_{(0)a}A_{(0)b}+\tau_{(0)b}A_{(0)a}+\Pi_{(0)ab}\,,\label{eq:h0abin4dsources}\\
\hat h_{(0)au} & = & \hat e^+_{(0)u}\hat e^+_{(0)a}=\tau_{(0)a}\,,\label{eq:h0auin4dsources}\\
\hat h_{(0)uu} & = & 0\,.
\end{eqnarray}
Likewise for the inverse boundary metric we can write\footnote{Below \eqref{eq:bdrygaugefield} we remarked that $A_{(0)a}$ is the leading component of $A_a-e^{-3\Phi/2}e_a^{\underline{t}}$. It can also be viewed as part of the leading component of $\hat h^{uu}=e^{\Phi}\left(h^{ab}A_aA_b+e^{-3\phi}\right)$ which starts at order $r^2$ with a coefficient given by $\hat h_{(0)}^{uu}$ whose value is determined by $A_{(0)a}$.}
\begin{eqnarray}
\hat h_{(0)}^{ab} & = & \Pi_{(0)}^{ab}\,,\label{eq:invh0ab}\\
\hat h_{(0)}^{au} & = & -v_{(0)}^a-\delta^{\underline{i}\underline{j}}e^a_{(0)\underline{i}}A_{(0)\underline{j}}=-v_{(0)}^a-\Pi_{(0)}^{ab}A_{(0)b}\,,\label{eq:invh0au}\\
\hat h_{(0)}^{uu} & = & -2A_{(0)\underline{t}}+\delta^{\underline{i}\underline{j}}A_{(0)\underline{i}}A_{(0)\underline{j}}=2A_{(0)a}v_{(0)}^a+\Pi_{(0)}^{ab}A_{(0)a}A_{(0)b}\,.\label{eq:invh0uu}
\end{eqnarray}
In these expressions we have defined
\begin{eqnarray}
\Pi_{(0)ab} & = & \delta_{\underline{i}\underline{j}}e^{\underline{i}}_{(0)a}e^{\underline{j}}_{(0)b}\,,  \label{Pidown} \\
\Pi_{(0)}^{ab} & = & \delta^{\underline{i}\underline{j}}e^{a}_{(0)\underline{i}}e^{b}_{(0)\underline{j}}\,. \label{Piup}
\end{eqnarray}

We have thus identified the most general boundary conditions compatible with \eqref{eq:hath00} using the relation between the 4- and 5-dimensional frame fields given in section \ref{subsec:framefields}. In other words, we have obtained the most general
4-dimensional boundary conditions corresponding to the Lifshitz UV as defined in section \ref{subsec:AlLif}.


As we will see in the next subsection there is no Lorentzian boundary metric to raise and lower indices. This means that $\tau_{(0)a}$ and $v_{(0)}^a$ are two unrelated quantities apart from the condition that $\tau_{(0)a}v_{(0)}^a = -1$. This is especially clear from a 5-dimensional perspective. Comparing \eqref{eq:h0abin4dsources}--\eqref{eq:invh0uu} with the parametrization \eqref{eq:parametrizationh0} we obtain
\begin{eqnarray}
v_{(0)}^a & = & \hat N_{(0)}^{a}\,,\\
\tau_{(0)a} & = & \hat H_{(0)a}\,.\label{eq:tau0a}
\end{eqnarray}

We conclude by expressing $\hat R_{(0)uu}$ in terms of the 4-dimensional sources. To this end we compute $\hat R_{(0)uu}$ using the metric \eqref{eq:parametrizationh0} giving
\begin{equation}\label{eq:5DexpressionRuu}
\hat R_{(0)uu}=\frac{1}{2}\left(\varepsilon_{(0)}^{abc}\hat H_{(0)a}\partial_b\hat H_{(0)c}\right)^2\,,
\end{equation}
where
\begin{equation}\label{eq:bdrydeterminant}
\varepsilon_{(0)}^{abc}=\frac{\epsilon^{abc}}{H_{(0)}\sqrt{\text{det}\,\Sigma_{(0)}}}=\epsilon^{\underline{a}\underline{b}\underline{c}}e^a_{(0)\underline{a}}e^b_{(0)\underline{b}}e^c_{(0)\underline{c}}=e_{(0)}^{-1}\epsilon^{abc}\,,
\end{equation}
with $\epsilon^{abc}$ totally antisymmetric, $\epsilon^{txy}=-1$ and $\text{det}\,\Sigma_{(0)}$ the determinant of $\Sigma_{(0)ij}$. Using \eqref{eq:tau0a} we obtain for $\hat R_{(0)uu}$ the expression
\begin{equation}
\hat R_{(0)uu}=\frac{1}{2}\left(\varepsilon_{(0)}^{abc}\tau_{(0)a}\partial_b \tau_{(0)c}\right)^2\,.
\end{equation}
We thus see that $\hat R_{(0)uu}=0$ is equivalent to hypersurface orthogonality of $\tau_{(0)a}$ and since nothing depends on $u$ and one has that $\hat H_{(0)u}=0$ it follows that $\hat H_{(0)[\hat a}\partial_{\hat b}\hat H_{(0)\hat c]}=0$.

Now that we have defined the boundary conditions for the Lifshitz UV completion and we have obtained all the 4-dimensional sources it is possible to compute the variation of the on-shell action (using the reduced counterterms) and study the Ward identities. This analysis will be performed in section \ref{sec:sourcesvevs}.  We will first study the boundary geometry in the next section.

\section{Boundary Geometry}\label{sec:bdrygeom}

In this section we examine in detail  the boundary geometry of the
$z=2$  Lifshitz space-times of our model \eqref{eq:4Daction}. This
will enable us to identify for example a boundary covariant
derivative so that we can write covariant expressions for the Ward
identities. The boundary geometry will also play an important role
in our expression for the anomaly. In particular, we will show that
in the case of a $\tau_{(0)a}$ satisfying
$\partial_a\tau_{(0)b}-\partial_b\tau_{(0)a}=0$ the boundary
geometry is Newton--Cartan \cite{Cartan1,escidoc:153604}
(see also \cite{Misner1973}) and that a nonzero
$\partial_a\tau_{(0)b}-\partial_b\tau_{(0)a}$ corresponds to adding
torsion.

The appearance of Newton-Cartan structures is expected as the boundary geometry is obtained by null-dimensional reduction of the AdS boundary geometry. The relation between null dimensional reduction along a circle parametrized by $u$ of a space-time with a parallel (covariantly constant) $\partial_u$ and Newton--Cartan geometry has been studied in \cite{Eisenhart}. The covector $\hat h_{(0)\hat a\hat b}(\partial_u)^{\hat b}$ is equal to $\delta_{\hat a}^a\tau_{(0)a}$ as given by \eqref{eq:h0auin4dsources}. Since the 5-dimensional boundary metric $\hat h_{(0)\hat a\hat b}$ is only defined up to conformal rescalings we naturally need to be able to deal with various $\tau_{(0)a}$ that are all locally proportional to each other. Since the condition $\partial_a\tau_{(0)b}-\partial_b\tau_{(0)a}=0$ is not invariant under rescalings of $\tau_{(0)a}$ in such a way that the rescaled $\tau_{(0)a}$ is also curl free we are naturally confronted with studying geometries obtained by dimensional reduction along a null circle generated by a hypersurface orthogonal null Killing vector which is not necessarily parallel. Such cases have been looked at in \cite{Julia:1994bs} and from the work of \cite{Kuenzle:1972zw} it is expected that the connection on the reduced 3-dimensional boundary will have torsion. We will also study the more general case where $\partial_u$ is a null Killing vector but not necessarily hypersurface orthogonal.

\subsection{Contraction of the local Lorentz group}\label{subsec:contractionLorentzgroup}

To get an idea about the boundary geometry described by $\tau_{(0)a}$ and $e_{(0)a}^{\underline{i}}$ we study how bulk local Lorentz transformations act on the leading components of the frame fields. To this end we consider local Lorentz transformations transforming the $e^{\underline{a}}_a$ into each other, i.e. the group of $SO(2,1)$ rotations leaving $e^3\equiv e^{\Phi/2}\tfrac{dr}{r}$ invariant. Here $e^3$ is the radial part of \eqref{eq:KKansatzmetric}. The local Lorentz transformations on the 4D vielbeins read
\begin{eqnarray}
e^{\underline{t}}_a & = & \Lambda^{\underline{t}}{}_{\underline{t}'}e^{\underline{t}'}_a+\Lambda^{\underline{t}}{}_{\underline{i}'}e^{\underline{i}'}_a\,,\label{eq:LTreta}\\
e^{\underline{i}}_a & = &
\Lambda^{\underline{i}}{}_{\underline{t}'}e^{\underline{t}'}_a+\Lambda^{\underline{i}}{}_{\underline{i}'}e^{\underline{i}'}_a\,,\label{eq:LTreia}
\end{eqnarray}
where
\begin{eqnarray}
- \Lambda^{\underline{t}}{}_{\underline{t}'} \Lambda^{\underline{t}}{}_{\underline{t}'}+\delta_{\underline{i}\underline{j}}\Lambda^{\underline{i}}{}_{\underline{t}'}\Lambda^{\underline{j}}{}_{\underline{t}'} & = & -1\,,\label{eq:Lorentztrafo1}\\
- \Lambda^{\underline{t}}{}_{\underline{t}'} \Lambda^{\underline{t}}{}_{\underline{i}'}+\delta_{\underline{i}\underline{j}}\Lambda^{\underline{i}}{}_{\underline{t}'}\Lambda^{\underline{j}}{}_{\underline{i}'} & = & 0\,,\label{eq:Lorentztrafo2}\\
- \Lambda^{\underline{t}}{}_{\underline{i}'} \Lambda^{\underline{t}}{}_{\underline{j}'}+\delta_{\underline{i}\underline{j}}\Lambda^{\underline{i}}{}_{\underline{i}'}\Lambda^{\underline{j}}{}_{\underline{j}'} & = & \delta_{\underline{i}'\underline{j}'}\,.\label{eq:Lorentztrafo3}
\end{eqnarray}
The $r$-expansion of the 4D vielbein is given in \eqref{etexp} and \eqref{eiexp} and after the local Lorentz transformation
we have the same expansion, i.e.
\begin{eqnarray}
e^{\underline{t}'}_a & = & r^{-2}e^{-\Phi_{(0)}/2}\tau'_{(0)a}+\ldots\,,\label{eq:bdryconditionet}\\
e^{\underline{i}'}_a & = & r^{-1}e^{\Phi_{(0)}/2}e^{\underline{i}'}_{(0)a}+\ldots\,,\label{eq:bdryconditione1}\\
e^{\underline{t}}_a & = & r^{-2}e^{-\Phi_{(0)}/2}\tau_{(0)a}+\ldots\,,\label{eq:bdryconditione2}\\
e^{\underline{i}}_a & = & r^{-1}e^{\Phi_{(0)}/2}e^{\underline{i}}_{(0)a}+\ldots\,,\label{eq:bdryconditione3}
\end{eqnarray}
where we note that $\Phi_{(0)}$ does not transform. From
\eqref{eq:LTreta}--\eqref{eq:LTreia} it follows that we need to
require
\begin{eqnarray}
\Lambda^{\underline{t}}{}_{\underline{t}'} & = & \Lambda^{\underline{t}}_{(0)\underline{t}'}+\ldots\,,\\
\Lambda^{\underline{t}}{}_{\underline{i}'} & = & r^{-1}\Lambda^{\underline{t}}_{(0)\underline{i}'}+\ldots\,,\\
\Lambda^{\underline{i}}{}_{\underline{t}'} & = & r\Lambda^{\underline{i}}_{(0)\underline{t}'}+\ldots\,,\\
\Lambda^{\underline{i}}{}_{\underline{i}'} & = & \Lambda^{\underline{i}}_{(0)\underline{i}'}+\ldots\,.
\end{eqnarray}
Plugging this into \eqref{eq:Lorentztrafo1}--\eqref{eq:Lorentztrafo3} we get the following conditions
\begin{eqnarray}
\Lambda^{\underline{t}}_{(0)\underline{t}'}\Lambda^{\underline{t}}_{(0)\underline{t}'} & = & 1\,,\\
\Lambda^{\underline{t}}_{(0)\underline{i}'} & = & 0\,,\\
\delta_{\underline{i}\underline{j}}\Lambda^{\underline{i}}_{(0)\underline{i}'}\Lambda^{\underline{j}}_{(0)\underline{j}'} & = & \delta_{\underline{i}'\underline{j}'}\,,
\end{eqnarray}
on the leading components of $\Lambda_{(0)\underline{b}}^{\underline{a}}$.  We will choose
\begin{equation}
\Lambda^{\underline{t}}_{(0)\underline{t}'}=1\,,
\end{equation}
so that we can recover the identity.

We thus find the following transformation of the leading components of the frame field expansions
\begin{eqnarray}
\tau_{(0)a} & = & \tau'_{(0)a}\,,\label{eq:bdrytangentspacetrafo1}\\
e^{\underline{i}}_{(0)a} & = & \Lambda^{\underline{i}}_{(0)\underline{t}'}\tau'_{(0)a}+\Lambda^{\underline{i}}_{(0)\underline{i}'}e^{\underline{i}'}_{(0)a}\,,\label{eq:bdrytangentspacetrafo2}
\end{eqnarray}
where $\Lambda^{\underline{i}}_{(0)\underline{t}'}$ are two free parameters\footnote{The three generators of these transformations are $J, G_1, G_2$ whose nonzero commutators are $[J,G_1]=G_2$ and $[J,G_2]=-G_1$. We can think of this as the contraction of the Lorentz group $SO(1,2)$ in which the $G_i$ play the role of Galilean boost generators.}. The corresponding transformation acting on the leading components of the inverse frame fields reads
\begin{eqnarray}
v_{(0)}^a & = & v'{}_{(0)}^a+e^a_{(0)\underline{i}'}\Lambda_{(0)\underline{i}}^{\underline{i}'}\Lambda_{(0)\underline{t}'}^{\underline{i}}\,,\label{eq:bdrytangentspaceinvtrafo1}\\
e^a_{(0)\underline{i}} & = & e^a_{(0)\underline{i}'}\Lambda_{(0)\underline{i}}^{\underline{i}'}\,.  \label{eq:bdrytangentspaceinvtrafo2}
\end{eqnarray}
For use below, we also note the infinitesimal versions of the transformations \eqref{eq:bdrytangentspacetrafo1}--\eqref{eq:bdrytangentspaceinvtrafo2}, (also denoted by $\Lambda_{(0)}$ with the notation $\Lambda_{(0)\underline{t}}^{\underline{i}}=\Lambda_{(0)}^{\underline{i}}$) yielding
\begin{eqnarray}
\delta \tau_{(0)a} & = & 0\,,\label{eq:invariancetau0_a}\\
\delta e^{\underline{i}}_{(0)a} & = & \tau_{(0)a}\Lambda_{(0)}^b e_{(0)b}^{\underline{i}}+\Lambda_{(0)\underline{j}}^{\underline{i}}e^{\underline{j}}_{(0)a}\,,\label{eq:booste0_a}\\
\delta v_{(0)}^a & = & \Lambda_{(0)}^a\,,\label{eq:boosttau0^a}\\
\delta e^a_{(0)\underline{i}} & = & -\Lambda_{(0)\underline{i}}^{\underline{j}}e_{(0)\underline{j}}^a\,,\label{eq:rotatione0^a}
\end{eqnarray}
where
\begin{equation}
\Lambda_{(0)}^a=\Lambda_{(0)}^{\underline{i}}e_{(0)\underline{i}}^a\,,
\end{equation}
and
\begin{equation}
\Lambda_{(0)\underline{i}\underline{j}}=-\Lambda_{(0)\underline{j}\underline{i}}\,.
\end{equation}
The flat index $\underline{i}$ can be raised and lowered with $\delta_{\underline{i}\underline{j}}$.

We can build two degenerate metrics out of these vielbeins that are invariant under the local tangent space group. These are $\tau_{(0)a}\tau_{(0)b}$ and $\Pi_{(0)}^{ab}$. On top of that we have that the boundary determinant $e_{(0)}$ as defined in equation \eqref{eq:bdrydeterminant} is an invariant as well.

The fact that we see a contraction of the local Lorentz group can be understood by observing that the vielbein boundary conditions \eqref{eq:bdryconditionet}--\eqref{eq:bdryconditione3} lead to a flattening of the tangent space light cones as one approaches the boundary so that the effective speed of light approaches infinity, leading to a contraction of the tangent space Lorentz
transformations. A similar analysis which also leads to a contraction of the local Lorentz group was performed in the case of 3-dimensional asymptotically locally Schr\"odinger space-times in
\cite{Hartong:2013cba}.

\subsection{Covariant derivative and vielbein postulate}\label{subsec:vielbeinpostulate}

We will now construct covariant derivatives that transform covariantly with respect to the local tangent space transformations \eqref{eq:invariancetau0_a}--\eqref{eq:rotatione0^a}. We will denote these covariant derivatives by $\mathcal{D}_{(0)}^T$. The meaning of the superscript $T$ will become clear later. By covariance we mean that the following transformation rules must be obeyed
\begin{eqnarray}
\delta\left(\mathcal{D}_{(0)a}^T\tau_{(0)b}\right) & = & 0\,,\label{eq:cov1}\\
\delta\left(\mathcal{D}_{(0)a}^T e_{(0)b}^{\underline{i}}\right) & = & \tau_{(0)b}\Lambda_{(0)}^c\left(\mathcal{D}^T_{(0)a}e_{(0)c}^{\underline{i}}\right)+\Lambda_{(0)\underline{j}}^{\underline{i}}\left(\mathcal{D}^T_{(0)a}e_{(0)b}^{\underline{j}}\right)\,,\label{eq:cov2}\\
\delta\left(\mathcal{D}_{(0)a}^Tv_{(0)}^b\right) & = & 0\,,\label{eq:cov3}\\
\delta\left(\mathcal{D}_{(0)a}^T e_{(0)\underline{i}}^b\right) & = & -\Lambda_{(0)\underline{i}}^{\underline{j}}\left(\mathcal{D}_{(0)a}^T e_{(0)\underline{j}}^b\right)\,.\label{eq:cov4}
\end{eqnarray}

In order to construct $\mathcal{D}_{(0)a}^T$ we introduce the connections $\Gamma_{(0)bc}^{Ta}$ (not assumed to be symmetric), $\omega_{(0)b}{}^{\underline{i}}$ and $\omega_{(0)b}{}^{\underline{i}}{}_{\underline{j}}$ in the following way\footnote{The appearance of the $\omega_{(0)b}{}^{\underline{i}}$ and $\omega_{(0)b}{}^{\underline{i}}{}_{\underline{j}}$ connections correlates with the transformations \eqref{eq:invariancetau0_a}--\eqref{eq:rotatione0^a}.}
\begin{eqnarray}
\mathcal{D}^T_{(0)a} \tau_{(0)b} & = & \partial_a\tau_{(0)b}-\Gamma_{(0)ab}^{Tc}\tau_{(0)c}\,,\\
\mathcal{D}^T_{(0)a} e^{\underline{i}}_{(0)b} & = & \partial_a e^{\underline{i}}_{(0)b}-\Gamma_{(0)ab}^{Tc}e^{\underline{i}}_{(0)c}+\omega_{(0)a}{}^{\underline{i}}\tau_{(0)b}+\omega_{(0)a}{}^{\underline{i}}{}_{\underline{j}}e^{\underline{j}}_{(0)b}\,,\\
\mathcal{D}^T_{(0)a} e^b_{(0)\underline{i}} & = & \partial_a e^b_{(0)\underline{i}}+\Gamma_{(0)ac}^{Tb} e^c_{(0)\underline{i}}-\omega_{(0)a}{}^{\underline{j}}{}_{\underline{i}}e^b_{(0)\underline{j}}\,,\\
\mathcal{D}^T_{(0)a} v_{(0)}^b & = & \partial_av_{(0)}^b+\Gamma_{(0)ac}^{Tb}v_{(0)}^c+\omega_{(0)a}{}^{\underline{i}}e^b_{(0)\underline{i}}\,.
\end{eqnarray}
We will denote by $\nabla^T_{(0)a}$ the covariant derivative containing only the connection $\Gamma_{(0)ab}^{Tc}$. In order that \eqref{eq:cov1} is obeyed we need that $\Gamma_{(0)ab}^{Tc}$ is $SO(2)$ invariant and that under boosts it transforms such that 
\begin{equation}\label{eq:Gammaboost}
\tau_{(0)c}\delta\Gamma_{(0)ab}^{Tc}=0\,.
\end{equation}
In order that \eqref{eq:cov2} holds we need that $\omega_{(0)a}{}^{\underline{i}}{}_{\underline{j}}$ transforms as
\begin{equation}\label{eq:omegatrafo1}
\delta\omega_{(0)a}{}^{\underline{i}}{}_{\underline{j}}=-\partial_a\Lambda_{(0)\underline{j}}^{\underline{i}}+\Lambda_{(0)\underline{k}}^{\underline{i}}\omega_{(0)a}{}^{\underline{k}}{}_{\underline{j}}-\omega_{(0)a}{}^{\underline{i}}{}_{\underline{k}}\Lambda_{(0)\underline{j}}^{\underline{k}}\,,
\end{equation}
under local $SO(2)$ transformations and as 
\begin{equation}
\delta\omega_{(0)a}{}^{\underline{i}}{}_{\underline{j}} = -\Lambda_{(0)}^{\underline{i}}e^c_{(0)\underline{j}}\nabla^T_{(0)a}\tau_{(0)c}+e^c_{(0)\underline{j}}e^{\underline{i}}_{(0)d}\delta\Gamma_{(0)ac}^{Td}\,,
\end{equation}
under local boosts while $\omega_{(0)a}{}^{\underline{i}}$ must transform as
\begin{equation}
\delta\omega_{(0)a}{}^{\underline{i}}=\Lambda_{(0)\underline{j}}^{\underline{i}}\omega_{(0)a}{}^{\underline{j}}
\end{equation}
under local $SO(2)$ transformations and as
\begin{equation}\label{eq:omegatrafo4}
\delta\omega_{(0)a}{}^{\underline{i}}= \Lambda_{(0)}^c\mathcal{D}^T_{(0)a}e_{(0)c}^{\underline{i}}-\partial_a\Lambda_{(0)}^{\underline{i}}-\omega_{(0)a}{}^{\underline{i}}{}_{\underline{j}}\Lambda_{(0)}^{\underline{j}}+\Lambda_{(0)}^{\underline{i}} v_{(0)}^b\nabla^T_{(0)a}\tau_{(0)b}-e_{(0)d}^{\underline{i}}v_{(0)}^c\delta\Gamma_{(0)ac}^{Td}\,,
\end{equation}
under local boosts. With these transformations one can then show that we have
\begin{eqnarray}
\hspace{-.5cm} \delta\left(\mathcal{D}^T_{(0)a}e^b_{(0)\underline{i}}- v_{(0)}^b\left(\nabla^T_{(0)a}\tau_{(0)c}\right)e^c_{(0)\underline{i}}\right) & = & -\Lambda_{(0)\underline{i}}^{\underline{j}}\left(\mathcal{D}^T_{(0)a}e^b_{(0)\underline{j}}- v_{(0)}^b\left(\nabla^T_{(0)a}\tau_{(0)c}\right)e^c_{(0)\underline{j}}\right)\,,\\
\hspace{-.5cm} \delta\left(\mathcal{D}^T_{(0)a} v_{(0)}^b- v_{(0)}^b\left(\nabla^T_{(0)a}\tau_{(0)c}\right)v_{(0)}^c\right) & = & 0\,.
\end{eqnarray}
Hence in order to obey \eqref{eq:cov3} and \eqref{eq:cov4} we need that
\begin{equation}\label{eq:VP1}
\nabla^T_{(0)a}\tau_{(0)c}=0\,.
\end{equation}
This equation implies that in general we need a connection with torsion, hence the superscript $T$. We split $\Gamma_{(0)ab}^{Tc}$ into a symmetric and an anti-symmetric part as
\begin{equation}
\Gamma_{(0)ab}^{Tc}=\Gamma_{(0)ab}^{c}+T_{(0)ab}^c\,,
\end{equation}
where we denote torsion by $T_{(0)ab}^c$ and where $\Gamma_{(0)ab}^{c}$ is symmetric. We will denote by $\nabla_{(0)a}$ the covariant derivative containing the connection $\Gamma_{(0)ab}^{c}$. Taking the symmetric part of \eqref{eq:Gammaboost} we see that under boosts
\begin{equation}\label{eq:boostGamma}
\tau_{(0)c}\delta\Gamma_{(0)ab}^c=0\,.
\end{equation}

The vielbein postulate\footnote{We thank Matthias Blau for useful discussions on the meaning of the vielbein postulate in relation to demanding covariance with respect to local tangent, coordinate and frame-to-coordinate transformations.} that we will impose on top of \eqref{eq:VP1} is
\begin{eqnarray}
\mathcal{D}_{(0)}^T e^{\underline{i}}_{(0)b} & = & 0\,,\\
\mathcal{D}_{(0)a}^T e^b_{(0)\underline{i}} & = & 0\,,\\
\mathcal{D}_{(0)a}^T v_{(0)}^b & = & 0\,.
\end{eqnarray}
These conditions imply that
\begin{eqnarray}
\omega_{(0)a}{}^{\underline{i}} & = & -e_{(0)b}^{\underline{i}}\nabla^T_{(0)a} v_{(0)}^b\,,\\
\omega_{(0)a}{}^{\underline{i}}{}_{\underline{j}} & = & e_{(0)b}^{\underline{i}}\nabla^T_{(0)a}e_{(0)\underline{j}}^b\,,
\end{eqnarray}
which are compatible with the transformations \eqref{eq:omegatrafo1}--\eqref{eq:omegatrafo4}.

It follows from the vielbein postulates above, as well as the specific tangent space group and the symmetry of $\Gamma_{(0)ab}^c$ that the $\Gamma_{(0)ab}^c$ connection must satisfy
\begin{equation}\label{eq:conditionGamma0}
\Pi_{(0)b}^d\Pi_{(0)c}^e\nabla_{(0)a}\Pi_{(0)de}=0\,,
\end{equation}
where we defined the projector $\Pi_{(0)a}^b$ via
\begin{equation}
\Pi_{(0)a}^b=\delta_a^b+\tau_{(0)a} v_{(0)}^b=\Pi_{(0)ac}\Pi_{(0)}^{cb}\,.
\end{equation}
If we differentiate the completeness relation
\begin{equation}\label{eq:completeness}
\Pi_{(0)ab}\Pi_{(0)}^{bc}-\tau_{(0)a}v_{(0)}^c = \delta_a^c\,,
\end{equation}
we obtain from \eqref{eq:conditionGamma0} the relation
\begin{equation}
\Pi_{(0)d}^b\Pi_{(0)e}^c\nabla_{(0)a}\Pi_{(0)}^{de}=0\,.
\end{equation}

\subsection{The choice of $\Gamma_{(0)ab}^c$}\label{subsec:Gamma0}

Our choice of $\Gamma_{(0)ab}^c$ will be inspired by the null dimensional reduction of the boundary geometry. Consider the Christoffel connection of the non-degenerate 5-dimensional boundary metric $\hat h_{(0)\hat a\hat b}$ possessing a null Killing vector $\partial_u$ and take all its legs in the directions of the three non-compact directions. Using \eqref{eq:h0abin4dsources}--\eqref{eq:invh0uu} we decompose this quantity as follows
\begin{equation}
\hat\Gamma_{(0)bc}^a = \Gamma_{(0)bc}^a-\frac{1}{2}\Pi_{(0)}^{ad}\left[\left(\partial_d\tau_{(0)b}-\partial_b\tau_{(0)d}\right)A_{(0)c}+\left(\partial_d\tau_{(0)c}-\partial_c\tau_{(0)d}\right)A_{(0)b}\right]\,,
\end{equation}
where $\Gamma_{(0)bc}^a$ is given by
\begin{eqnarray}
\Gamma_{(0)bc}^a & = & -\frac{1}{2}v_{(0)}^a\left(\partial_b\tau_{(0)c}+\partial_c\tau_{(0)b}\right)+\frac{1}{2}\Pi_{(0)}^{ad}\left(\partial_b \Pi_{(0)cd}+\partial_c \Pi_{(0)bd}-\partial_d \Pi_{(0)bc}\right)\nonumber\\
&&-\frac{1}{2}\Pi_{(0)}^{ad}\left(F_{(0)db}\tau_{(0)c}+F_{(0)dc}\tau_{(0)b}\right)\,,\label{eq:Gamma0}
\end{eqnarray}
with $F_{(0)ab}=\partial_a A_{(0)b}-\partial_b A_{(0)a}$. This choice for $\Gamma_{(0)bc}^a$ is such that it takes the same functional form as in Newton--Cartan but with the important difference that we do not impose any properties on $\tau_{(0)a}$. The connection $\Gamma_{(0)bc}^a$ satisfies the following properties
\begin{eqnarray}
\Gamma_{(0)ac}^a & = & e_{(0)}^{-1}\partial_c e_{(0)}-\frac{1}{2}v_{(0)}^a\left(\partial_a\tau_{(0)c}-\partial_c\tau_{(0)a}\right)\,,\label{eq:derivativedet e0}\\
\nabla_{(0)a}\tau_{(0)b} & = & \frac{1}{2}\left(\partial_a\tau_{(0)b}-\partial_b\tau_{(0)a}\right)\,, \label{nabt} \\
\nabla_{(0)a} v_{(0)}^b & = & \frac{1}{2} v_{(0)}^bv_{(0)}^c\left(\partial_a\tau_{(0)c}-\partial_c\tau_{(0)a}\right)+\frac{1}{2}\Pi_{(0)}^{bc}\mathcal{L}_{v_{(0)}}\Pi_{(0)ac}\nonumber\\
&&-\frac{1}{2}\Pi_{(0)}^{bc}v_{(0)}^d\left(F_{(0)ca}\tau_{(0)d}+F_{(0)cd}\tau_{(0)a}\right)\,,\\
\nabla_{(0)a}\Pi_{(0)}^{bc} & = & \frac{1}{2}\left(\partial_a\tau_{(0)d}-\partial_d\tau_{(0)a}\right)\left(\Pi_{(0)}^{bd}v_{(0)}^c+\Pi_{(0)}^{cd} v_{(0)}^b\right)\,,\\
\nabla_{(0)a}\Pi_{(0)bc} & = & \frac{1}{2}\tau_{(0)b}\mathcal{L}_{v_{(0)}}\Pi_{(0)ac}+\frac{1}{2}\tau_{(0)c}\mathcal{L}_{v_{(0)}}\Pi_{(0)ab}\nonumber\\
&&+\frac{1}{2}\left[\Pi_{(0)c}^eF_{(0)ea}\tau_{(0)b}+\Pi_{(0)b}^eF_{(0)ea}\tau_{(0)c}+\Pi_{(0)c}^eF_{(0)eb}\tau_{(0)a}+\Pi_{(0)b}^eF_{(0)ec}\tau_{(0)a}\right]\nonumber\\
&=& \left(\tau_{(0)b}\Pi_{(0)cd}+\tau_{(0)c}\Pi_{(0)bd}\right)\nabla_{(0)a}v_{(0)}^d\,, \label{nabF}
\end{eqnarray}
where $\mathcal{L}_{v_{(0)}}$ is the Lie derivative along $v_{(0)}^a$. Equation \eqref{nabt} implies that $\Gamma_{(0)ab}^c$ is compatible with \eqref{eq:VP1}
while the last equation implies that $\Gamma_{(0)ab}^c$ is compatible with \eqref{eq:conditionGamma0}.

The connection \eqref{eq:Gamma0} is not boost invariant. To see this we need to first know how $A_{(0)a}$ transforms under boosts. This follows from \eqref{eq:h0abin4dsources} or \eqref{eq:invh0au} and the boost transformations of the boundary vielbeins given in \eqref{eq:invariancetau0_a}--\eqref{eq:rotatione0^a}. This leads to the transformation
\begin{equation}
\delta A_{(0)a}=-\Lambda_{(0)a}\,,
\end{equation}
where we remind the reader that $v_{(0)}^a\Lambda_{(0)a}=0$. It then follows that under a boost $\Gamma_{(0)bc}^a$ transforms as
\begin{equation}
\delta\Gamma_{(0)bc}^a=\frac{1}{2}\Pi_{(0)}^{ad}\left[\Lambda_{(0)c}\left(\partial_b\tau_{(0)d}-\partial_d\Lambda_{(0)b}\right)+\Lambda_{(0)b}\left(\partial_c\tau_{(0)d}-\partial_d\Lambda_{(0)c}\right)\right]\,.
\end{equation}
Taking for example $\Gamma_{(0)bc}^a=\hat \Gamma_{(0)bc}^a$ would be boost invariant, but it would not be compatible with \eqref{eq:conditionGamma0}. What this means in other words is that the vielbein postulates do not impose that the connection $\Gamma_{(0)bc}^a$ is boost invariant but only that it obeys \eqref{eq:boostGamma} and \eqref{eq:conditionGamma0}.

\subsection{Newton--Cartan}\label{subsec:NewtonCartan}

The boundary geometry becomes Newton--Cartan \cite{Cartan1,escidoc:153604} (see also \cite{Misner1973}) if and only if $\tau_{(0)a}$ is taken to be closed. With this additional assumption we get using the $\Gamma_{(0)ab}^c$ as given in \eqref{eq:Gamma0} the vielbein postulates (dropping the superscript $T$ as there is no torsion) \cite{Andringa:2010it}
\begin{eqnarray}
\mathcal{D}_{(0)a} \tau_{(0)b} & = & 0\,,\\
\mathcal{D}_{(0)a}  v_{(0)}^b & = & 0\,,\\
\mathcal{D}_{(0)a} e^{\underline{i}}_{(0)b} & = & 0\,,\\
\mathcal{D}_{(0)a} e^{b}_{(0)\underline{i}} & = & 0\,.
\end{eqnarray}
This implies
\begin{eqnarray}
\nabla_{(0)a}\Pi_{(0)}^{bc} & = & 0\,,\label{eq:NC1}\\
\nabla_{(0)a}\tau_{(0)b} & = & 0\,.\label{eq:NC2}
\end{eqnarray}
Provided we have $\partial_a\tau_{(0)b}-\partial_b\tau_{(0)a}=0$ the $\Gamma_{(0)bc}^a$ is of the form given in \cite{Dautcourt,Andringa:2010it} and of the form used in \cite{Son:2013rqa} if furthermore $F_{(0)ab}=0$.

We remark that one should not confuse Newton--Cartan geometry with Newtonian gravity\footnote{On the boundary of our Lifshitz UV completion the Newton--Cartan geometry is not dynamical. If we however consider it as a dynamical theory it becomes equivalent to Newtonian gravity when we impose the Ehlers conditions.\label{footnote:dynamicalNC}}. This requires additional conditions such as the so-called Ehlers \cite{escidoc:153604} conditions. In the next subsection we will discuss the deformation of Newton--Cartan geometry by adding a specific torsion tensor to it that is proportional to $\partial_a\tau_{(0)b}-\partial_b\tau_{(0)a}$. This will turn out to be a very natural extension of the Newton--Cartan framework.

\subsection{Torsional Newton--Cartan}\label{subsec:torsionalNC}

When $\tau_{(0)a}$ is not closed we find a more general structure which is torsional Newton-Cartan.
To see this we define a torsion tensor $T_{(0)ab}^c$ as
\begin{equation}
T_{(0)ab}^c=-\frac{1}{2}v_{(0)}^c\left(\partial_a\tau_{(0)b}-\partial_b\tau_{(0)a}\right)\,. \label{torsiontensor}
\end{equation}
Next consider the covariant derivative $\nabla_{(0)a}^T$ of section \ref{subsec:vielbeinpostulate} which is defined as
\begin{eqnarray}
\nabla_{(0)a}^TX_{(0)}^b & = & \nabla_{(0)a}X_{(0)}^b+T_{(0)ac}^bX_{(0)}^c\,,\\
\nabla_{(0)a}^TX_{(0)b} & = & \nabla_{(0)a}X_{(0)b}-T_{(0)ab}^cX_{(0)c}\,.
\end{eqnarray}
The relations \eqref{nabt}-\eqref{nabF} of section \ref{subsec:Gamma0} can then be written as
\begin{eqnarray}
\nabla^T_{(0)a}\tau_{(0)b} & = & 0\,,\label{eq:torsionalNC1}\\
\nabla^T_{(0)a} v_{(0)}^b & = & \frac{1}{2}\Pi_{(0)}^{bc}\mathcal{L}_{v_{(0)}}\Pi_{(0)ac}-\frac{1}{2}\Pi_{(0)}^{bc}v_{(0)}^d\left(F_{(0)ca}\tau_{(0)d}+F_{(0)cd}\tau_{(0)a}\right)\,,\\
\nabla^T_{(0)a}\Pi_{(0)}^{bc} & = & 0\,,\label{eq:torsionalNC2}\\
\nabla^T_{(0)a}\Pi_{(0)bc} & = & \left(\tau_{(0)b}\Pi_{(0)cd}+\tau_{(0)c}\Pi_{(0)bd}\right)\nabla^T_{(0)a}v_{(0)}^d\,,
\end{eqnarray}
where \eqref{eq:torsionalNC1} is compatible with \eqref{eq:VP1}. Equations \eqref{eq:torsionalNC1} and \eqref{eq:torsionalNC2} are the defining equations for the torsion of torsional Newton--Cartan geometry\footnote{Loosely speaking one can think of torsional Newton--Cartan geometry as the non-relativistic analogue of a Riemann--Cartan space-time. One could consider more general torsion tensors but we have no need for that here.}. We note that with this
definition, equation \eqref{eq:derivativedet e0} implies that $\nabla_{(0)a}^TX_{(0)}^a$ is not a total derivative.

\subsubsection*{Twistless torsional Newton--Cartan}

An important special case of torsional Newton--Cartan (TNC) geometry is obtained when we impose $\tau_{(0)}\wedge d\tau_{(0)}=0$ but $d\tau_{(0)}\neq 0$. This allows us to write
\begin{equation}\label{eq:dtau0}
\partial_a\tau_{(0)b}-\partial_b\tau_{(0)a}=\tau_{(0)a}\sigma_{(0)b}-\tau_{(0)b}\sigma_{(0)a}\,,
\end{equation}
where
\begin{equation}\label{eq:defsigma0}
\sigma_{(0)a}=-v_{(0)}^c\left(\partial_c\tau_{(0)a}-\partial_a\tau_{(0)c}\right)=-\mathcal{L}_{v_{(0)}}\tau_{(0)a}\,.
\end{equation}
We define the twist tensor $\omega_{(0)ab}$ as
\begin{equation}\label{eq:TTNC}
\omega_{(0)ab}=\frac{1}{2}\Pi_{(0)a}^c\Pi_{(0)b}^d\left(\partial_c\tau_{(0)d}-\partial_d\tau_{(0)c}\right)\,.
\end{equation}
This quantity vanishes for the case where we impose \eqref{eq:dtau0} . We will refer to this as twistless torsional Newton--Cartan geometry (TTNC). This explains the last column in table \ref{table}. The property \eqref{eq:TTNC} can also be used as a definition of TTNC as $\tau_{(0)}\wedge d\tau_{(0)}=0$ follows from it.

Since the torsion tensor $T_{(0)ab}^c$ is proportional to $v_{(0)}^c$ when it is twistless it is also temporal in the sense that projecting all its components with $\Pi_{(0)b}^a$ gives zero. Hence there is no torsion in the spatial directions and one could read TTNC equally as temporal torsional Newton--Cartan.

We will see later in section \ref{subsec:Wardidentities4D} that we can rescale $\tau_{(0)a}$ and $e_{(0)a}^{\underline{i}}$ by a bulk diffeomorphism (known as an anisotropic Weyl transformation). Hence in the case of TTNC there is always a boundary structure in the class of anisotropically conformally equivalent boundary geometries that is Newton--Cartan. However to treat the whole class of anisotropically conformally equivalent boundary geometries in a unified way we need to add torsion.

We will later see in section \ref{subsec:anomaly} that TTNC as a dynamical theory (hence moving away from our setting in which the geometry is non-dynamical, see also footnote \ref{footnote:dynamicalNC}) has striking similarities with Ho\v{r}ava--Lifshitz theories of gravity with one very important difference namely that the underlying geometry does not admit Lorentzian metrics which is assumed to be the case in Ho\v{r}ava--Lifshitz theories. For further comments we refer the reader to sections \ref{subsec:anomaly} and \ref{sec:discussion}.

\subsection{Curvature}\label{subsec:curvature}

Now that we have defined two notions of covariant derivatives $\nabla_{(0)a}$ and $\nabla^T_{(0)a}$ it is natural to consider their associated curvature tensors. We define the Riemann tensor $R^a_{(0)bcd}$ as usual by
\begin{eqnarray}
\left[\nabla_{(0)a}, \nabla_{(0)b}\right]Y^c_{(0)} &=& R^c_{(0)dab}Y^d_{(0)}\,,\label{eq: covcovY1}\\
\left[\nabla_{(0)a}, \nabla_{(0)b}\right]Y_{(0)c} &=& - R^d_{(0)cab}Y_{(0)d}\,.\label{eq: covcovY2}
\end{eqnarray}
It is then given explicitly in terms of $\Gamma_{(0)}$ as
\begin{equation}\label{eq: Riem tensor def gam}
R^c_{(0)dab}= \partial_a\Gamma^c_{(0)bd} - \partial_b\Gamma^c_{(0)ad} + \Gamma^c_{(0)ae}\Gamma^e_{(0)bd} - \Gamma^c_{(0)be}\Gamma^e_{(0)ad} \,.
\end{equation}
Note that because one cannot raise and lower indices it is useful to
have \eqref{eq: covcovY1} and \eqref{eq: covcovY2}. Moreover, this
Riemann tensor does not have all the usual symmetries that one
normally associates with a Riemann tensor. A property that might
appear unusual is the non-vanishing of $R^c_{(0)cab}$ due to the
fact that the curl of $\Gamma_{(0)ca}^c$ does not vanish in general,
as follows from \eqref{eq:derivativedet e0}. Another direct
consequence of \eqref{eq:derivativedet e0} is that the Ricci tensor
defined as $R_{(0)ab}=R^c_{(0)acb}$ is not symmetric in general. It
can be seen from our definition of the Riemann tensor
\eqref{eq: Riem tensor def gam} and \eqref{eq:derivativedet e0} that
the combination $R_{(0)ab} + \frac{1}{2}\nabla_{(0)b}\sigma_{(0)a}$
is symmetric.

We define the extrinsic curvature tensor $K_{(0)ab}$ in analogy with its definition in Lorentzian geometry as follows
\begin{equation}\label{eq:defextrinsiccurv}
K_{(0)ab} = \frac{1}{2}\mathcal{L}_{v_{(0)}}\Pi_{(0)ab}\,.
\end{equation}
By contracting equation \eqref{eq:conditionGamma0} with $v_{(0)}^a$, the extrinsic curvature tensor can also be written as
\begin{equation}
K_{(0)ab}=\frac{1}{2}\mathcal{L}_{v_{(0)}}\Pi_{(0)ab}=\frac{1}{2}\left(\Pi_{(0)a}^c\Pi_{(0)bd}+\Pi_{(0)b}^c\Pi_{(0)ad}\right)\nabla_{(0)c}v_{(0)}^d\,.
\end{equation}
We will use the convention that whenever a tensor $X_{(0)a}$ is orthogonal to $v_{(0)}^a$ we raise its index with $\Pi^{ab}_{(0)}$ and whenever a tensor $X_{(0)}^a$ is orthogonal to $\tau_{(0)a}$ we lower its index with $\Pi_{(0)ab}$. So we will write for example
\begin{equation}
K_{(0)}^{ab} = \Pi_{(0)}^{ac}\Pi_{(0)}^{bd}K_{(0)cd}\,,
\end{equation}
and also
\begin{equation}\label{eq:defsigmaup}
\sigma_{(0)}^a=\Pi_{(0)}^{ab}\sigma_{(0)b}\,,
\end{equation}
where $\sigma_{(0)a}$ is defined in \eqref{eq:defsigma0}. The extrinsic curvature scalar $K_{(0)}$ is given by
\begin{equation}
K_{(0)} = \Pi^{ab}_{(0)}K_{(0)ab}\,.
\end{equation}

We can also define a curvature tensor for the connection including torsion. We denote this Riemann tensor by $R^{T}_{(0)}{}^c{}_{dab}$ and it is defined via
\begin{equation}\label{eq: Riemtorsion}
\left[\nabla^T_{(0)a}, \nabla^T_{(0)b}\right]Y^c_{(0)} = R^{T}_{(0)}{}^c{}_{dab}Y^d_{(0)} - 2T^d_{(0)ab}\nabla^T_{(0)d}Y^c_{(0)}\,.
\end{equation}
The relation between this Riemann tensor and the one in \eqref{eq: Riem tensor def gam}  is
\begin{eqnarray}
R^{T}_{(0)}{}^c{}_{dab} &=& R^c_{(0)dab} +
\frac{1}{4}\left(F_{(\tau)ad} K_{(0)be} - F_{(\tau)bd}
K_{(0)ae}+F_{(0)ae} F_{(\tau)bd} - F_{(0)be}F_{(\tau)ad} \right.
\nonumber\\&& \left. + F_{(0)ef} F_{(\tau)bd}\tau_{(0)a}v_{(0)}^f
- F_{(0)ef} F_{(\tau)ad}
\tau_{(0)b}v_{(0)}^f\right)\Pi_{(0)}^{ce} -
v_{(0)}^c\nabla_{(0)[a}F_{(\tau)b]d}\,,\label{eq compare Riem}
\end{eqnarray}
where $F_{(\tau)ab}=\partial_a\tau_{(0)b}-\partial_b\tau_{(0)a}$. Note that \eqref{eq compare Riem} can also be obtained by replacing
$\Gamma^c_{(0)ab} \rightarrow \Gamma^c_{(0)ab} + T^c_{(0)ab}$ in \eqref{eq: Riem tensor def gam} (with
$T^c_{(0)ab}$  the torsion tensor \eqref{torsiontensor}), as it should.
 By construction $R^{T}_{(0)}{}^c{}_{dab}\tau_{(0)c}$ should vanish as follows from \eqref{eq:torsionalNC1} and this can easily be verified. The corresponding Ricci tensor is given by
\begin{equation}
R^{T}_{(0)ab} =R^{Tc}_{(0)\,acb}= R_{(0)ab}
+\frac{1}{2}\nabla_{(0)c}\left(F_{(\tau)ab}v_{(0)}^c\right) -
\frac{1}{2}\nabla_{(0)b}\sigma_{(0)a} + \frac{1}{4}\sigma_{(0)a}
\sigma_{(0)b}\,,\label{eq RicciTors}
\end{equation}
and is also not symmetric in general.

We often work with projected quantities using our projector
$\Pi^a_{(0)b} = \delta^a_b +v_{(0)}^a\tau_{(0)b}$ and so it will
be useful to define projected covariant derivatives and their
curvatures. This will play an important role later in the discussion
of the anisotropic Weyl anomaly in section \ref{subsec:anomaly}. We define the
projected covariant derivative acting on vectors of the form
$X^b_{(0)}= \Pi^b_{(0)d}Y^d_{(0)}$ as
\begin{equation}\label{eq: def projcovderiv}
\mathcal{D}_{(0)a}X^b_{(0)} \equiv \Pi^c_{(0)a} \Pi^b_{(0)d}
\nabla^T_{(0)c}X^d_{(0)} = \Pi^c_{(0)a} \Pi^b_{(0)d}
\nabla_{(0)c}X^d_{(0)}\,.
\end{equation}
We can then define an
associated projected Riemann tensor $\mathcal{R}_{(0)}{}^c{}_{dab}$
from the equation
\begin{equation}\label{eqprojectec}
\left[\mathcal{D}_{(0)a}, \mathcal{D}_{(0)b}\right]X^c_{(0)} =
\mathcal{R}_{(0)}{}^c{}_{dab}X^d_{(0)} -
2\Pi^{e}_{(0)a}\Pi^{f}_{(0)b}\Pi^c_{(0)g}T^{d}_{(0)ef}\nabla_{(0)d}X^{g}_{(0)}\,.
\end{equation}
From \eqref{eqprojectec} it follows, after computation, that
actually
\begin{equation}\label{eq: sol project Riem}
\mathcal{R}_{(0)}{}^c{}_{dab} =
\Pi^{e}_{(0)a}\Pi^{f}_{(0)b}\Pi^c_{(0)g}\Pi^{h}_{(0)d}
R^{T}_{(0)}{}^{g}{}_{hef}\,.
\end{equation}
For the associated Ricci tensor we obtain
\begin{equation}
\mathcal{R}_{(0)ab}=
\Pi^{e}_{(0)c}\Pi^{f}_{(0)b}\Pi^c_{(0)g}\Pi^{h}_{(0)a}
R^{T}_{(0)}{}^{g}{}_{hef} = \Pi^{e}_{(0)a}\Pi^{f}_{(0)b} R^{T}_{(0)ef} \,,\label{eq:
ProjRicci}
\end{equation}
using that $R^{T}_{(0)}{}^c{}_{dab}\tau_{(0)c}=0$. The relation among the Ricci scalars is
\begin{equation}
\mathcal{R}_{(0)} = \Pi_{(0)}^{ab}\mathcal{R}_{(0)ab}=R^T_{(0)} = R_{(0)} - \frac{1}{2}\nabla_{(0)a}\sigma_{(0)}^a\,.
\end{equation}
We note that $\mathcal{R}_{(0)ab}$ is in general not
symmetric either but it will be in the case of TTNC which we turn to next.

\subsubsection*{Riemannian geometry for simultaneity hypersurfaces in TTNC}

By construction TTNC has no torsion in the spatial directions (see equation \eqref{eq:TTNC}). We therefore expect that the projected geometry in this case has the usual properties of Riemannian geometry. The first indication for this comes from the property
\begin{equation}
\mathcal{R}_{(0)cab}^c=0\,.
\end{equation}
The Ricci tensor is symmetric since we have
\begin{equation}
\mathcal{R}_{(0)[ab]}=\Pi_{(0)a}^c\Pi_{(0)b}^d\partial_{[c}\sigma_{(0)d]}=0\,.
\end{equation}
Then one can show that
\begin{equation}
\mathcal{R}_{(0)ghef}=\Pi_{(0)cg}\mathcal{R}_{(0)hef}^c=\Pi_{(0)g}^c\Pi_{(0)h}^d\Pi_{(0)e}^a\Pi_{(0)f}^bS_{(0)cdab}\,,
\end{equation}
where
\begin{eqnarray}
S_{(0)cdab} & = & \frac{1}{2}\partial_a\partial_d\Pi_{(0)bc}-\frac{1}{2}\partial_a\partial_c\Pi_{(0)bd}\nonumber\\
&&+\frac{1}{4}\Pi_{(0)}^{ef}\left(\partial_a\Pi_{(0)de}+\partial_d\Pi_{(0)ae}-\partial_e\Pi_{(0)ad}\right)\left(\partial_b\Pi_{(0)cf}+\partial_c\Pi_{(0)bf}-\partial_f\Pi_{(0)bc}\right)\nonumber\\
&&+\frac{1}{2}K_{(0)ad}\left(\partial_b\tau_{(0)c}+\partial_c\tau_{(0)b}\right)+\frac{1}{2}K_{(0)bc}\left(\partial_a\tau_{(0)d}+\partial_d\tau_{(0)a}\right)-(a\leftrightarrow b)\,.
\end{eqnarray}
Hence it follows that $S_{(0)cdab}$ and thus $\mathcal{R}_{(0)ghef}$ has all the usual symmetry properties of the Riemann tensor. In particular ($a$ only takes three values)
\begin{eqnarray}
\mathcal{R}_{(0)[gh]ef} & = & 0\,,\\
\mathcal{R}_{(0)efgh} & = & \mathcal{R}_{(0)ghef}\,.
\end{eqnarray}
Since $v_{(0)}^a$ contracted with any component of $\mathcal{R}_{(0)ghef}$ gives zero there is only one free component $\mathcal{R}_{(0)ghef}$ as expected for a 2-dimensional Riemann tensor. This implies that the projected Ricci tensor satisfies the property
\begin{equation}
\mathcal{R}_{(0)ab}=\frac{1}{2}\mathcal{R}_{(0)}\Pi_{(0)ab}\,.
\end{equation}
We conclude that in TTNC the hypersurfaces orthogonal to $\tau_{(0)a}$ which describe surfaces of absolute simultaneity are still described by ordinary Riemannian geometry.

\section{Boundary Stress-Energy Tensor and Ward Identities}\label{sec:sourcesvevs}

In this section we turn our attention to the boundary stress-energy tensor and the associated Ward identities
of our model. To this end we will employ again the relation between the 5D and 4D theory, the identification of the sources in
section \ref{sec:UVcompletion}  along with the structure of the boundary geometry that was described in the previous section.

\subsection{The action with counterterms and its variation}\label{sec:actioncounterterms}

The complete 4-dimensional action is given by \eqref{eq:4Daction} where the counterterm action is obtained by dimensional reduction of the 5-dimensional counterterm action \eqref{eq:Sct1}. This was done in \cite{Chemissany:2012du} and the result is
\begin{eqnarray}
        S_{\text{ct}} & = & \frac{2\pi L}{\kappa_{5}^{2}}\int_{\partial \mathcal{M}} d^3
x\sqrt{-h}\left[-3e^{-\Phi/2} - \frac{1}{4}e^{\Phi/2}\left(R_{(h)}
         - \frac{3}{2}\partial_a \Phi\partial^a\Phi - \frac{1}{4}e^{3\Phi}F_{ab}F^{ab}
\right.\right.\nonumber \\
        & &\left.\left. - \frac{1}{2}\partial_a\phi\partial^a\phi-
\frac{1}{2}e^{2\phi}D_a\chi D^a\chi - \frac{k^2}{2}e^{2\phi-3\Phi} \right)
\right]+ \log r \frac{2\pi L}{\kappa_{5}^{2}}\int_{\partial \mathcal{M}} d^3x\sqrt{-h}\mathcal{A} \,,\label{eq:reducedcountertermaction}
\end{eqnarray}
where
\begin{equation}
\mathcal{A}=e^{-\Phi /2}\hat{\mathcal{A}}
\end{equation}
with $\hat{\mathcal{A}}$ given in \eqref{eq:anomalycounterterm} in which the reduction ansatz should
be substituted. The resulting expression is given in \cite{Chemissany:2012du}.

The total variation of the (renormalized) action can be written as
\begin{eqnarray}
 \delta S_{ren} & = & \frac{2\pi L}{2\kappa_{5}^{2}}\int_{\mathcal{M}}d^{4}x
\sqrt{-g}\left( \mathcal{E}_{\mu \nu} \delta g^{\mu \nu} + \mathcal{E}^{\mu}\delta A_{\mu}+ \mathcal{E}_{\Phi}\delta
\Phi + \mathcal{E}_{\phi} \delta \phi +\mathcal{E}_\chi\delta\chi \right)
\label{eq:totalvariation} \\
 && - \frac{2\pi L}{2\kappa_{5}^{2}}\int_{\partial
\mathcal{M}}d^{3}x\sqrt{-h}\left(T_{ab}\delta h^{ab} + 2T^{a} \delta
A_{a}+ 2T_{\Phi} \delta \Phi + 2T_{\phi} \delta \phi +2T_\chi
\delta\chi -2\frac{\delta r}{r}\mathcal{A}\right)\,,\nonumber
\end{eqnarray}
with the equations of motion given by \eqref{eq:Einsteineqs4D}--\eqref{eq:eomchi} where $T_{ab}$, $T^{a}$, $T_{\Phi}$, $T_{\phi}$ and $T_\chi$ can in principle be computed straightforwardly. However we will prefer to relate them to their 5-dimensional counterparts given in \eqref{eq:Tab}--\eqref{eq:Tchi}. This can be done by dimensionally reducing \eqref{eq:deltaS}. We write
\begin{equation}
\sqrt{-\hat h}\hat T_{\hat a\hat b}\delta\hat h^{\hat a\hat
b}=\sqrt{-h}\left(T_{ab}\delta h^{ab}+2T^{a}\delta
A_a+2T_\Phi\delta\Phi\right)\,,
\end{equation}
and thus we obtain
\begin{eqnarray}
T_{ab} & = & (\hat h_{uu})^{-7/4}\left[(\hat h_{uu})^2\hat T_{ab}-\hat h_{uu}\hat h_{au}\hat T_{bu}-\hat h_{uu}\hat h_{bu}\hat T_{au}+\hat h_{au}\hat h_{bu}\hat T_{uu}\right]\,,\label{eq:Tab4D}\\
T^{a} & = & -(\hat h_{uu})^{-1/4}\hat h^{a\hat b}\hat T_{\hat bu}\,,\\
T_\Phi & = & \frac{1}{2}(\hat h_{uu})^{-1/4}\hat h^{\hat a\hat b}\hat T_{\hat a\hat b}-\frac{3}{2}(\hat h_{uu})^{-5/4}\hat T_{uu}\,.
\end{eqnarray}
This implies that we also have
\begin{eqnarray}
T_a & = & (\hat h_{uu})^{-3/4}\left(\hat h_{au}\hat T_{uu}-\hat h_{uu}\hat T_{au}\right)\,,\\
T_{ab}-T_{b}A_{a} & = & (\hat h_{uu})^{-3/4}\left(\hat h_{uu}\hat T_{ab}-\hat h_{bu}\hat T_{au}\right)\,.
\end{eqnarray}
In a similar manner we obtain
\begin{eqnarray}
T_\phi & = & (\hat h_{uu})^{-1/4}\hat T_{\hat\phi}\,,\\
T_\chi & = & (\hat h_{uu})^{-1/4}\hat T_{\hat\chi}\,,\label{eq:Tchi4D}
\end{eqnarray}
in terms of $\hat T_{\hat\phi}$, $\hat T_{\hat\chi}$ appearing in \eqref{eq:deltaS}. We also note that
in the variation of the 5-dimensional axion $\delta\hat\chi=\delta\chi+k\delta u=\delta'\chi$ we have absorbed the gauge transformation $k\delta u$ into the variation of the 4-dimensional axion and dropped the prime.

As we will see, a more useful quantity to compute in our case is the HIM
 boundary stress-energy tensor  \cite{Hollands:2005ya}.
In order to compute this,  we vary the action with respect to the inverse frame field $e^{a}_{\underline{a}}$, defined via
\begin{equation}
h^{ab} = \eta^{\underline{a}\underline{b}}e^{a}_{\underline{a}}e^{b}_{\underline{b}}\,.
\end{equation}
We thus have
\begin{eqnarray}
 T_{ab}\delta h^{ab} + 2T^{a}\delta A_{a} =2S^{\underline{a}}{}_{a}\delta e^{a}_{\underline{a}} + 2T^{\underline{a}}\delta A_{\underline{a}}\,,
\end{eqnarray}
where $A_{\underline{a}}=A_a e^{a}_{\underline{a}}$ and
\begin{equation}\label{eq:S}
S^{\underline{a}}{}_{a}=\left(T_{ab}- T_{b}A_{a}\right)e^{b\underline{a}}\,.
\end{equation}

\subsection{Variation of the on-shell action}\label{subsec:variationonshellaction}

Since we have observed that we need frame fields for a proper definition of the 4D sources we now
 write the total variation of the action \eqref{eq:totalvariation} in a frame field basis, yielding
\begin{eqnarray}
 \delta S_{\text{ren}} & = & - \frac{2\pi L}{\kappa_{5}^{2}}\int_{\partial \mathcal{M}}\!\!\!\!d^{3}xe\left(S^{\underline{t}}_{a}\delta e^{a}_{\underline{t}}+S^{\underline{i}}_{a}\delta e^{a}_{\underline{i}}
 +T^{\underline{i}}\delta A_{\underline{i}}+T_{\varphi}\delta\varphi+T_{\psi}\delta\psi+ T_{\phi} \delta \phi +T_\chi\delta\chi  -\mathcal{A}\frac{\delta r}{r}  \right)\,,\nonumber\\
\end{eqnarray}
where
\begin{eqnarray}
\varphi & = & \left(A_a-e^{-3\Phi/2}e_a^{\underline{t}}\right)e^a_{\underline{t}} \,,\label{eq:varphi}\\
\psi & = & \left(A_a+e^{-3\Phi/2}e_a^{\underline{t}}\right)e^a_{\underline{t}} \,,\\
T_{\varphi} & = & \frac{1}{2}T^{\underline{t}}+\frac{1}{3}e^{3\Phi/2}T_\Phi\,,\label{eq:Tvarphi4D}\\
T_{\psi} & = & \frac{1}{2}T^{\underline{t}}-\frac{1}{3}e^{3\Phi/2}T_\Phi\,,\label{eq:Tpsi4D}
\end{eqnarray}
 where we left out the equations of motion since we are going to put the variation on-shell.

Using the expressions \eqref{eq:invframe1}--\eqref{eq:invframe4} and the boundary conditions \eqref{eq:expinvframe1}--\eqref{eq:expinvframe5} as well as \eqref{eq:varphi0-}-\eqref{eq:inve0i}  we find that the 4D fields have the expansions
\begin{eqnarray}
e^{a}_{\underline{t}} & = & -r^2e^{\Phi_{(0)}/2}v_{(0)}^a+\ldots\,,\\
e^{a}_{\underline{i}} & = & r e^{-\Phi_{(0)}/2}e^{a}_{(0)\underline{i}}+\ldots\,,\\
A_{\underline{i}} & = & re^{-\Phi_{(0)}/2}A_{(0)\underline{i}}+\ldots\,,\\
\varphi & = &  r^2e^{\Phi_{(0)}/2}A_{(0)\underline{t}}+\ldots\,,\label{eq:varphi-exp}\\
\psi & = &  2e^{-3\Phi_{(0)}/2}+\ldots\,,\\
\phi & = & \phi_{(0)}+\ldots\,,\\
\chi & = & \chi_{(0)}+\ldots\,,
\end{eqnarray}
along with
\begin{equation}
e=\sqrt{-h}=r^{-4}e^{\Phi_{(0)}/2}e_{(0)}+\ldots\,,
\end{equation}
where $e_{(0)}=\text{det}\,e^{\underline{a}}_{(0)a}$ with $e_{(0)a}^{\underline{t}}=\tau_{(0)a}$. Further, using equations  \eqref{eq:Tab4D}--\eqref{eq:Tchi4D}, \eqref{eq:S}, \eqref{eq:Tvarphi4D} and \eqref{eq:Tpsi4D} and the results of section \ref{subsec:onepointfunctions} we have
\begin{eqnarray}
S^{\underline{t}}_a & = & r^2e^{-\Phi_{(0)}}S^{\underline{t}}_{(0)a}+\ldots\,,\label{eq:SEt'a}\\
S^{\underline{i}}_a & = & r^3S^{\underline{i}}_{(0)a}+\ldots\,,\\
T^{\underline{i}} & = & r^3T^{\underline{i}}_{(0)}+\ldots\,,\\
T_{\varphi} & = & r^2 e^{-\Phi_{(0)}}T_{(0)}^{\underline{t}}+\ldots\,,\\
T_{\psi} & = & -\frac{1}{3}r^4e^{\Phi_{(0)}}\langle\mathcal{O}_\Phi\rangle+\ldots\,,\\
T_{\phi} & = & r^4e^{-\Phi_{(0)}/2}\langle\mathcal{O}_\phi\rangle+\ldots\,,\\
T_{\chi} & = & r^4e^{-\Phi_{(0)}/2}\langle\mathcal{O}_\chi\rangle+\ldots\,,\label{eq:TEchi}
\end{eqnarray}
where we indicated the first non-vanishing component. This allows us to put the variation of the action on-shell giving
\begin{eqnarray}
 \delta S_{\text{ren}} & = & - \frac{2\pi L}{\kappa_{5}^{2}}\int_{\partial \mathcal{M}}d^{3}x e_{(0)}\left(-S^{\underline{t}}_{(0)a}\delta v_{(0)}^a+S^{\underline{i}}_{(0)a}\delta e^{a}_{(0)\underline{i}}
 +T_{(0)}^{\underline{t}}\delta A_{(0)\underline{t}}+ T^{\underline{i}}_{(0)}\delta A_{(0)\underline{i}} +\langle\mathcal{O}_\chi\rangle\delta\chi_{(0)}\right.\nonumber\\
 &&\left.+ \langle\mathcal{O}_\phi\rangle \delta \phi_{(0)}+\frac{1}{2}\left(S_{(0)\underline{t}}^{\underline{t}}-S_{(0)\underline{i}}^{\underline{i}}+A_{(0)\underline{t}}T_{(0)}^{\underline{t}}-A_{(0)\underline{i}}T_{(0)}^{\underline{i}}+2\langle\mathcal{O}_\Phi\rangle\right)\delta\Phi_{(0)}  - \mathcal{A}_{(0)}\frac{\delta r}{r}\right)\,,\nonumber\\
 &&\label{eq:4dRen_Action}
\end{eqnarray}
where
\begin{equation}\label{eq:4Danomaly}
\mathcal{A}_{(0)}=\hat{\mathcal{A}}_{(0)}\,,
\end{equation}
with $\hat{\mathcal{A}}_{(0)}$ given in \eqref{eq:hatA0}.

\subsection{Ward identities}\label{subsec:Wardidentities4D}

We will next study the Ward identities by varying the action with respect to local symmetries of the theory. In the 5-dimensional theory the local symmetries are those diffeomorphisms that preserve the radial gauge choice for the metric, also known as PBH transformations (see section \ref{subsec:Wardidentities}). The 4-dimensional Ward identities result from the 4-dimensional versions of the 5-dimensional PBH transformations as well as the local symmetries acting on the frame fields given in \eqref{eq:invariancetau0_a}--\eqref{eq:rotatione0^a}.

The PBH-transformations of the 5-dimensional frame fields are
\begin{eqnarray}
    \delta \hat{e}^{u}_{(0)+} &=& 2\hat\xi^{r}_{(0)}\hat{e}^{u}_{(0)+} + \hat\xi^{a}_{(0)}\partial_{a}\hat{e}^{u}_{(0)+} - \hat{e}^{a}_{(0)+}\partial_{a}\hat\xi^{u}_{(0)} \,,\\
    \delta \hat{e}^{a}_{(0)-} = \delta \hat{e}^{a}_{(0)+} &=& 2\hat\xi^{r}_{(0)}\hat{e}^{a}_{(0)+} + \hat\xi^{b}_{(0)}\partial_{b}\hat{e}^{a}_{(0)+} - \hat{e}^{b}_{(0)+}\partial_{b}\hat\xi^{a}_{(0)}\,, \\
    \delta \hat{e}^{u}_{(0)-} &=&\hat \xi^{a}_{(0)}\partial_{a}\hat{e}^{u}_{(0)-}\,, \\
    \delta \hat{e}^{u}_{(0)\underline{i}} &=& \hat\xi^{r}_{(0)}\hat{e}^{u}_{(0)\underline{i}} +\hat \xi^{a}_{(0)}\partial_{a}\hat{e}^{u}_{(0)\underline{i}} - \hat{e}^{a}_{(0)\underline{i}}\partial_{a}\hat\xi^{u}_{(0)}\,, \\
    \delta \hat{e}^{a}_{(0)\underline{i}} &=&\hat \xi^{r}_{(0)}\hat{e}^{a}_{(0)\underline{i}} + \hat\xi^{b}_{(0)}\partial_{b}\hat{e}^{a}_{(0)\underline{i}} - \hat{e}^{b}_{(0)\underline{i}}\partial_{b}\hat\xi^{a}_{(0)}\,.
\end{eqnarray}
Using the map between the 4- and 5-dimensional frame fields that is given at the beginning of section \ref{subsec:4Dsources} the PBH-transformations of the 4-dimensional frame fields are
\begin{eqnarray}
    \delta v_{(0)}^a &=& 2\xi^{r}_{(0)} v_{(0)}^a +\xi^{b}_{(0)}\partial_{b} v_{(0)}^a - v_{(0)}^b \partial_{b}\xi^{a}_{(0)}\,, \label{eq:PBH1}\\
    \delta e^{a}_{(0)\underline{i}} &=& \xi^{r}_{(0)}e^{a}_{(0)\underline{i}} +\xi^{b}_{(0)}\partial_{b}e^{a}_{(0)\underline{i}} - e^{b}_{(0)\underline{i}}\partial_{b}\xi^{a}_{(0)}\,, \label{eq:PBH2}\\
    \delta A_{(0)\underline{t}} &=& 2\xi^{r}_{(0)}A_{(0)\underline{t}} + \xi^{a}_{(0)}\partial_{a}A_{(0)\underline{t}} -  v_{(0)}^a\partial_{a}\Sigma_{(0)}\,, \label{eq:PBH3}\\
    \delta A_{(0)\underline{i}} &=& \xi^{r}_{(0)}A_{(0)\underline{i}} + \xi^{a}_{(0)}\partial_{a}A_{(0)\underline{i}} + e^{a}_{(0)\underline{i}}\partial_{a}\Sigma_{(0)}\,, \label{eq:PBH4}\\
    \delta \Phi_{(0)} &=& \xi^{a}_{(0)}\partial_{a}\Phi_{(0)}\,, \label{eq:PBH5}\\
    \delta \phi_{(0)} &=& \xi^{a}_{(0)}\partial_{a}\phi_{(0)}\,, \label{eq:PBH6}\\
    \delta \chi_{(0)} &=& \xi^{a}_{(0)}\partial_{a}\chi_{(0)} + k\Sigma_{(0)}\,,\label{eq:PBH7}
\end{eqnarray}
where we defined $\xi_{(0)}^r=\hat\xi_{(0)}^r$, $\xi_{(0)}^a=\hat\xi_{(0)}^a$ and $\Sigma_{(0)}=\hat\xi_{(0)}^u$ for $u$-independent PBH transformation generators. We thus see that there will be Ward identities associated with the generators, $\hat\xi^{r}_{(0)}$, $\hat\xi^{a}_{(0)}$ and $\hat\xi^{u}_{(0)}$, corresponding to anisotropic Weyl \cite{Horava:2009vy}, boundary diffeomorphisms and gauge transformations, respectively. The Ward identities are
\begin{eqnarray}
   0 & = & 2S^{\underline{t}}_{(0)\underline{t}} + 2T^{\underline{t}}_{(0)}A_{(0)\underline{t}}+ S^{\underline{i}}_{(0)\underline{i}}+T^{\underline{i}}_{(0)}A_{(0)\underline{i}}-\mathcal{A}_{(0)} \label{eq:anisoWeylWardidentity} \\
0 & = & - \frac{1}{e_{(0)}}\partial_{a}\left( e_{(0)}T_{(0)}^{a}\right) + k \left\langle \mathcal{O}_{\chi} \right\rangle  \label{eq:gaugeWardidentity}\\
0 & = &  -S^{\underline{t}}_{(0)b}\partial_{a} v_{(0)}^b + S^{\underline{i}}_{(0)b}\partial_{a}e^{b}_{(0)\underline{i}} + \frac{1}{e_{(0)}}\partial_{b}\left( e_{(0)}S_{(0)a}^b \right)+T_{(0)}^{\underline{t}} \partial_{a}A_{(0)\underline{t}} + T^{\underline{i}}_{(0)}\partial_{a}A_{(0)\underline{i}}+ \left\langle \mathcal{O}_{\chi} \right\rangle \partial_{a}\chi_{(0)} \nonumber\\
&&+ \left\langle \mathcal{O}_{\phi} \right\rangle \partial_{a}\phi_{(0)} +\frac{1}{2}\left(S_{(0)\underline{t}}^{\underline{t}}-S_{(0)\underline{i}}^{\underline{i}}+A_{(0)\underline{t}}T_{(0)}^{\underline{t}}-A_{(0)\underline{i}}T_{(0)}^{\underline{i}}+2\left\langle \mathcal{O}_{\Phi} \right\rangle\right)\partial_{a}\Phi_{(0)}    \,.\label{eq:diffeoWardidentity}
\end{eqnarray}
where $T_{(0)}^a=-T^{\underline{t}}_{(0)} v_{(0)}^a+T^{\underline{i}}_{(0)}e^{a}_{(0)\underline{i}}$ and $S_{(0)a}^b=-S^{\underline{t}}_{(0)a} v_{(0)}^b+S^{\underline{i}}_{(0)a}e^{b}_{(0)\underline{i}}$.

\subsection{Dimensional reduction of the vevs and additional Ward Identities}\label{subsec:dimredvevs}

By considering the relation between the 4D and 5D vevs, a number of additional Ward identities can be found, as we now show.
Using \eqref{eq:SEt'a}--\eqref{eq:TEchi} and the expressions of section \ref{sec:actioncounterterms} as well as \eqref{eq:bdrystresstensor}--\eqref{eq:vevOdualtochi0} we find the following expressions for the 4-dimensional vevs 
\begin{eqnarray}
    e^{-\Phi_{(0)}}S^{\underline{t}}_{(0)a} &=& -\frac{1}{\sqrt{2}}\hat{e}^{u}_{(0)-}\hat t_{au}=e^{-\Phi_{(0)}}\hat{t}_{au}\,,\label{eq:4D in 5D vevs 1} \\
    S^{\underline{i}}_{(0)a} &=& \hat{e}^{b\underline{i}}_{(0)}\hat{t}_{ab}+ \hat{e}^{u\underline{i}}_{(0)}\hat{t}_{au}=e_{(0)}^{\underline{i}\,b}\hat t_{ab}-A_{(0)}^{\underline{i}}\hat t_{au} \,,\label{eq:4D in 5D vevs 2}\\
    e^{-\Phi_{(0)}}T^{\underline{t}}_{(0)}  &=& \frac{1}{\sqrt{2}}\hat{e}^{u}_{(0)-}\hat{t}_{uu}=-e^{-\Phi_{(0)}}\hat t_{uu}\,, \label{eq:4D in 5D vevs 3}\\
    T^{\underline{i}}_{(0)} &=& -\hat{e}^{u\underline{i}}_{(0)}\hat{t}_{uu} - \hat{e}^{a\underline{i}}_{(0)}\hat t_{au}=A_{(0)}^{\underline{i}}\hat t_{uu}-e_{(0)}^{\underline{i}\,a}\hat t_{au}\,, \label{eq:4D in 5D vevs 4}\\
    -\frac{1}{3}e^{\Phi_{(0)}}\left\langle \mathcal{O}_{\Phi} \right\rangle &=& -\frac{\sqrt{2}}{6}\hat{e}^{+}_{(0)u}\hat{t}^{\hat{a}}{}_{\hat{a}} + \frac{1}{\sqrt{2}}\hat{e}^{u}_{(0)+}\hat{t}_{uu} + \frac{1}{\sqrt{2}}\hat{e}^{a}_{(0)+}\hat{t}_{au}\nonumber\\
    &=&-\frac{1}{6}e^{\Phi_{(0)}}\mathcal{A}_{(0)}-\frac{1}{2}e^{\Phi_{(0)}}A_{(0)\underline{t}}\hat t_{uu}-\frac{1}{2}e^{\Phi_{(0)}}v_{(0)}^a\hat t_{au}\,,\label{eq:4D in 5D vevs 5}\\
    \langle O_\phi\rangle & = & \langle\hat O_{\hat\phi}\rangle\,,\\
    \langle O_\chi\rangle & = & \langle\hat O_{\hat\chi}\rangle\,,\label{eq:4D in 5D vevs 7}
\end{eqnarray}
in terms of the 5-dimensional vevs $\hat t_{\hat a\hat b}, \langle\hat O_{\hat\phi}\rangle, \langle\hat O_{\hat\phi}\rangle$. Hence using \eqref{eq:4D in 5D vevs 1}--\eqref{eq:4D in 5D vevs 3} it follows that
\begin{eqnarray}
\hat t_{au} & = & S^{\underline{t}}_{(0)a}\,,\label{eq:tau}\\
e_{(0)}^{\underline{i}\,b}\hat t_{ab} & = & S^{\underline{i}}_{(0)a}+A_{(0)}^{\underline{i}}S^{\underline{t}}_{(0)a}\,,\label{eq:hat t ab in 4D vevs}\\
\hat t_{uu} & = & -T_{(0)}^{\underline{t}} \label{eq:tuu}\,.
\end{eqnarray}
Substituting these relations in \eqref{eq:4D in 5D vevs 4} and \eqref{eq:4D in 5D vevs 5} we obtain
\begin{eqnarray}
0 & = & A_{(0)}^{\underline{i}}T_{(0)}^{\underline{t}}+e_{(0)}^{\underline{i}\,a}S_{(0)a}^{\underline{t}}+T_{(0)}^{\underline{i}} \,,\label{eq:vevrelation1}\\
0 & = & S_{(0)\underline{t}}^{\underline{t}}-S_{(0)\underline{i}}^{\underline{i}}+A_{(0)\underline{t}}T_{(0)}^{\underline{t}}-A_{(0)\underline{i}}T_{(0)}^{\underline{i}}+2\langle O_\Phi\rangle\,,\label{eq:vevrelation2}
\end{eqnarray}
where we used \eqref{eq:anisoWeylWardidentity} to remove $\mathcal{A}_{(0)}$ from \eqref{eq:4D in 5D vevs 5}. Further by contracting \eqref{eq:hat t ab in 4D vevs} with $e_{(0)}^{\underline{j}\,a}$ and antisymmetrizing in $(\underline{i},\underline{j})$ we obtain the relation
\begin{equation}\label{eq:vevrelation3}
0 = S_{(0)}^{\underline{i}\underline{j}}+A_{(0)}^{\underline{i}}S_{(0)a}^{\underline{t}\underline{j}}-(\underline{i}\leftrightarrow\underline{j})\,,
\end{equation}
where $S_{(0)}^{\underline{i}\underline{j}}=e_{(0)}^{\underline{j}\,a}S_{(0)a}^{\underline{i}}$ and $S_{(0)}^{\underline{t}\underline{j}}=e_{(0)}^{\underline{j}\,a}S_{(0)a}^{\underline{t}}$.

\subsection{Local tangent space transformations of the sources and vevs}\label{subsec:localtangentspacetrafos}

To see where the relations \eqref{eq:vevrelation1}--\eqref{eq:vevrelation3} come from,  consider the inverse boundary metric $\hat h_{(0)}^{\hat a\hat b}$ written in terms of the 4-dimensional sources, equations \eqref{eq:invh0ab}--\eqref{eq:invh0uu}. We now look for transformations of the sources that leave these expressions invariant.

\subsubsection*{Local Galilean boost transformations}

The first such transformation is the boost transformation \eqref{eq:boosttau0^a}
\begin{eqnarray}
\delta v_{(0)}^a & = & \Lambda_{(0)}^{\underline{i}}e^a_{(0)\underline{i}}\,,\label{eq:Galileanboost1}\\
\delta A_{(0)\underline{i}} & = & -\Lambda_{(0)\underline{i}}\,,\label{eq:Galileanboost2}\\
\delta A_{(0)\underline{t}} & = & -\Lambda_{(0)}^{\underline{i}}A_{(0)\underline{i}}\,. \label{eq:Galileanboost3}
\end{eqnarray}
Using \eqref{eq:4dRen_Action} the associated Ward identity is \eqref{eq:vevrelation1}. These transformations imply that the boundary gauge field $A_{(0)a}$ transforms under local tangent space boosts as
\begin{equation}
\delta A_{(0)a} = -\Lambda_{(0)a}\,,
\end{equation}
where $v_{(0)}^a\Lambda_{(0)a}=0$ and the $v_{(0)}^a$ transformation \eqref{eq:Galileanboost1} can be written as
\begin{equation}\label{eq:Galileanboost4}
\delta v_{(0)}^a=\Lambda_{(0)}^a\,,
\end{equation}
where $\Lambda^a_{(0)}=\Pi_{(0)}^{ab}\Lambda_{(0)b}$ (see also section \ref{subsec:contractionLorentzgroup}). The parameter $\Lambda_{(0)a}$ is such that $\Lambda_{(0)\underline{i}}=e_{(0)\underline{i}}^a\Lambda_{(0)a}$ and hence is only defined up to shifts by terms proportional to $\tau_{(0)a}$. This is required in order for $\hat h_{(0)ab}$ in \eqref{eq:h0abin4dsources} to remain invariant when using the fact that $\Pi_{(0)ab}$ transforms under boosts as
\begin{equation}
\delta\Pi_{(0)ab}=\Lambda_{(0)a}\tau_{(0)b}+\Lambda_{(0)b}\tau_{(0)a}\,,
\end{equation}
as follows from \eqref{eq:booste0_a}. Using the expressions \eqref{eq:4D in 5D vevs 1}--\eqref{eq:4D in 5D vevs 7} it is straightforward to work out that the vevs transform under Galilean boosts as
\begin{eqnarray}
\delta S_{(0)a}^{\underline{i}} & = & \Lambda_{(0)}^{\underline{i}}S_{(0)a}^{\underline{t}}\,,\\
\delta T_{(0)}^{\underline{i}} & = & \Lambda_{(0)}^{\underline{i}}T_{(0)}^{\underline{t}}\,,\\
\delta\langle O_\Phi\rangle & = & -\frac{3}{2}\Lambda_{(0)}^{\underline{i}}T_{(0)\underline{i}}\,,
\end{eqnarray}
where we left out those vevs that are inert.

Comparing \eqref{eq:h0abin4dsources}--\eqref{eq:invh0uu} with the parametrization \eqref{eq:parametrizationh0} we obtain
\begin{equation}
A_{(0)a}=-\hat N_{(0)a}\,.
\end{equation}
This means that in the parametrization \eqref{eq:hatN0} the spatial components of $A_{(0)a}$ have been put equal to zero. This can be understood as fixing the freedom to perform a boost. Because of the restriction $v_{(0)}^a\Lambda_{(0)a}=0$ there is one component in $A_{(0)a}$ that cannot be removed. This component is essentially $\hat h_{(0)}^{uu}=2A_{(0)a}v_{(0)}^a+\Pi_{(0)}^{ab}A_{(0)a}A_{(0)b}$.

Some of the geometric definitions such as the connection \eqref{eq:Gamma0} and the extrinsic curvature \eqref{eq:defextrinsiccurv} are not boost invariant. For the extrinsic curvature we will later in equation \eqref{eq:boostinvextrinsiccurv} define a manifestly boost invariant expression. Regarding the covariant derivative not being boost invariant one has to treat separately derivatives along $v_{(0)}^a$ and $\Pi_{(0)b}^a$ and build boost invariant objects out of them as for example done in section \ref{subsec:anomaly}.

\subsubsection*{Local $SO(2)$ rotations}

The next symmetry leaving $\hat h_{(0)}^{\hat a\hat b}$ invariant is given by
\begin{eqnarray}
\delta e^a_{(0)\underline{i}} & = & -\Lambda_{(0)}^{\underline{j}}{}_{\underline{i}}e^a_{(0)\underline{j}}\,,\label{eq:localSO(2)1}\\
\delta A_{(0)\underline{i}} & = & -\Lambda_{(0)}^{\underline{j}}{}_{\underline{i}}A_{(0)\underline{j}}\,,\label{eq:localSO(2)2}
\end{eqnarray}
where $\Lambda_{(0)\underline{i}\underline{j}}=-\Lambda_{(0)\underline{j}\underline{i}}$. This symmetry gives rise to the Ward identity \eqref{eq:vevrelation3}. In section \ref{subsec:contractionLorentzgroup} we have shown that these are the local $SO(2)$ transformations that together with the Galilean boosts that we have just discussed are induced by bulk local Lorentz transformations acting on the boundary frame fields $e^{\underline{a}}_{(0)a}$. The vevs transform in the obvious way as
\begin{eqnarray}
\delta S_{(0)a}^{\underline{i}} & = & -\Lambda_{(0)\underline{j}}{}^{\underline{i}}S_{(0)a}^{\underline{j}}\,,\\
\delta T_{(0)}^{\underline{i}} & = & -\Lambda_{(0)\underline{j}}{}^{\underline{i}}T_{(0)}^{\underline{j}}\,,
\end{eqnarray}
with the other vevs remaining inert.

\subsubsection*{Local dilatations}\label{page:localdilatations}

There is one more local transformation leaving $\hat h_{(0)}^{\hat a\hat b}$ trivially invariant. It is given by\footnote{The transformation acts on the 5-dimensional vielbein sources as follows
$\delta\hat e^+_{(0)a}  =  -\Lambda_{(0)}\hat e^+_{(0)a}$,
$\delta\hat e^-_{(0)a}  =  \Lambda_{(0)}\hat e^-_{(0)a} $,
$ \delta\hat e^+_{(0)u}  =  \Lambda_{(0)}\hat e^+_{(0)u}$,
$\delta\hat e^{\underline{i}}_{(0)a}  =  0$ leaving $\hat h_{(0)\hat a\hat b}$ invariant.}
\begin{equation}
\delta\Phi_{(0)} = \Lambda_{(0)}\,,\label{eq:omega5}
\end{equation}
leading to the relation \eqref{eq:vevrelation2}. This transformation takes the form of a local dilatation shifting $\Phi_{(0)}$. We have defined the 4-dimensional sources in section \ref{subsec:4Dsources} and the vevs in \eqref{eq:SEt'a}--\eqref{eq:TEchi} such that they all have zero weight with respect to these local dilatations.

In distinction to  the other local symmetries, this dilatation symmetry is only there at leading order. For example $\hat h_{(2)uu}$ which is given in \eqref{eq:h2uu} is not invariant under it. Since the $\Lambda_{(0)}$ rescaling is not a local symmetry of the full Fefferman--Graham expansion we are not able to use it to remove a source component such as $\Phi_{(0)}$. It does however produce the  additional Ward identity \eqref{eq:vevrelation2} which can be used to remove $\delta\Phi_{(0)}$ from the variation of the on-shell action \eqref{eq:4dRen_Action}. This is very convenient as after doing so the variations in \eqref{eq:4dRen_Action} are unconstrained while in the case in which we do not remove the term in front of $\delta\Phi_{(0)}$ the variations are constrained by \eqref{eq:constraint4D2}. We will always choose to remove the term proportional to $\delta\Phi_{(0)}$ so that the variation of the on-shell action is now given in terms of 14 sources and 14 vevs. Of either set we can remove 8 by local symmetries (diffeomorphisms, anisotropic Weyl, gauge and local tangent space boost and SO(2) transformations) and their associated Ward identities. The boundary field $\Phi_{(0)}$ is no longer a source and is simply given by \eqref{eq:constraint4D2}. We thus count 6+6 sources and vevs. In section \ref{subsec:holographicreconstruction} we will see that the full 4-dimensional Fefferman--Graham expansion obtained by dimensional reduction of the 5-dimensional Fefferman--Graham expansion contains on top of these sources and vevs one additional free scalar function.

\subsection{Gauge transformations and scaling dimensions of the vevs}\label{subsec:gaugeandscalevevs}

Using the transformations of $v_{(0)}^a$ and $e^a_{(0)\underline{i}}$ under local boosts, rotation and dilatations we find that the quantity $S_{(0)a}^b=-S^{\underline{t}}_{(0)a} v_{(0)}^b+S^{\underline{i}}_{(0)a}e^{b}_{(0)\underline{i}}$ appearing prominently in \eqref{eq:diffeoWardidentity} transforms as
\begin{equation}
\delta S_{(0)a}^b=0\,.
\end{equation}
We conclude that $S_{(0)a}^b$ is invariant under the local tangent space transformations. In this subsection we ask how
the quantity $S_{(0)a}^b$ as well as the other vevs transform under gauge transformations with parameter $\Sigma_{(0)}$ and anisotropic Weyl transformations generated by $\xi^r_{(0)}$ where we take $\xi^r_{(0)}$ to be constant in which case they are referred to as scale transformations and we compute the associated scaling dimensions of the vevs.

\subsubsection*{Gauge transformations}

The gauge transformation is described by the PBH transformations of \eqref{eq:PBH1}--\eqref{eq:PBH7} with the parameter $\Sigma_{(0)}$. This gauge transformation only acts on the source $A_{(0)a}$ and transforms it as
\begin{equation}
\delta A_{(0)a}=\partial_a\Sigma_{(0)}\,.
\end{equation}
To work out the gauge transformations of the vevs we use that the action of the PBH transformations on the 5-dimensional vevs is given by
\begin{equation}\label{eq:PBHtab}
\delta\hat t_{\hat a\hat b}=\hat\xi_{(0)}^{\hat c}\partial_{\hat c}\hat t_{\hat a\hat b}+\hat t_{\hat c\hat b}\partial_{\hat a}\hat\xi_{(0)}^{\hat c}+\hat t_{\hat a\hat c}\partial_{\hat b}\hat\xi_{(0)}^{\hat c}+\delta_{\hat\xi_{(0)}^r}\hat t_{\hat a\hat b}\,.
\end{equation}
Taking $\hat\xi_{(0)}^{\hat a}=\delta^{\hat a}_u\Sigma_{(0)}$ and $\hat\xi_{(0)}^r=0$ and using \eqref{eq:4D in 5D vevs 1}--\eqref{eq:4D in 5D vevs 5} we obtain the following gauge transformations of the vevs
\begin{eqnarray}
\delta S^{\underline{t}}_{(0)a} & = & -T_{(0)}^{\underline{t}}\partial_a\Sigma_{(0)}\,,\\
\delta S^{\underline{i}}_{(0)a} & = & \left(e_{(0)}^{\underline{i}\,b}S_{(0)b}^{\underline{t}}+A_{(0)}^{\underline{i}}T_{(0)}^{\underline{t}}\right)\partial_a\Sigma_{(0)}=-T_{(0)}^{\underline{i}}\partial_a\Sigma_{(0)}\,,\label{eq:gaugeS0i}
\end{eqnarray}
with the other vevs gauge invariant and where we used \eqref{eq:vevrelation1} in \eqref{eq:gaugeS0i}. With these transformations one can show
\begin{equation}
\delta S_{(0)a}^b=-T_{(0)}^b\partial_a\Sigma_{(0)}\,,
\end{equation}
so that we find that $S_{(0)a}^b$ is not gauge invariant. If we define the shifted vev
\begin{equation}
\mathcal{T}_{(0)a}^b=S_{(0)a}^b+T_{(0)}^b\frac{1}{k}\partial_a\chi_{(0)}\,, \label{Tshifted}
\end{equation}
it follows that the quantity $\mathcal{T}_{(0)a}^b$ is both gauge invariant as well as invariant under local tangent space transformations.

\subsubsection*{Scaling dimensions of the vevs}

If we consider PBH transformations with $\hat\xi^a_{(0)}=0$ and $\hat\xi^r_{(0)}=\text{cst}$ the 5-dimensional boundary stress-energy tensor $\hat t_{\hat a\hat b}$ has scaling dimension two meaning that it transforms as
\begin{equation}
\delta\hat t_{\hat a\hat b}=2\hat\xi^r_{(0)}\hat t_{\hat a\hat b}\,.
\end{equation}
Using the scaling dimensions of the 4-dimensional sources given in \eqref{eq:PBH1}--\eqref{eq:PBH7} and the relation between the 5- and 4-dimensional vevs given in \eqref{eq:4D in 5D vevs 1}--\eqref{eq:4D in 5D vevs 7} we obtain the set of scaling dimensions given in table \ref{table:scalingdimensions}.
\begin{table}[h!]
      \centering
      \begin{tabular}{|c|c|c|c|c|c|c|c|}
      \hline
 & $S_{(0)a}^{\underline{t}}$ & $S_{(0)a}^{\underline{i}}$ & $T_{(0)}^{\underline{t}}$ & $T_{(0)}^{\underline{i}}$ &$\langle O_\Phi\rangle$ & $\langle O_\phi\rangle$ & $\langle O_\chi\rangle$\\
  \hline
 scaling dimension &2&3&2&3&4&4&4    \\
         \hline
           \end{tabular}
      \caption{Scaling dimensions of the 4-dimensional vevs.}\label{table:scalingdimensions}
\end{table}
These are the vevs as they appear in the variation of the on-shell action \eqref{eq:4dRen_Action}. Other vevs that we encounter such as $\mathcal{T}_{(0)\underline{t}}^{\underline{t}}$, and $\mathcal{T}_{(0)}^{\underline{t}\underline{i}}$ have the scaling dimensions given in table \ref{table:scalingdimensions2}.
\begin{table}[h!]
      \centering
      \begin{tabular}{|c|c|c|c|c|c|c|}
      \hline
 & $\mathcal{T}_{(0)}^{\underline{t}\underline{t}}$ & $\mathcal{T}_{(0)}^{\underline{t}\underline{i}}$ & $\mathcal{T}_{(0)}^{\underline{i}\underline{t}}$ & $\mathcal{T}_{(0)}^{\underline{i}\underline{j}}$ & $\mathcal{T}_{(0)a}^{b}$ & $T_{(0)}^a$ \\
  \hline
 scaling dimension &4&3&5&4&4&4    \\
         \hline
           \end{tabular}
      \caption{Scaling dimensions of some derived vevs.}\label{table:scalingdimensions2}
\end{table}
Following \cite{Ross:2009ar} we call $\mathcal{T}_{(0)}^{\underline{t}\underline{t}}$ the energy density, $\mathcal{T}_{(0)}^{\underline{t}\underline{i}}$ the momentum density, $\mathcal{T}_{(0)}^{\underline{i}\underline{t}}$ the energy flux and $\mathcal{T}_{(0)}^{\underline{i}\underline{j}}$ the stress. We point out that even though the energy flux has scaling dimension 5 and would thus appear to be an irrelevant operator\footnote{Since $e_{(0)}$ has dimension -4 (which is $z=2$ plus 2 spatial dimensions) an operator is irrelevant when its dimension is larger than 4.} this is not a problem since the operators in table \ref{table:scalingdimensions} are all either relevant or marginal and it is these that we source.
Necessary and sufficient conditions for the existence of conserved boundary currents such as continuity equations are discussed in section \ref{subsec:bdrycurrents}.

\subsection{Covariantizing the Ward identities}\label{subsec:covWard}

We conclude this section by presenting the Ward identities in a covariant form with respect to the 
boundary geometry that we described in section \ref{sec:bdrygeom}.
Using our vielbein postulate and choice of $\Gamma_{(0)ab}^c$ the gauge Ward identity \eqref{eq:gaugeWardidentity} can be written as
\begin{equation}\label{eq:gaugeWard}
  k \left\langle \mathcal{O}_{\chi} \right\rangle = \nabla_{(0)a}T_{(0)}^a-\frac{1}{2} v_{(0)}^b\left(\partial_b\tau_{(0)a}-\partial_a\tau_{(0)b}\right)T_{(0)}^a\,,
\end{equation}
where $T_{(0)}^a=-T_{(0)}^{\underline{t}}  v_{(0)}^a +T^{\underline{i}}_{(0)} e^{a}_{(0)\underline{i}}$ while the diffeomorphism Ward identity \eqref{eq:diffeoWardidentity} can be rewritten as
\begin{eqnarray}
0 & = & \nabla_{(0)b}S_{(0)a}^b+\frac{1}{2}S_{(0)a}^bv_{(0)}^c\left(\partial_c\tau_{(0)b}-\partial_b\tau_{(0)c}\right)-S_{(0)b}^{\underline{t}}\nabla_{(0)a} v_{(0)}^b+S_{(0)b}^{\underline{i}}\nabla_{(0)a}e_{(0)\underline{i}}^b\nonumber\\
&&  + T_{(0)}^{\underline{t}} \partial_{a}A_{(0)\underline{t}}+ T^{\underline{i}}_{(0)}\partial_{a}A_{(0)\underline{i}} + \left\langle \mathcal{O}_{\phi} \right\rangle \partial_{a}\phi_{(0)} + \left\langle \mathcal{O}_{\chi} \right\rangle \partial_{a}\chi_{(0)} \,.\label{eq:diffeoWardrewritten3}
\end{eqnarray}

Expressing the Ward identities \eqref{eq:anisoWeylWardidentity}, \eqref{eq:vevrelation1}--\eqref{eq:vevrelation3}, \eqref{eq:gaugeWard} and \eqref{eq:diffeoWardrewritten3} in terms of gauge invariant vevs we find
\begin{eqnarray}
\mathcal{A}_{(0)} & = & 2\mathcal{T}_{(0)\underline{t}}^{\underline{t}}+2B_{(0)\underline{t}}T_{(0)}^{\underline{t}}+\mathcal{T}_{(0)\underline{i}}^{\underline{i}}+B_{(0)\underline{i}}T_{(0)}^{\underline{i}}\,,\label{eq:traceWardidentity}\\
k\left\langle O_\chi\right\rangle & = & \nabla_{(0)a}T_{(0)}^a-\frac{1}{2} v_{(0)}^b\left(\partial_b\tau_{(0)a}-\partial_a\tau_{(0)b}\right)T_{(0)}^a\,,\label{eq:LifshitzgaugeWard}\\
\nabla_{(0)b}\mathcal{T}_{(0)a}^b & = & -\mathcal{T}_{(0)b}^c\left(-\tau_{(0)c}\nabla_{(0)a} v_{(0)}^b+e_{(0)c}^{\underline{i}}\nabla_{(0)a}e_{(0)\underline{i}}^b\right)+\frac{1}{2}\mathcal{T}_{(0)a}^bv_{(0)}^c\left(\partial_c\tau_{(0)b}-\partial_b\tau_{(0)c}\right)\nonumber\\
&&-T_{(0)}^{\underline{t}}\partial_aB_{(0)\underline{t}}-T_{(0)}^{\underline{i}}\partial_aB_{(0)\underline{i}}-\langle O_\phi\rangle\partial_a\phi_{(0)}\,,\label{eq:LifshitzdiffeoWard}\\
\mathcal{T}_{(0)}^{\underline{t}\underline{i}}+B_{(0)}^{\underline{i}}T_{(0)}^{\underline{t}}&=&-T_{(0)}^{\underline{i}} \,,\label{eq:boostWardidentity}\\
0 & =& \mathcal{T}_{(0)}^{\underline{i}\underline{j}}-B_{(0)}^{\underline{i}}T_{(0)}^{\underline{j}}-(\underline{i}\leftrightarrow\underline{j})\,,\label{eq:SO(2)Wardidentity}\\
\left\langle O_\Phi\right\rangle & = & -\frac{1}{2}\left(\mathcal{T}_{(0)\underline{t}}^{\underline{t}}+B_{(0)\underline{t}}T_{(0)}^{\underline{t}}-\mathcal{T}_{(0)\underline{i}}^{\underline{i}}-B_{(0)\underline{i}}T_{(0)}^{\underline{i}}\right)\,,\label{eq:vevPhi}
\end{eqnarray}
where we wrote
\begin{eqnarray}
B_{(0)\underline{t}} & = & A_{(0)\underline{t}}+\frac{1}{k}v_{(0)}^a\partial_a\chi_{(0)}\,,\\
B_{(0)\underline{i}} & = & A_{(0)\underline{i}}-\frac{1}{k}e^a_{(0)\underline{i}}\partial_a\chi_{(0)}\,.
\end{eqnarray}

If we use the torsional covariant derivative of section \ref{subsec:torsionalNC} we can write the Ward identities \eqref{eq:LifshitzgaugeWard} and \eqref{eq:LifshitzdiffeoWard} as
\begin{eqnarray}
k\left\langle O_\chi\right\rangle & = & \nabla^T_{(0)a}T_{(0)}^a-2T_{(0)ab}^aT_{(0)}^b\,,\\
\nabla^T_{(0)b}\mathcal{T}_{(0)a}^b & = & -\mathcal{T}_{(0)b}^c\left(-\tau_{(0)c}\nabla^T_{(0)a} v_{(0)}^b+e_{(0)c}^{\underline{i}}\nabla^T_{(0)a}e_{(0)\underline{i}}^b\right)+2T_{(0)ac}^b\mathcal{T}_{(0)b}^c+2T_{(0)bc}^b\mathcal{T}_{(0)a}^c\nonumber\\
&&-T_{(0)}^{\underline{t}}\partial_aB_{(0)\underline{t}}-T_{(0)}^{\underline{i}}\partial_aB_{(0)\underline{i}}-\langle O_\phi\rangle\partial_a\phi_{(0)}\,.\label{eq:LifshitzdiffeoWardtorsion}
\end{eqnarray}
The form of the diffeomorphism Ward identity \eqref{eq:LifshitzdiffeoWardtorsion} is similar to the one given in \cite{Hollands:2005ya} with the differences that here i) the vielbeins do not transform under the Lorentz group but rather under the contracted Lorentz group,
ii)  we cannot raise and lower indices and iii)  in general we have a torsion term $T_{(0)ac}^b$.

\section{Further Physical Properties}\label{sec:properties}

In this section we continue our analysis of the physical properties of the boundary theory described via our holographic
prescription. We first consider the construction of conserved boundary currents for the case of a boundary geometry described by TNC and define the corresponding conserved charges
for the case when $\tau_{(0)a}$ is hypersurface orthogonal, i.e. for TTNC. Then we turn to a detailed analysis of the anisotropic Weyl anomaly density $\mathcal{A}_{(0)}$. Finally, we comment on the appearance of an undetermined function in the Fefferman--Graham expansion and the interpretation of this in terms of a second UV completion of our IR theory.

\subsection{Conserved boundary currents}\label{subsec:bdrycurrents}

To construct conserved boundary currents from our gauge invariant stress-energy tensor \eqref{Tshifted}
we start
by contracting equation \eqref{eq:LifshitzdiffeoWard} with a vector $K_{(0)}^a$. The resulting equation can be written as
\begin{eqnarray}
\hspace{-1.2cm} \nabla_{(0)b}\left(K_{(0)}^a\mathcal{T}_{(0)a}^b\right) & = & \mathcal{T}_{(0)a}^b\left(\tau_{(0)b}\mathcal{L}_{K_{(0)}}v_{(0)}^a-e_{(0)b}^{\underline{i}}\mathcal{L}_{K_{(0)}}e_{(0)\underline{i}}^a+\frac{1}{2}K_{(0)}^a\mathcal{L}_{v_{(0)}}\tau_{(0)b}\right)\nonumber\\
&&-T_{(0)}^{\underline{t}}\mathcal{L}_{K_{(0)}}B_{(0)\underline{t}}-T_{(0)}^{\underline{i}}\mathcal{L}_{K_{(0)}}B_{(0)\underline{i}}-\langle O_{\phi}\rangle\mathcal{L}_{K_{(0)}}\phi_{(0)}\,,\label{eq:diffeoWardKilling}
\end{eqnarray}
where $\mathcal{L}_{K_{(0)}}$ denotes the Lie derivative along $K_{(0)}^a$. Let us next subtract the term $-T_{(0)bc}^bK_{(0)}^a\mathcal{T}_{(0)a}^c=\tfrac{1}{2}\mathcal{T}_{(0)a}^bK_{(0)}^a\mathcal{L}_{v_{(0)}}\tau_{(0)b}$ from both sides. We thus find the conserved current
\begin{equation}
e_{(0)}^{-1}\partial_{b}\left(e_{(0)}K_{(0)}^a\mathcal{T}_{(0)a}^b\right)= \nabla_{(0)b}\left(K_{(0)}^a\mathcal{T}_{(0)a}^b\right)-T_{(0)bc}^bK_{(0)}^a\mathcal{T}_{(0)a}^c=0\,,
\end{equation}
if and only if
\begin{eqnarray}
0 & = & \mathcal{T}_{(0)a}^b\left(\tau_{(0)b}\mathcal{L}_{K_{(0)}}v_{(0)}^a-e_{(0)b}^{\underline{i}}\mathcal{L}_{K_{(0)}}e_{(0)\underline{i}}^a+K_{(0)}^a\mathcal{L}_{v_{(0)}}\tau_{(0)b}\right)\nonumber\\
&&-T_{(0)}^{\underline{t}}\mathcal{L}_{K_{(0)}}B_{(0)\underline{t}}-T_{(0)}^{\underline{i}}\mathcal{L}_{K_{(0)}}B_{(0)\underline{i}}-\langle O_{\phi}\rangle\mathcal{L}_{K_{(0)}}\phi_{(0)}\,.\label{eq:diffeoWardKillingb}
\end{eqnarray}
We will not impose any conditions on the vevs other than the Ward identities.

To find the necessary and sufficient conditions for the right hand side of \eqref{eq:diffeoWardKillingb} to vanish upon use of the Ward identities we proceed as follows. We use equation \eqref{eq:boostWardidentity} to remove $\mathcal{T}_{(0)}^{\underline{t}\underline{i}}$ and the $SO(2)$ Ward identity \eqref{eq:SO(2)Wardidentity} is used to eliminate the antisymmetric part $\mathcal{T}_{(0)}^{[\underline{i}\underline{j}]}$. This leaves us with an equation involving the following vevs: $\mathcal{T}_{(0)}^{(\underline{i}\underline{j})}+B_{(0)}^{(\underline{i}}T_{(0)}^{\underline{j})}$, $\mathcal{T}_{(0)\underline{t}}^{\underline{t}}+B_{(0)\underline{t}}T_{(0)}^{\underline{t}}$, $\mathcal{T}_{(0)}^{\underline{i}\underline{t}}$, $T_{(0)}^{\underline{i}}$, $T_{(0)}^{\underline{t}}$ and $\langle O_\phi\rangle$. We have by now used up all the Ward identities except for \eqref{eq:traceWardidentity} which we then use to remove $\mathcal{T}_{(0)\underline{t}}^{\underline{t}}+B_{(0)\underline{t}}T_{(0)}^{\underline{t}}$. We finally demand that each term in front of these remaining vevs vanishes by itself,  in order for \eqref{eq:diffeoWardKillingb} to hold without imposing any constraints on the vevs other than the Ward identities. This gives the following set of conditions for the matter fields
\begin{eqnarray}
\mathcal{L}_{K_{(0)}}\phi_{(0)} & = & 0\,,\\
v_{(0)}^a\mathcal{L}_{K_{(0)}}B_{(0)a} & = & 0\,,\\
\Pi_{(0)c}^a\mathcal{L}_{K_{(0)}}B_{(0)a} & = & -\Pi_{(0)ac}\mathcal{L}_{K_{(0)}}v_{(0)}^a+B_{(0)a}K_{(0)}^a\sigma_{(0)c}\,,\label{eq:KillingPiB}
\end{eqnarray}
resulting from the terms proportional to $\langle O_\phi\rangle$, $T_{(0)}^{\underline{t}}$ and $T_{(0)}^{\underline{i}}$, respectively, and
\begin{eqnarray}
\hspace{-1cm}0 & = & \Pi_{(0)c}^b\left(\mathcal{L}_{K_{(0)}}\tau_{(0)b}-\tau_{(0)a}K_{(0)}^a\sigma_{(0)b}\right)\,,\label{eq:Tit}\\
\hspace{-1cm}0 & = & \mathcal{A}_{(0)}v_{(0)}^a\mathcal{L}_{K_{(0)}}\tau_{(0)a}\,,\label{eq:anomalyandconservedcurrent}\\
\hspace{-1cm}0 & = & v_{(0)}^a\mathcal{L}_{K_{(0)}}\tau_{(0)a}-\frac{1}{2}\Pi_{(0)ab}\mathcal{L}_{K_{(0)}}\Pi_{(0)}^{ab}-K_{(0)}^a\sigma_{(0)a}\,,\label{eq:tracepart}\\
\hspace{-1cm}0 & = & \left(\Pi_{(0)ac}\Pi_{(0)bd}-\frac{1}{2}\Pi_{(0)ab}\Pi_{(0)cd}\right)\mathcal{L}_{K_{(0)}}\Pi_{(0)}^{cd}\nonumber\\
&&-\left(\Pi_{(0)ab}\sigma_{(0)e}-\Pi_{(0)eb}\sigma_{(0)a}-\Pi_{(0)ea}\sigma_{(0)b}\right)K_{(0)}^e\,,\label{eq:tracelesspart}
\end{eqnarray}
for the boundary vielbeins where we recall that $\sigma_{(0)a}=-\mathcal{L}_{v_{(0)}}\tau_{(0)a}$. Equation \eqref{eq:anomalyandconservedcurrent} comes from the term proportional to $\mathcal{T}_{(0)\underline{t}}^{\underline{t}}+B_{(0)\underline{t}}T_{(0)}^{\underline{t}}$ and equation \eqref{eq:Tit} from the term proportional to $\mathcal{T}_{(0)}^{\underline{i}\underline{t}}$. The last two equations result from the trace part and the trace free part of the term proportional to $\mathcal{T}_{(0)}^{(\underline{i}\underline{j})}+B_{(0)}^{(\underline{i}}T_{(0)}^{\underline{j})}$. From equation \eqref{eq:anomalyandconservedcurrent} we see that if the anomaly density $\mathcal{A}_{(0)}$ is non-vanishing we get an extra condition on the boundary vielbeins. The term $\mathcal{L}_{v_{(0)}}\tau_{(0)a}$ appearing at various places is related to the presence of torsion. Note that equation \eqref{eq:KillingPiB} also contains non-trivial information about the existence of a boundary conserved current from the point of view of the boundary vielbeins through the term $\Pi_{(0)ac}\mathcal{L}_{K_{(0)}}v_{(0)}^a$.

Equation \eqref{eq:Tit} implies that there is a function $\lambda_{(0)}$ such that
\begin{equation}\label{eq:LieK0tau0_a}
\mathcal{L}_{K_{(0)}}\tau_{(0)a}=\lambda_{(0)}\tau_{(0)a}+\tau_{(0)b}K_{(0)}^b\sigma_{(0)a}\,,
\end{equation}
which via equation \eqref{eq:anomalyandconservedcurrent} is constrained to satisfy
\begin{equation}
\mathcal{A}_{(0)}\lambda_{(0)}=0\,.
\end{equation}
Continuing like this we find from \eqref{eq:tracepart} that we have
\begin{equation}
\mathcal{L}_{K_{(0)}}\Pi_{(0)}^{ab}=-\left(\lambda_{(0)}+K_{(0)}^c\sigma_{(0)c}\right)\Pi_{(0)}^{ab}+v_{(0)}^a\chi_{(0)}^b+ v_{(0)}^b\chi_{(0)}^a\,,
\end{equation}
for some vector $\chi_{(0)}^a$. It follows that \eqref{eq:tracelesspart} becomes
\begin{equation}
\left(\Pi_{(0)ab}\sigma_{(0)e}-\Pi_{(0)eb}\sigma_{(0)a}-\Pi_{(0)ea}\sigma_{(0)b}\right)K_{(0)}^e=0\,,
\end{equation}
which implies upon contraction with $K_{(0)}^a$
\begin{equation}
\Pi_{(0)ab}K_{(0)}^aK_{(0)}^b\sigma_{(0)c}=0\,,
\end{equation}
so that we either must have
\begin{equation}
\sigma_{(0)a}=0\,,\qquad\text{or}\qquad K_{(0)}^a=\kappa_{(0)}v_{(0)}^a\,,
\end{equation}
for some function $\kappa_{(0)}$.

To summarize, the conditions for the existence of a boundary conserved current split into two cases depending on whether $\sigma_{(0)a}=0$ or $\sigma_{(0)a}\neq 0$. When $\sigma_{(0)a}\neq 0$ the conditions become
\begin{eqnarray}
K_{(0)}^a & = & \kappa_{(0)}v_{(0)}^a\,,\\
\mathcal{L}_{K_{(0)}}\phi_{(0)} & = & 0\,,\\
\mathcal{L}_{K_{(0)}}B_{(0)a} & = & B_{(0)c}K_{(0)}^c\sigma_{(0)a}\,,\\
\partial_a\kappa_{(0)} & = & -\lambda_{(0)}\tau_{(0)a}\,,\\
0 & = & \mathcal{A}_{(0)}\lambda_{(0)}\,,\\
\mathcal{L}_{K_{(0)}}\Pi_{(0)}^{ab} & = & -\lambda_{(0)}\Pi_{(0)}^{ab}+v_{(0)}^a\chi_{(0)}^b+ v_{(0)}^b\chi_{(0)}^a\,,
\end{eqnarray}
where we used $\mathcal{L}_{K_{(0)}}\tau_{(0)a}=-\partial_a\kappa_{(0)}-\kappa_{(0)}\sigma_{(0)a}$ for $K_{(0)}^a=\kappa_{(0)}v_{(0)}^a$ in \eqref{eq:LieK0tau0_a} and when $\sigma_{(0)a}=0$ the conditions become
\begin{eqnarray}
\mathcal{L}_{K_{(0)}}\phi_{(0)} & = & 0\,,\\
\mathcal{L}_{K_{(0)}}B_{(0)a} & = & -\Pi_{(0)ac}\mathcal{L}_{K_{(0)}}v_{(0)}^c\,,\\
\mathcal{L}_{K_{(0)}}\tau_{(0)a} & = & \lambda_{(0)}\tau_{(0)a}\,,\\
0 & = & \mathcal{A}_{(0)}\lambda_{(0)}\,,\\
\mathcal{L}_{K_{(0)}}\Pi_{(0)}^{ab} & = & -\lambda_{(0)}\Pi_{(0)}^{ab}+v_{(0)}^a\chi_{(0)}^b+ v_{(0)}^b\chi_{(0)}^a\,,
\end{eqnarray}
where in both cases we also solved for $\mathcal{L}_{K_{(0)}}B_{(0)a}$. In general $\sigma_{(0)a}=0$ does not imply that the torsion is vanishing, but in the case of TTNC it does via equation \eqref{eq:dtau0}.

When $\tau_{(0)a}$ is hypersurface orthogonal which can happen for both $\sigma_{(0)a}\neq 0$ and $\sigma_{(0)a}=0$ a natural definition of a conserved charge $Q[K_{(0)}]$ is
\begin{equation}
Q[K_{(0)}]=\int_{\Sigma}d^2x\sqrt{\gamma_{(0)}}K_{(0)}^b\mathcal{T}_{(0)b}^a\tau_{(0)a}\,,
\end{equation}
where $\Sigma$ is the hypersurface to which $\tau_{(0)a}$ is normal and with $\sqrt{\gamma_{(0)}}$ the metric induced on this hypersurface. For example if we choose coordinates such that $\tau_{(0)i}=0$ we can write $\tau_{(0)a}=\tfrac{e_{(0)}}{\sqrt{\gamma_{(0)}}}\partial_a t$ and $\Sigma$ will be the surface $t=\text{cst}$.

It would be interesting to study further the possible choices for $K_{(0)}^a$, the algebra of the vectors $K_{(0)}^a$ and charges $Q[K_{(0)}]$ and how $K_{(0)}^a$ and $Q[K_{(0)}]$ transform under local boosts.

\subsection{Anisotropic Weyl anomaly}\label{subsec:anomaly}

In this subsection we will express the 4-dimensional anomaly density
$\mathcal{A}_{(0)}$ in terms of the natural curvature objects of
torsional Newton--Cartan. The 4-dimensional anomaly density is
simply equal to the 5-dimensional anomaly density
$\hat{\mathcal{A}}_{(0)}$ and was computed in appendix
\ref{app:reducedanomaly} by dimensional reduction (see equation
\eqref{eq: full hatA0}). For simplicity we restrict ourselves to the
case of TTNC in this subsection, i.e. we assume hypersurface
orthogonality for $\tau_{(0)a}$. Our goal is thus to take \eqref{eq:
full hatA0} and rewrite it using the geometry worked out in section
\ref{subsec:curvature}. It will prove convenient to use the
projected Riemannian geometry on the hypersurfaces to which
$\tau_{(0)a}$ is orthogonal and their extrinsic curvature. The main
challenge in rewriting \eqref{eq: full hatA0} is to identify an
appropriate total derivative term such that the remaining terms take
a simple form that are furthermore invariant under the anisotropic
conformal rescalings generated by $\xi^r_{(0)}$. Without giving any
further details we find that for TTNC the anomaly can be written as
\begin{eqnarray}
\mathcal{A}_{(0)} &=&
\frac{1}{4}k^4e^{4\phi_{(0)}}I_{(0)}^2+\frac{1}{8}k^2e^{2\phi_{(0)}}K'_{(0)ab}K'_{(0)cd}\left(\Pi_{(0)}^{ac}\Pi_{(0)}^{bd}-\frac{1}{2}
\Pi_{(0)}^{ab}\Pi_{(0)}^{cd} \right) \nonumber\\
&&+\frac{1}{48}\left(\mathcal{R}_{(0)}
-\mathcal{D}_{(0)a}\sigma_{(0)}^a-\frac{1}{2}\Pi_{(0)}^{ab}\partial_a\phi_{(0)}\partial_b\phi_{(0)}+2k^2e^{2\phi_{(0)}}I_{(0)}\right)^2\nonumber\\
&&+\frac{5}{16}k^2e^{2\phi_{(0)}}\left(v_{(0)}^a\partial_a\phi_{(0)}+\Pi_{(0)}^{ab}B_{(0)a}\partial_b\phi_{(0)}\right)^2
\nonumber\\
&&
-\frac{1}{2}k^2e^{2\phi_{(0)}}I_{(0)}\left(\mathcal{D}_{(0)c}\left(\Pi_{(0)}^{cd}\partial_d\phi_{(0)}\right)+\Pi_{(0)}^{cd}\partial_c\phi_{(0)}\partial_d\phi_{(0)}\right)\nonumber\\
&&
+\frac{1}{16}\left(\mathcal{D}_{(0)a}\left(\Pi_{(0)}^{ab}\partial_b\phi_{(0)}\right)\right)^2+\frac{1}{64}\left(\Pi_{(0)}^{ab}\partial_a\phi_{(0)}\partial_b\phi_{(0)}\right)^2
+e_{(0)}^{-1}\partial_a\left(e_{(0)}J^a_{(0)}\right)\,,
\end{eqnarray}
where
\begin{eqnarray}
\Pi_{(0)}^{ac}\Pi_{(0)}^{bd}K'_{(0)ab} & = & \Pi_{(0)}^{ac}\Pi_{(0)}^{bd}\left(K_{(0)ab}-\sigma_{(0)(a}\Pi_{(0)b)}^cB_{(0)c}+\mathcal{D}_{(0)(a}\left(\Pi_{(0)b)}^cB_{(0)c}\right)\right)\,,\\
\label{eq:defK'0ab}
I_{(0)} &=&v_{(0)}^aB_{(0)a} +
\frac{1}{2}\Pi^{ab}_{(0)}B_{(0)a}B_{(0)b}\,.
\end{eqnarray}
Equation \eqref{eq:defK'0ab} for a constant axion is equal to $\Pi_{(0)}^{ac}\Pi_{(0)}^{bd}\tilde K_{(0)ab}$ where $\tilde K_{(0)ab}$ is the boost invariant extrinsic curvature given by
\begin{equation}\label{eq:boostinvextrinsiccurv}
\tilde K_{(0)ab} =\frac{1}{2}\mathcal{L}_{v_{(0)}^c+\Pi_{(0)}^{cd}A_{(0)d}}\left(\Pi_{(0)ab}+\tau_{(0)a}A_{(0)b}+\tau_{(0)b}A_{(0)a}\right)\,.
\end{equation}
The current $J_{(0)}^a$ is\footnote{In deriving the expressions for $\mathcal{A}_{(0)}$ and $J_{(0)}^a$ many identities from TTNC have been used that can all be derived using the formulas of sections \ref{subsec:torsionalNC} and \ref{subsec:curvature} including, to mention a few,
\begin{align}
&\frac{1}{2}\Pi_{(0)}^{ac}D_{(0)a}\chi_{(0)}D_{(0)c}\chi_{(0)}\mathcal{R}_{(0)}=\Pi_{(0)}^{bc}\Pi_{(0)d}^aD_{(0)c}\chi_{(0)}[\nabla_{(0)a},\nabla_{(0)b}]\left(\Pi_{(0)}^{de}D_{(0)e}\chi_{(0)}\right)\,,\\
&\Pi_{(0)b}^a \nabla_{(0)a}\sigma_{(0)}^b=e_{(0)}^{-1}\partial_a\left(e_{(0)}\sigma_{(0)}^a\right)+\sigma_{(0)a}\sigma_{(0)}^a\,.
\end{align}}
\begin{eqnarray}
J^a_{(0)} &=&
\frac{1}{8}k^2e^{2\phi_{(0)}}I_{(0)}\left(\sigma^a_{(0)}
+ 2\Pi_{(0)}^{ab}\partial_b\phi_{(0)}\right)+\frac{1}{8}\Pi_{(0)}^{cd}\sigma_{(0)d}\Pi_{(0)c}^e\Pi_{(0)f}^a\nabla_{(0)e}\Pi_{(0)}^{bf}\sigma_{(0)b} \nonumber\\
&&-\frac{1}{8}\Pi_{(0)}^{ad}\sigma_{(0)d}\Pi_{(0)b}^c\nabla_{(0)c}\Pi_{(0)}^{be}\sigma_{(0)e}-\frac{1}{8}\Pi_{(0)}^{ac}\Pi_{(0)}^{bd}\sigma_{(0)d}\partial_b\phi_{(0)}\partial_c\phi_{(0)} \nonumber\\
&& +\frac{1}{16}\Pi_{(0)}^{ad}\Pi_{(0)}^{bc}\sigma_{(0)d}\partial_b\phi_{(0)}\partial_c\phi_{(0)}+\frac{1}{8}k^2e^{2\phi_{(0)}}K'_{bc}\Pi_{(0)}^{bc}\left( v_{(0)}^a+ \Pi_{(0)}^{ad}B_{(0)d} \right)\nonumber\\
&& - \frac{1}{8}k^2e^{2\phi_{(0)}}\left(v_{(0)}^a\Pi_{(0)}^{bc} -
 v_{(0)}^b\Pi_{(0)}^{ac}
\right)\left(2\partial_b\phi_{(0)}\Pi^e_{(0)c}B_{(0)e} +\nabla_{(0)c}B_{(0)b}\right)\nonumber\\
&& +\frac{1}{8}k^2e^{2\phi_{(0)}}\Pi_{(0)e}^d\Pi_{(0)b}^a\nabla_{(0)d}\left(\Pi_{(0)}^{bc}B_{(0)c}\right)\Pi_{(0)}^{ef}B_{(0)f} + \frac{1}{8}k^2e^{2\phi_{(0)}}\Pi_{(0)}^{ac}
\nabla_{(0)b}\left( v_{(0)}^bB_{(0)c}\right)\nonumber\\
&&-\frac{1}{4}k^2e^{2\phi_{(0)}}\nabla_{(0)c}\left(\Pi_{(0)}^{ac}I_{(0)}\right)\,.
\end{eqnarray}
We have written the result for $e_{(0)}\mathcal{A}_{(0)}-\partial_a\left(e_{(0)}J_{(0)}^a\right)$ in a manifestly boost invariant manner. This requires some work as the reduction discussed in appendix \ref{app:reducedanomaly} breaks manifest boost invariance. We did not bother to do the same for the current term because we expect that the term $\partial_a\left(e_{(0)}J_{(0)}^a\right)$ can be removed by adding finite counterterms to the action just like in the context of the AdS/CFT correspondence \cite{Henningson:1998gx}.

As remarked at the beginning of this section the only assumption that we made was that $\tau_{(0)a}$ is hypersurface orthogonal. This means that in the language of section \ref{subsec:AlLif} this result applies to the case of AlLif boundary conditions. If we take $\phi_{(0)}=\log g_s$ equal to a constant and furthermore take $D_{(0)a}\chi_{(0)}=0$ we find
\begin{eqnarray}
\mathcal{A}_{(0)} & = &
\frac{1}{8}k^2g_s^2\left(K_{(0)}^{ab}K_{(0)ab} -
\frac{1}{2}K_{(0)}^2\right) + \frac{1}{48}\left( \mathcal{R}_{(0)}-\Pi_{(0)b}^a \nabla_{(0)a}\sigma_{(0)}^b \right)^2 \nonumber\\
&&+ e_{(0)}^{-1}\partial_{a}\left(e_{(0)}J^a_{(0)}\right)\,,\label{eq:anomalysimplified}
\end{eqnarray}
as the equivalent vielbein way of writing the result found in \cite{Chemissany:2012du}.

The terms contained in $e_{(0)}\mathcal{A}_{(0)}-\partial_a\left(e_{(0)}J_{(0)}^a\right)$ are all invariant under anisotropic conformal rescalings. This can be seen by noting that the combinations
\begin{align}
&\mathcal{R}_{(0)}-\Pi_{(0)b}^a\nabla_{(0)a}\sigma_{(0)}^b\,,\\
&K'_{(0)ab}K'_{(0)cd}\left(\Pi_{(0)}^{ac}\Pi_{(0)}^{bd}-
\frac{1}{2} \Pi_{(0)}^{ab}\Pi_{(0)}^{cd} \right)\,, \\
&\Pi_{(0)}^{ab}\nabla_{(0)a}\left(\Pi_{(0)b}^c\partial_c\phi_{(0)}\right)\,,
\end{align}
transform with weights 2, 4 and 2, respectively ($e_{(0)}$ has weight $-4$) under anisotropic Weyl rescalings.

Interestingly, the part $e_{(0)}\mathcal{A}_{(0)}-\partial_a\left(e_{(0)}J_{(0)}^a\right)$ takes the form of a Lagrangian. Let us entertain the possibility that we can read it as an actual Lagrangian. The boost invariant kinetic terms are
\begin{align}
&\Pi_{(0)}^{ac}\Pi_{(0)}^{bd}K'_{(0)cd}\,,\\
&kv_{(0)}^a\partial_a\phi_{(0)}-\Pi_{(0)}^{bc}D_{(0)b}\chi_{(0)}\partial_c\phi_{(0)}\,,
\end{align}
and appear in the action as second order in time derivatives. Note that these terms are proportional to $k^2$ so that it is crucial to perform a Scherk--Schwarz reduction in order to obtain them. We read the term $v_{(0)}^aB_{(0)a}+ \frac{1}{2}\Pi_{(0)}^{ab}B_{(0)a}B_{(0)b}$ as a non-derivative term because $\partial_a\chi_{(0)}$ has been eaten by $A_{(0)a}$ via a St\"uckelberg mechanism. There thus appears a non-derivative term at order $k^4$ in the action. This term is essentially $\hat h_{(0)}^{uu}$ made gauge invariant, which is already there for AlLif boundary conditions. It has not appeared in the literature so far because of too restrictive parametrizations of the various ADM gauges that have been used. In all cases one simply took $A_{(0)a}=0$. At order $k^0$ the action contains fourth order derivative terms built out of curvatures and projected covariant derivatives. These are thus gradient potential terms.

This Lagrangian has striking similarities with Ho\v{r}ava--Lifshitz (HL) type Lagrangians \cite{Horava:2008ih}. For example pushing this analogy we would call TTNC with furthermore $\partial_a\tau_{(0)b}-\partial_b\tau_{(0)a}=0$ projectable Ho\v{r}ava--Lifshitz and TTNC with nonzero $\partial_a\tau_{(0)b}-\partial_b\tau_{(0)a}$ non-projectable HL gravity. Furthermore, the object $\sigma_{(0)}^a$ corresponds to the acceleration vector of the foliation defined by the hypersurface orthogonality of $\tau_{(0)a}$. The action \eqref{eq:anomalysimplified} is precisely of the form of a 3-dimensional $z=2$ conformal HL gravity with nonzero potential term \cite{Griffin:2011xs}. However, in the most general case we notice one absolutely crucial difference. In HL gravity one assumes the existence of an underlying Lorentzian geometry. In other words the tangent space is described by Minkowski space-time. Here, on the other hand, this is not the case since we have a non-relativistic metric structure and the tangent space group contains Galilean boosts, which is the origin of the boundary gauge field $A_{(0)a}$. Ultimately the action is therefore, despite its functional form, not of a HL type. It is nevertheless an interesting question to ask what kind of dynamics is described by an action of a HL type defined on a TTNC geometry.

Going back to $\mathcal{A}_{(0)}$ being an anomaly density, based on anisotropic conformal symmetry arguments, one expects in general two different types of central charges for Lifshitz field theories \cite{Griffin:2011xs,Baggio:2011ha}. One is proportional to the coefficient in front of the extrinsic curvature term and one to the coefficient in front of the spatial curvature term. In the notation of \cite{Baggio:2011ha} these are denoted by $C_1$ and $C_2$, respectively (see \cite{Chemissany:2012du} for their appropriately normalized values). For other examples of Lifshitz anisotropic Weyl anomalies see \cite{Adam:2009gq,Baggio:2011ha}. When $I_{(0)}\neq 0$ there is one more term in the anomaly density. This is the term at order $k^4$. It would be interesting to understand better the role of  this non-derivative term.

\subsection{Irrelevant deformations and a second UV completion}\label{subsec:holographicreconstruction}

In section \ref{subsec:dimredvevs} we wrote the 4-dimensional vevs in terms of the 5-dimensional ones. If we try to do this the other way around we find for $\hat t_{ab}$
\begin{eqnarray}
\hat t_{ab} & = &  \left(v_{(0)}^cv_{(0)}^d\hat t_{cd}\right)\tau_{(0)a}\tau_{(0)b}-\left(S_{(0)}^{\underline{i}\underline{t}}+A_{(0)}^{\underline{i}}S_{(0)}^{\underline{t}\underline{t}}\right)\left(e_{(0)\underline{i}a}\tau_{(0)b}+e_{(0)\underline{i}b}\tau_{(0)a}\right)\nonumber\\
&&+\left(S_{(0)}^{\underline{i}\underline{j}}+A_{(0)}^{\underline{i}}S_{(0)}^{\underline{t}\underline{j}}\right)e_{(0)\underline{i}a}e_{(0)\underline{j}b}\,.\label{eq:hattab}
\end{eqnarray}
Due to equation \eqref{eq:vevrelation3} the right hand side of $\hat t_{ab}$ is symmetric in $a$ and $b$. Because of the appearance of the function $v_{(0)}^cv_{(0)}^d\hat t_{cd}$ it follows that $\hat t_{ab}$ is not fully determined by the 4-dimensional vevs. It can be shown that when reducing the 5-dimensional Ward identities \eqref{eq:tracet}--\eqref{eq:divt} the term $v_{(0)}^cv_{(0)}^d\hat t_{cd}$ drops out and is after reduction not in any sense coupled to any one of the sources and vevs. This is consistent with the fact that it does not appear in the variation of the on-shell action \eqref{eq:4dRen_Action}. Nevertheless, it appears in the 4-dimensional Fefferman--Graham expansion that we give in appendix \ref{app:reconstruction}, where it shows up in the expansion of the metric at order $r^2$. The Fefferman--Graham expansion thus contains 6+6 sources and vevs and the free function $v_{(0)}^cv_{(0)}^d\hat t_{cd}$.

We have noted before that the same theory also admits another branch of solutions that are briefly discussed in appendix \ref{subsec:h0uuneq0}. These solutions are asymptotic to a hyperscaling violating geometry with $\theta=-1$ and $z=1$. The UV expansions \eqref{eq:UVexpansion1}--\eqref{eq:UVexpansionlast} are controlled by 7+7 sources and vevs (we omitted the expansions for the axion-dilaton field). This is the same number as in the 5-dimensional theory without the constraint that $\hat h_{(0)uu}=0$. Hence, in this case all components of the 5-dimensional boundary stress-energy tensor after reduction of $\hat t_{\hat a\hat b}$ have a dual source.

We also noted that the solution \eqref{eq:IRperspective1}--\eqref{eq:IRperspective3} asymptotes to a $z=2$ Lifshitz space-time in the IR. This means that when studying linearized perturbations around the $z=2$ Lifshitz space-time we expect to see one mode going like $\epsilon r^{-2}$ where $\epsilon$ controls the linearized perturbation. Going to higher orders in $\epsilon$ means that we are going to see a series in $\epsilon r^{-2}$ and in order for this to remain small $r$ runs large. In other words this corresponds to a mode that is sourcing an irrelevant operator. Indeed if we expand \eqref{eq:IRperspective3} around $r=\infty$ we notice a perturbation going like $r^{-2}$. Hence the spectrum of linearized perturbations around the $z=2$ Lifshitz space-time of our 4-dimensional model contains (after removing gauge redundancy) 7+7 parameters with one of them corresponding to an irrelevant perturbation. If we switch off this mode (our constraint $\hat h_{(0)uu}=0$) and turn on the remaining relevant perturbations we flow to the UV that we referred to as the Lif UV. If we turn on this irrelevant perturbation (the case $\hat h_{(0)uu}>0$) and then additionally turn on the relevant perturbations we flow towards the other UV that is asymptotic to a hyperscaling violating geometry with $\theta=-1$ and $z=1$.
The presence of the extra free function in the expansion of the Lif UV theory thus signals that there is an irrelevant operator whose source has been turned off\footnote{We thank Elias Kiritsis for useful discussions on this point.}.

A similar phenomenon has been observed in the context of $\theta=1$ and $z=3$ hyperscaling violating geometries that can be uplifted to 5-dimensional $z=-1$ Schr\"odinger space-times. These are asymptotically AdS solutions of AdS gravity without any matter added. This reduces to an Einstein--Maxwell-dilaton theory in 4-dimensions. There is a similar issue there in that the solutions depend on whether the reduction is along a circle that becomes asymptotically null or one that is asymptotically spacelike leading to two different UV completions from a 4-dimensional point of view \cite{Singh:2010cj}. It would in fact be interesting to work out the details of the computation of the sources and vevs in that case.

The fact that for our Lifshitz UV completion we count in total 6+6+1 free functions is in strong contrast with what one has observed for the massive vector model (2 scalar fields less than our model). In that model for $z=2$ we have 5+5 free functions in the expansion. The way we came to this answer is as follows. Using the equations for the linearized perturbation analysis\footnote{Linearized perturbations of $z=2$ Lifshitz solutions of the massive vector model have also been studied in \cite{Danielsson:2009gi,Ross:2009ar,Baggio:2011cp,Cheng:2009df,Baggio}.} of \cite{Gath:2012pg} (setting the parameters $a$ and $b$ defined in \cite{Gath:2012pg} equal to zero and truncating the scalar field) we observe by looking at purely radial perturbations around Lifshitz that there are 4 integration constants in the tensor modes, 8 in the vector modes and 4 in the scalar modes (in the radial gauge of \cite{Gath:2012pg} one actually encounters 5 parameters but one can be removed by a rescaling of the radial coordinate). One can remove 6 parameters using diffeomorphisms (3 off-shell and another 3 on-shell) leading to 10 parameters. One of these constants corresponds to a marginal deformation. It turns out that this deformation is marginally relevant \cite{Danielsson:2009gi,Cheng:2009df,Baggio,Holsheimer:2013ula} and hence we do not set it to zero (see also the discussion at the end of section \ref{subsec:radialgaugeEframe}). Our point of view is that in order to get the full Lifshitz UV completion one should allow for all deformations around Lifshitz that are not irrelevant. The main difference between the massive vector model and our model is that in our case there is an irrelevant deformation of the Lifshitz geometry that is absent in the massive vector model.

\section{Discussion and Outlook}\label{sec:discussion}

We conclude by summarizing some of the main points and lessons.

{\bf The $z=2$ model and a second UV completion}: We have discussed holographic properties of a specific model admitting $z=2$ Lifshitz solutions that can be obtained by dimensional reduction from AdS. This circumvents having to work out a Fefferman--Graham expansion for the massive vector model which is currently still lacking beyond results obtained using linearized perturbation theory. The limitation of our approach is that it works only for $z=2$. However, it should be stressed that this is  a special value, which must be treated separately anyway. From what we know about the $z=2$ case in the massive vector model we see from our analysis that having additional scalars in the theory can markedly change many qualitative features such as the UV structure of the theory. In our case, starting at an IR Lifshitz fixed point there are two possible UV completions depending on whether or not we turn on a certain irrelevant operator. From the higher-dimensional perspective this corresponds to performing a reduction with a null or spacelike circle on the AdS boundary. In the bulk the circle is always spacelike. In the case where the boundary circle is null, we get a Lifshitz UV with no hyperscaling violation and $z=2$ and in the case where the circle on the AlAdS$_5$ boundary is spacelike we get the $\theta=-1$ and $z=1$ UV completion of appendix \ref{subsec:h0uuneq0}. This should be  contrasted with the massive vector model for $z=2$ where there is just one UV completion which allows for a marginally relevant deformation. In this paper we focussed our attention on the holographic setup for the case of the Lifshitz UV completion. In general whenever one studies Kaluza--Klein holography \cite{Kanitscheider:2008kd} there are typically assumptions concerning the leading components of the KK dilaton. Interesting additional branches of solutions may occur when different choices are made for the fall-off of the KK dilaton.

{\bf Vielbeins, sources and torsional Newton--Cartan}: In order to identify the sources and in order for these sources to be the leading component of some field it proved very convenient to use a vielbein decomposition of the metric and vector field in the model. By the vector we mean here the field that transforms under gauge transformations and thus not the massive vector that has eaten the axion. Since the vector and the timelike vielbein are proportional to each other at leading order it was useful to consider specific linear combinations of these two quantities, such that for the new field variables the leading terms are independent sources. This allowed us to identify the boundary gauge field $A_{(0)a}$. The boundary geometry is thus described by the sources appearing in the vielbeins and the bulk gauge field. This geometry turns out to be torsional Newton--Cartan (TNC) with a specific torsion tensor that is zero if and only if $\tau_{(0)a}dx^a$, the leading component of the timelike vielbein, is closed in which case the boundary geometry is ordinary Newton--Cartan. To the best of our knowledge this geometric structure has not been studied before. An important special case is where $\tau_{(0)a}dx^a$ is hypersurface orthogonal but not exact. In this case we call the boundary geometry TTNC for temporal or twistless torsional Newton--Cartan. An added bonus of using vielbeins is that one does not need to resort to a specific gauge choice on top of radial gauge such as the ADM gauge that is often used in the Lifshitz literature. Such a gauge choice can of course always be made but one must be careful not to miss any sources such as the boundary gauge field $A_{(0)a}$ and not to make too strong assumptions such as imposing hypersurface orthogonality of $\tau_{(0)a}$ before starting to solve the equations of motion if one's goal is to find the most general solution. We expect that when studying other holographic models for Lifshitz invariant field theories with some $z>1$ the boundary geometry will always be described by TNC. This is because for any $z>1$ the local tangent space group induced from the bulk onto the boundary will be the contracted Lorentz group and subsequently there will be a degenerate metric structure. It is then natural to choose the same connections as here. 

{\bf Boundary gauge field}: The boundary gauge field $A_{(0)a}$ transforms under boosts such that only the combination $2v_{(0)}^aA_{(0)a}+\Pi_{(0)}^{ab}A_{(0)a}A_{(0)b}$ is boost invariant. Further it transforms under gauge transformations but not in such a way that one can gauge this boost invariant combination away. Associated with the gauge symmetry we have the Ward identity $\partial_a\left(e_{(0)}T_{(0)}^a\right)=e_{(0)}k\langle O_\chi\rangle$, where $T^a_{(0)}$ is a boost and gauge invariant current whose (non-)\linebreak conservation is controlled by the vev of the axion. Associated with the boost symmetry we have the Ward identity \eqref{eq:boostWardidentity}. The boundary gauge field differs from what one usually encounters in AdS/CFT (or from what we would find for the other $\theta=-1$ and $z=1$ UV completion) because it transforms under boosts whereas this would not happen for $z=1$. It would therefore be interesting to get a better understanding of the nature of the currents in the boundary theory that $A_{(0)a}$ is sourcing. It could for example be informative to add a second Maxwell term to the bulk Lagrangian and to study the sources for this additional gauge field and contrast it with our $A_{(0)a}$. We expect this second gauge field to behave qualitatively different from $A_{(0)a}$ as only one boundary gauge field will be part of the boundary TNC geometry. In more general 4-dimensional models supporting Lifshitz geometries than the one studied here it is known that one always needs one Maxwell term to support the Lifshitz geometry. Here we see from a boundary perspective why this is so. The bulk Maxwell field together with the bulk vielbeins are both needed to describe the boundary TNC geometry. This gives a rationale for why one usually separates out one Maxwell field from the others in solutions of charged Lifshitz black holes as in \cite{Brynjolfsson:2009ct,Tarrio:2011de}. 

{\bf Properties of the boundary stress-energy tensor}: One of the central results of this paper is derivation of the boundary stress-energy tensor and its corresponding Ward identities, including their covariant form in terms of the non-relativistic boundary geometry that we uncovered. We also note that we have computed the scaling dimensions of the vevs (see table \ref{table:scalingdimensions}) demonstrating that these correspond to
relevant and marginal operators. From these one can compute the scaling dimensions of the energy density, momentum density, energy flux and stress all of which are composite operators in terms of the vevs and sources. In particular, this showed that even though the energy flux has dimension 5, and thus appears to be an irrelevant deformation, this is not a problem since it is a product of a dimension 2 source with a dimension 3 vev.

{\bf Conserved boundary currents and anomaly}: The Ward identities for the boundary stress-energy tensor, namely the diffeomorphism Ward identity and the $z=2$ trace Ward identity due to anisotropic Weyl symmetries are not generally of the form of a divergence of some current. We have studied the existence of boundary conserved currents in section \ref{subsec:bdrycurrents} by postulating the existence of some kind of TNC analogue of a (conformal) `Killing vector'. It would be interesting to study further the conditions for the existence of such conserved currents and the associated conserved charges. In particular this might be useful for a general study of Lifshitz thermodynamics in terms of the boundary charges. The $z=2$ trace Ward identity contains an anomaly related to the $z=2$ anisotropic Weyl anomaly. We observe that even though it takes the form of a Ho\v{r}ava--Lifshitz action this analogy is not perfect because the underlying geometry is TNC and not Lorentzian. Furthermore the anomaly contains zeroth order derivative terms involving the boundary gauge field which have not been observed before. These terms become second order in derivatives if we set $A_{(0)a}=0$ in which case they can be seen as axion kinetic terms. When $A_{(0)a}\neq 0$ it is more natural to read them as zeroth order in derivatives for the massive vector $B_{(0)a}=A_{(0)a}-\tfrac{1}{k}\partial_a\chi_{(0)}$. It would be interesting to understand their origin better, e.g. by using the techniques of \cite{Baggio:2011ha}.

{\bf Constraint on the sources}: Another noteworthy aspect is that in the reduction from five to four dimensions a constraint on the sources appeared, but that we could deal with this constraint explicitly since it is paired with a leading order symmetry. By this we mean that this additional symmetry, which corresponds to local dilatations, is only there at leading order in the FG expansion.  By appropriately redefining the boundary vielbeins (involving rescaling with a power of $e^{\Phi_{(0)}}$) we were then able to use the Ward identity corresponding to this additional symmetry to remove the constrained source from the variation of the on-shell action, leaving a variation with respect to unconstrained sources.

{\bf Radial gauge}: We also note that, motivated by the dimensional reduction, our analysis naturally involves a non-radial gauge in which the holographic expansion seemingly takes its simplest form. It is possible, in principle, to go to radial gauge, but the results could be much more difficult to obtain. One may thus wonder whether going to an appropriate non-radial gauge may be likewise preferred in other models. In general a recipe for obtaining a FG expansion in radial gauge could be the following. Consider purely radial linearized perturbations in radial gauge and distill from this an asymptotic expansion by looking at the $r$-dependence of the higher order $\epsilon$ terms where $\epsilon$ is the perturbation parameter describing the Lifshitz perturbations. With this information one can trade the $\epsilon$ expansion for a radial asymptotic expansion. Ignoring marginally relevant/irrelevant perturbations this works provided we turn off the sources for the irrelevant deformations so that higher order in $\epsilon$ correlates with more subleading terms in the radial expansion. The next step would be to turn the coefficients into functions of the boundary coordinates. For this to work the corresponding sources must remain relevant after doing so. In section \ref{subsec:radialgaugeEframe} and appendix \ref{app:radialgauge} we have studied the question of constructing a radial gauge expansion by a coordinate transformation from our non-radial gauge. We have investigated this problem by looking at pure gauge perturbations around the non-radial gauge solution to second order in $\epsilon$. It was observed that one cannot trade the $\epsilon$ expansion for a radial one precisely because of the boundary dependence of the sources. This may suggest that radial gauge is not always the preferred choice to study asymptotic expansions for Lifshitz holography.

{\bf Open directions}: We conclude by mentioning a number of interesting open directions. First of all, it would be interesting to study the probes in the Lifshitz space-time that we have briefly considered in section \ref{subsec:Lifshitzspace} and the associated two-point functions. Using the relation between AdS and Lifshitz probes one may get another perspective on the interesting results of \cite{Keeler:2013msa}. An interesting generalization of our setup, which we leave for future work, will be to add charge to the five-dimensional theory and to compute the effects in the reduced theory. We also note that further insights into the holographic model we studied are likely to be gained by studying the reduction at the weak coupling side, i.e. by reducing the boundary D3-brane world-volume theory with an axion coupling. We expect this
to be described by a non-relativistic deformation of the D2-brane world-volume theory. Another point worth pursuing, motivated by the analysis of the anomaly in our model, is the connection of our results to Ho\v{r}ava--Lifshitz gravity. In particular, one may wonder what the dynamics is of a Ho\v{r}ava--Lifshitz type action, defined on a TTNC geometry. Finally, we remark that it would be very interesting to use our results in the context of Lifshitz black holes and Lifshitz hydrodynamics. In particular, it would be interesting to obtain a fluid/gravity type derivation of Lifshitz hydrodynamics \cite{Hoyos:2013eza} which has potential applications to holographic realizations of Son's model for the effective theory of the fractional quantum Hall effect that relies on Newton--Cartan geometry \cite{Son:2013rqa}.

\subsubsection*{Acknowledgements}

We thank Marco Baggio, Matthias Blau, Jan de Boer, Wissam Chemissany, Kristian Holsheimer, Elias Kiritsis, Ioannis Papadimitriou, Simon Ross, Marika Taylor and L\'arus Thorlacius for useful discussions. Some of the calculations have been performed using Cadabra \cite{Peeters:2006kp}. The work of JH and NO is supported in part by the Danish National Research Foundation project ``Black holes and their role in quantum gravity''. BR acknowledges support from the Swiss National Science Foundation through the fellowship PBBEP2\_144805. JH also thanks the Isaac Newton Institute, Cambridge, for financial support and hospitality during the later stages of this work. 

\appendix

\section{Holographic Renormalization of the 5-Dimensional Theory}\label{sec:holoren5D}

\noindent In this appendix we summarize the relevant  results  in the 5-dimensional model of AdS gravity coupled to an axion-dilaton system and review the holographic renormalization carried out in \cite{Papadimitriou:2011qb}.  However, instead of using the
Hamiltonian formalism of \cite{Papadimitriou:2011qb}, we will work within a Lagrangian framework.
We will give the solutions of the equations of motion up to NNLO and discuss the local and anomaly counterterms as well as the one-point functions for asymptotically locally AdS (AlAdS) boundary conditions \cite{deHaro:2000xn,Papadimitriou:2005ii}.

\subsection{Fefferman--Graham expansions and counterterms}\label{ref:FGexpansions}

The solution to equations \eqref{eq:Einsteineqs}--\eqref{eq:chieom} expressed as an asymptotic series in radial gauge, i.e. as a Fefferman--Graham (FG) expansion \cite{FeffermanGraham}, reads\footnote{We will denote here and further below by $a_{(n,m)}$ the coefficient at order $r^n(\log r)^m$ of the field $r^\Delta a$ where $r^{-\Delta}$ is the leading term in the expansion of $a$ with the exception of the $a_{(n,0)}$ term which we will simply denote as $a_{(n)}$.}
\begin{eqnarray}
\hat g_{\hat\mu\hat\nu}dx^{\hat\mu} dx^{\hat\nu} &=& \frac{dr^2}{r^2}+\hat h_{\hat a\hat b}dx^{\hat a}dx^{\hat b}\,, \label{eq: sol metric gauge} \\
\hat h_{\hat a\hat b} &=& \frac{1}{r^2}\left[\hat h_{(0)\hat a\hat b}+r^2\hat h_{(2)\hat a\hat b}+r^4\log r \hat h_{(4,1)\hat a\hat b}+r^4\hat h_{(4)\hat a\hat b}+O(r^6\log r)\right]\,, \label{eq: sol metric}\\
\hat\phi &=& \hat\phi_{(0)} + r^2\hat\phi_{(2)}+r^4\log r\hat\phi_{(4,1)} + r^4\hat\phi_{(4)}+O(r^6\log r)\,, \label{eq: sol phi}\\
\hat\chi &=& \hat\chi_{(0)} + r^2\hat\chi_{(2)}+r^4\log r\hat\chi_{(4,1)}+ r^4\hat\chi_{(4)}+O(r^6\log r)\label{eq: sol chi}\,,
\end{eqnarray}
where the coefficients are given by
\begin{eqnarray}
\hat h_{(2)\hat a\hat b} &=& -\frac{1}{2}\left(\hat R_{(0)\hat a\hat b}-\frac{1}{2}\partial_{\hat a}\hat\phi_{(0)}\partial_{\hat b}\hat\phi_{(0)}-\frac{1}{2}e^{2\hat\phi_{(0)}}\partial_{\hat a}\hat\chi_{(0)}\partial_{\hat b}\hat\chi_{(0)}\right) \nonumber\\
&&
+\frac{1}{12}\hat h_{(0)\hat a\hat b}\left(\hat R_{(0)}-\frac{1}{2}(\partial\hat\phi_{(0)})^2-\frac{1}{2}e^{2\hat\phi_{(0)}}(\partial\hat\chi_{(0)})^2\right)\,,\label{eq:h2ab}\\
\hat\phi_{(2)} &=& \frac{1}{4}\left(\hat\square_{(0)}\hat\phi_{(0)}-e^{2\hat\phi_{(0)}}\left(\partial\hat\chi_{(0)}\right)^2\right) \,,\\
\hat\chi_{(2)} &=& \frac{1}{4}\left(\hat\square_{(0)}\hat\chi_{(0)}+2\partial_{\hat a}\hat\phi_{(0)}\partial^{\hat a}\hat\chi_{(0)}\right)\label{eq:chi(2)}\,,
\end{eqnarray}
at second order and by
\begin{eqnarray}
\hat h_{(4,1)\hat a\hat b} &=& \hat h_{(2)\hat a\hat c}\hat h^{\hat c}_{(2)\hat b} + \frac{1}{4}\hat\nabla^{\hat c}_{(0)}\left(\hat\nabla_{(0)\hat a}\hat h_{(2)\hat b\hat c}+\hat\nabla_{(0)\hat b}\hat h_{(2)\hat a\hat c}-\hat\nabla_{(0)\hat c}\hat h_{(2)\hat a\hat b}\right) -\frac{1}{4}\hat\nabla_{(0)\hat a}\hat\nabla_{(0)\hat b}\hat h^{\hat c}_{(2)\hat c} \nonumber\\
&& -\frac{1}{2}\partial_{(\hat a}\hat\phi_{(0)}\hat\nabla_{(0)\hat b)}\hat\phi_{(2)} -
\frac{1}{2}e^{2\hat\phi_{(0)}}\partial_{(\hat a}\hat\chi_{(0)}\hat\nabla_{(0)\hat b)}\hat\chi_{(2)}
- \frac{1}{2}e^{2\hat\phi_{(0)}}
\hat\phi_{(2)}\partial_{\hat a}\hat\chi_{(0)}\partial_{\hat b}\hat\chi_{(0)}  \nonumber\\
&& -\hat h_{(0)\hat a\hat b}\left( \frac{1}{4}\hat h_{(2)}^{\hat c\hat d}\hat h_{(2)\hat c\hat d}+\frac{1}{2}\hat\phi_{(2)}^2+\frac{1}{2}e^{2\hat\phi_{(0)}}\hat\chi_{(2)}^2\right) \,,\\
\hat\phi_{(4,1)} &=& -\frac{1}{4}\left[\hat\square_{(0)}\hat\phi_{(2)}+2\hat\phi_{(2)}\hat h^{\hat a}_{(2)\hat a} - 4e^{2\hat\phi_{(0)}}\hat\chi^2_{(2)}+\frac{1}{2}\partial^{\hat a}\hat\phi_{(0)}\hat\nabla_{(0)\hat a}\hat h^{\hat b}_{(2)\hat b}-\hat h^{\hat a\hat b}_{(2)}\hat\nabla_{(0)\hat a}\partial_{\hat b}\hat\phi_{(0)} \right.\nonumber\\
&& \left. -\partial^{\hat a}\hat\phi_{(0)}\hat\nabla^{\hat b}_{(0)}\hat h_{(2)\hat a\hat b}+e^{2\hat\phi_{(0)}}\partial_{\hat a}\hat\chi_{(0)}\left(\partial_{\hat b}\hat\chi_{(0)}\hat h^{\hat a\hat b}_{(2)}-2\hat\phi_{(2)}\partial^{\hat a}\hat\chi_{(0)}-2\hat\nabla^{\hat a}_{(0)}\hat\chi_{(2)}\right)\right]\,,\\
&&\nonumber\\
\hat\chi_{(4,1)} &=& -\frac{1}{4}\left[ 8\hat\chi_{(2)}\hat\phi_{(2)} +
2\hat\chi_{(2)}\hat h^{\hat a}_{(2)\hat a} + \hat\square_{(0)}\hat\chi_{(2)} -
\hat h^{\hat a\hat b}_{(2)}\hat\nabla_{(0)\hat a}\partial_{\hat b}\hat\chi_{(0)} +
2\hat\nabla_{(0)\hat a}\hat\chi_{(2)}\partial^{\hat a}\hat\phi_{(0)} \right.\nonumber\\
&& \left. +\partial^{\hat a}\hat\chi_{(0)}\left(\frac{1}{2}
\hat\nabla_{(0)\hat a}\hat h^{\hat b}_{(2)\hat b}-\hat\nabla^{\hat b}_{(0)}\hat h_{(2)\hat a\hat b}  -
2\partial^{\hat b}\hat\phi_{(0)}\hat h_{(2)\hat a\hat b}+
2\hat\nabla_{(0)\hat a}\hat\phi_{(2)}\right)\right]\,,
\end{eqnarray}
at order $r^4\log r$. We note that the quantity $\hat h_{(4,1)\hat a\hat b}$ is traceless.  Indices of the expansion coefficients are raised and lowered with the AdS boundary metric $\hat h_{(0)\hat a\hat b}$. At order $r^4$ we have that $\hat h_{(4)\hat a\hat b}$ is constrained by
\begin{eqnarray}
\hat h_{(4)\hat a}^{\hat a} & = & \frac{1}{4}\hat h_{(2)\hat a\hat b}\hat h_{(2)}^{\hat a\hat b}-\frac{1}{2}\hat\phi_{(2)}^2-\frac{1}{2}e^{2\hat\phi_{(0)}}\hat\chi_{(2)}^2\,,\label{eq:traceh4}\\
\hat\nabla^{\hat b}_{(0)}\hat h_{(4)\hat a\hat b} & = & -e^{2\hat\phi_{(0)}}\hat\chi_{(2)}^2\partial_{\hat a}\hat\phi_{(0)}+\hat\phi_{(4)}\partial_{\hat a}\hat\phi_{(0)}+e^{2\hat\phi_{(0)}}\hat\chi_{(4)}\partial_{\hat a}\hat\chi_{(0)}+e^{2\hat\phi_{(0)}}\hat\phi_{(2)}\hat\chi_{(2)}\partial_{\hat a}\hat\chi_{(0)}\nonumber\\
&&-\frac{1}{2}\hat\phi_{(2)}\hat\nabla_{(0)\hat a}\hat\phi_{(2)}-\frac{1}{2}e^{2\hat\phi_{(0)}}\hat\chi_{(2)}\hat\nabla_{(0)\hat a}\hat\chi_{(2)}-\frac{1}{4}\hat h_{(2)}^{\hat b\hat c}\hat\nabla_{(0)\hat a}\hat h_{(2)\hat b\hat c}\nonumber\\
&&-\frac{1}{4}\hat h_{(2)\hat a\hat c}\hat\nabla^{\hat c}_{(0)}\hat h_{(2)\hat b}^{\hat b}+\frac{1}{2}\hat h_{(2)}^{\hat b\hat c}\hat\nabla_{(0)\hat b}\hat h_{(2)\hat a\hat c}+\frac{1}{2}\hat h_{(2)\hat a}^{\hat c}\hat\nabla^{\hat b}_{(0)}\hat h_{(2)\hat b\hat c}\,.\label{eq:divh4}
\end{eqnarray}
Following \cite{deHaro:2000xn} we write the coefficient $\hat h_{(4)\hat a\hat b}$ as
\begin{equation}
\hat h_{(4)\hat a\hat b} =\hat X_{\hat a\hat b} + \frac{1}{2}\hat t_{\hat a\hat b}\,,\label{eq: h4ab}
\end{equation}
where $\hat t_{\hat a\hat b}$ is the boundary energy-momentum tensor defined in \eqref{eq:bdrystresstensor}.
The trace and divergence of $\hat t_{\hat a\hat b}$ will be given below together with the explicit form of $\hat X_{\hat a\hat b}$. In the expansion for the scalars we have that $\hat\phi_{(4)}$ and $\hat\chi_{(4)}$ are fully arbitrary functions of the boundary coordinates.

A counterterm action that cancels all divergences of the on-shell action $S_{\text{bulk}}+S_{\text{GH}}$ is given by \cite{Papadimitriou:2011qb,Chemissany:2012du}
\begin{eqnarray}
S_{\text{ct}} & = &
\frac{1}{\kappa_5^2}\int_{\partial\mathcal{M}}d^4x\sqrt{-\hat h}\left(-3-\frac{1}{4}\hat Q+\hat{\mathcal{A}}\left(\lambda+\log r\right)\right)\,,\label{eq:Sct1}
\end{eqnarray}
where $\lambda$ is some scheme dependent parameter (minimal subtraction corresponds to $\lambda=0$) and where
\begin{eqnarray}
\hspace{-1cm}\hat Q\!\! &=&\!\! \hat h^{\hat a\hat b}\hat Q_{\hat a\hat b}\,, \qquad \hat Q_{\hat a\hat b}=\hat R_{(\hat h)\hat a\hat b} - \frac{1}{2}\partial_{\hat a}\hat\phi\partial_{\hat b}\hat\phi-\frac{1}{2}e^{2\hat\phi}\partial_{\hat a}\hat\chi\partial_{\hat b}\hat\chi\,,\\
\hspace{-1cm}\hat{\mathcal{A}}\!\! &=&\!\!
\frac{1}{8}\left(\hat Q^{\hat a\hat b}\hat Q_{\hat a\hat b}-\frac{1}{3}\hat Q^2+\frac{1}{2}\left(\hat\square_{(\hat h)}\hat\phi-e^{2\hat\phi}(\partial\hat\chi)^2\right)^2+\frac{1}{2}e^{2\hat\phi}\left(\hat\square_{(\hat h)}\hat\chi+2\partial_{\hat a}\hat\phi\partial^{\hat a}\hat\chi\right)^2\right)\,.\label{eq:anomalycounterterm}
\end{eqnarray}

\subsection{One-point functions}\label{subsec:onepointfunctions}

To compute one-point functions, we write the total variation of $S_{\text{ren}}=S_{\text{bulk}}+S_{\text{GH}}+S_{\text{ct}}$ as
\begin{eqnarray}
\delta S_{\text{ren}} &=&
\frac{1}{2\kappa_5^2}\int_{\mathcal{M}}d^5x\sqrt{-\hat g}\left(
\hat{\mathcal{E}}_{\hat \mu\hat \nu}\delta \hat g^{\hat\mu\hat\nu} + \hat{\mathcal{E}}_{\hat\phi}\delta\hat\phi
+ \hat{\mathcal{E}}_{\hat\chi}\delta\hat\chi \right) \nonumber\\ && -
\frac{1}{2\kappa_5^2} \int_{\partial\mathcal{M}}d^4x\sqrt{-\hat h} \left(
\hat T_{\hat a\hat b}\delta\hat h^{\hat a\hat b} + 2\hat T_{\hat\phi}\delta\hat\phi +
2\hat T_{\hat\chi}\delta\hat\chi \right)\,,\label{eq:deltaS}
\end{eqnarray}
where $\hat{\mathcal{E}}_{\hat\mu\hat\nu},\hat{\mathcal{E}}_{\hat\phi},\hat{\mathcal{E}}_{\hat\chi}$
are the equations of motion \eqref{eq:Einsteineqs} to \eqref{eq:chieom} and where
\begin{eqnarray}
\hat T_{\hat a\hat b} &=& (\hat K-3)\hat h_{\hat a\hat b} -\hat K_{\hat a\hat b} +
\frac{1}{2}\hat Q_{\hat a\hat b}- \frac{1}{4}\hat h_{\hat a\hat b}\hat Q + \left(\lambda +\log
r\right)\hat T^{(\hat A)}_{\hat a\hat b} \,,\label{eq:Tab}\\
\hat T_{\hat\phi} &=& \frac{1}{2}\hat n^{\hat\mu}\partial_{\hat\mu}\hat\phi +
\frac{1}{4}\left(\hat\square_{(\hat h)}\hat\phi - e^{2\hat\phi}(\partial\hat\chi)^2\right) +
\left(\lambda +\log
r\right)\hat T^{(\hat A)}_{\hat\phi} \,,\label{eq:Tphi}\\
\hat T_{\hat\chi} &=& \frac{1}{2}e^{2\hat\phi}\hat n^{\hat\mu}\partial_{\hat\mu}\hat\chi
+\frac{1}{4}e^{2\hat\phi}\left(\hat\square_{(\hat h)}\hat\chi
+2\partial_{\hat a}\hat\chi\partial^{\hat a}\hat\phi\right) + \left(\lambda +\log
r\right)\hat T^{(\hat A)}_{\hat\chi} \,.\label{eq:Tchi}
\end{eqnarray}
Here we defined
\begin{equation}
\hat T^{(A)}_{\hat a\hat b} = - \frac{2\kappa_5^2}{\sqrt{-\hat h}} \frac{\delta\hat A}{\delta
\hat h^{\hat a\hat b}}\,, \qquad\hat T^{(\hat A)}_{\hat\phi} = - \frac{\kappa_5^2}{\sqrt{-\hat h}}
\frac{\delta\hat A}{\delta\hat\phi}\,,\qquad\hat T^{(\hat A)}_{\hat\chi} = -
\frac{\kappa_5^2}{\sqrt{-\hat h}} \frac{\delta\hat A}{\delta\hat\chi}\,,
\end{equation}
with
\begin{equation}
\hat A=\frac{1}{\kappa_5^2}\int_{\partial\mathcal{M}}d^4x\sqrt{-\hat h}\hat{\mathcal{A}}\,.
\end{equation}

From the expansions it follows that
$\sqrt{-\hat h}=r^{-4}\sqrt{-\hat h_{(0)}}+O(r^{-2})$, $\delta
\hat h^{\hat a\hat b} = r^2\delta\hat h_{(0)}^{\hat a\hat b} + O(r^4)$, $\delta\hat \phi =
\delta\hat \phi_{(0)} + O(r^2)$ and $\delta \hat\chi = \delta
\hat\chi_{(0)} + O(r^2)$,
which is used to  obtain the following one-point functions (we take the cut-off boundary at $r=\epsilon$)
\begin{eqnarray}
\langle \hat T_{(0)\hat a\hat b} \rangle &=& - \frac{2\kappa_5^2}{\sqrt{-\hat h_{(0)}}} \frac{\delta S_{\text{ren}}^{\text{on-shell}}}{\delta\hat h_{(0)}^{\hat a\hat b}} = \lim_{\epsilon \rightarrow 0} \epsilon^{-2}\hat T_{\hat a\hat b} = 2\hat h_{(4)\hat a\hat b} - 2\hat X_{\hat a\hat b}=\hat t_{\hat a\hat b}\,,\label{eq:bdrystresstensor}\\
\langle \hat{\mathcal{O}}_{\hat\phi} \rangle &=& - \frac{\kappa_5^2}{\sqrt{-\hat h_{(0)}}} \frac{\delta S_{\text{ren}}^{\text{on-shell}}}{\delta\hat \phi_{(0)}} = \lim_{\epsilon \rightarrow 0} \epsilon^{-4}\hat T_{\hat\phi} = \nonumber\\
&&-2\hat\phi_{(4)}-\frac{1}{2}\hat\phi_{(2)}\hat h^{\hat a}_{(2)\hat a} +e^{2\hat\phi_{(0)}}\hat\chi^2_{(2)} -\frac{1}{2} \left(3 - 4\lambda\right)\hat\phi_{(4,1)}\,,\label{eq:vevOdualtophi0}\\
\langle \hat{\mathcal{O}}_{\hat\chi} \rangle &=& - \frac{\kappa_5^2}{\sqrt{-\hat h_{(0)}}}
\frac{\delta S_{\text{ren}}^{\text{on-shell}}}{\delta\hat \chi_{(0)}} = \lim_{\epsilon \rightarrow 0}
\epsilon^{-4} \hat T_{\hat\chi} =\nonumber\\
&& - 2e^{2\hat\phi_{(0)}}\hat\chi_{(4)} -\frac{1}{2}e^{2\hat\phi_{(0)}}\left(\hat\chi_{(2)} \hat h^{\hat a}_{(2)\hat a} +
4\hat\chi_{(2)}\hat\phi_{(2)} + (3-4\lambda)\hat\chi_{(4,1)}\right) \label{eq:vevOdualtochi0} \,,\qquad
\end{eqnarray}
where
\begin{equation}
\hat X_{\hat a\hat b} = \frac{1}{2}\hat h_{(2)\hat a\hat c}\hat h^{\hat c}_{(2)\hat b}-\frac{1}{4}\hat h^{\hat c}_{(2)\hat c}\hat h_{(2)\hat a\hat b} -
\frac{1}{4}\hat h_{(0)\hat a\hat b}\hat{\mathcal{A}}_{(0)} - \frac{1}{4}\left(3-4\lambda\right)\hat h_{(4,1)\hat a\hat b} \,,\label{eq: Xab}
\end{equation}
with
\begin{equation}\label{eq:hatA0}
\hat{\mathcal{A}}_{(0)}=\lim_{\epsilon \rightarrow 0}
\epsilon^{-4}\hat{\mathcal{A}}
=\frac{1}{2}\left(\hat h_{(2)}^{\hat a\hat b}\hat h_{(2)\hat a\hat b}-(\hat h_{(2)\hat a}^{\hat a})^2\right)+\hat\phi_{(2)}^2+e^{2\hat\phi_{(0)}}\hat\chi_{(2)}^2\,.
\end{equation}

All  the contributions to the one-point functions from the $r^4\log r$ terms in the FG expansions can be removed by choosing $\lambda = \frac{3}{4}$. The boundary energy-momentum tensor is identified with $\hat t_{\hat a\hat b}$ in \eqref{eq: h4ab}.
 Using equations \eqref{eq:traceh4} and \eqref{eq:divh4} we can compute for any choice of $\lambda$
its trace and divergence
\begin{eqnarray}
\hat t^{\hat a}_{\;\;\hat a} &=& \hat{\mathcal{A}}_{(0)}\,,\label{eq:tracet}\\
\hat\nabla_{(0)\hat a}\hat t^{\hat a}{}_{\hat b} &=& -\langle \hat{\mathcal{O}}_{\hat\phi} \rangle\partial_{\hat b}\hat\phi_{(0)} -\langle \hat{\mathcal{O}}_{\hat\chi} \rangle\partial_{\hat b}\hat\chi_{(0)} \,.\label{eq:divt}
\end{eqnarray}

We thus have $\hat t_{\hat a\hat b}$ (10 components) plus $\langle \hat{\mathcal{O}}_{\hat\phi} \rangle$ and $\langle \hat{\mathcal{O}}_{\hat\chi} \rangle$ vevs minus the 5 constraints \eqref{eq:tracet} and \eqref{eq:divt} leading to 7 independent vevs. These correspond to 7 independent sources coming from $\hat h_{(0)\hat a\hat b}$, $\hat\phi_{(0)}$ and $\hat\chi_{(0)}$ (12 in total) minus the freedom to perform coordinate transformations that preserve the FG gauge removing 5 components (1 because of local Weyl rescalings and 4 coming from diffeomorphisms acting on $\hat h_{(0)\hat a\hat b}$).

\subsection{Boundary foliations}\label{subsec:boundaryfoliations}

We now choose a parametrization for $\hat h_{(0)\hat a\hat b}$ such that there is a coordinate $u$ with the property $\hat h_{(0)uu}=0$. To this end we first use a null bein basis
\begin{equation}\label{eq:nullbeinbasish0}
\hat h_{(0)\hat a\hat b} = -\hat H_{(0)\hat a}\hat N_{(0)\hat b} -\hat H_{(0)\hat b}\hat N_{(0)\hat a} + \hat\Pi_{(0)\hat a\hat b}\,,
\end{equation}

Consider the following parametrization of the boundary metric with $\hat h_{(0)uu}=0$
\begin{eqnarray}
\hat h_{(0)\hat a\hat b}dx^{\hat a}dx^{\hat b} & = & 2\left(\hat H_{(0)}dt+\hat H_{(0)i}\left(dx^i+\hat H_{(0)}\hat N_{(0)}^idt\right)\right)\left(du-\hat H_{(0)}\hat N_{(0)}dt\right)\nonumber\\
&&+\hat\Sigma_{(0)ij}\left(dx^i+\hat H_{(0)}\hat N_{(0)}^idt\right)\left(dx^j+\hat H_{(0)}\hat N_{(0)}^jdt\right)\,,\label{eq:parametrizationh0}
\end{eqnarray}
where $\hat N_{(0)}=0 \Leftrightarrow \hat h^{uu}_{(0)}=0$. The gauge choice can be obtained by taking
\begin{eqnarray}
\hat N_{(0)\hat a} &=& (-1,\hat H_{(0)}\hat N_{(0)},0)\,,\label{eq:hatN0} \\
\hat H_{(0)\hat a} &=& (0, \hat H_{(0)}(1+\hat H_{(0)i}\hat N^i_{(0)}), \hat H_{(0)i})\,,\\
\hat\Pi_{(0)\hat a\hat b} &=&
\left(\begin{array}{ccc}
0 & 0 & 0 \\
0 & \hat H_{(0)}^2\hat\Sigma_{(0)ij}\hat N^i_{(0)}\hat N^j_{(0)} & \hat H_{(0)}\hat\Sigma_{(0)ij}\hat N^j_{(0)} \\
0 & \hat H_{(0)}\hat\Sigma_{(0)ij}\hat N^i_{(0)} & \hat\Sigma_{(0)ij}
\end{array}\right)=\delta_{\underline{i}\underline{j}}\hat e_{(0)\hat a}^{\underline{i}}\hat e_{(0)\hat b}^{\underline{j}}\,,\\
\hat e_{(0)\hat a}^{\underline{i}} & = & \left(\begin{array}{c}
                            0\\
                            \hat H_{(0)}\hat e_{(0)i}^{\underline{i}}\hat N_{(0)}^i\\
                            \hat e_{(0)i}^{\underline{i}}
                            \end{array}\right)\,.
\end{eqnarray}
We then have
\begin{eqnarray}
\hat N_{(0)}^{\hat a} = \left(\begin{array}{c}
                    -\hat N_{(0)}\\
                    -\hat H_{(0)}^{-1}\\
                    \hat N_{(0)}^i
            \end{array}\right)\,,\qquad
\hat H_{(0)}^{\hat a} = \left(\begin{array}{c}
                    1\\
                    0\\
                    0
            \end{array}\right)\,,
\end{eqnarray}
\begin{equation}
\hat e_{(0)\underline{i}}^{\hat a} = \left(\begin{array}{c}
                            \hat e^u_{(0)\underline{i}}\\
                            \hat e^t_{(0)\underline{i}}\\
                            \hat e_{(0)\underline{i}}^i
                            \end{array}\right)\,,
\end{equation}
where the components are given by
\begin{eqnarray}
\hat e^u_{(0)\underline{i}} & = & -\hat N_{(0)}\left(1+\hat H_{(0)j}\hat N_{(0)}^j\right)^{-1}\hat H_{(0)i}\hat e_{(0)\underline{i}}^i\,,\label{eq:A0underlinei=0}\\
\hat e^t_{(0)\underline{i}} & = & -\hat H_{(0)}^{-1}\left(1+\hat H_{(0)j}\hat N_{(0)}^j\right)^{-1}\hat H_{(0)i}\hat e_{(0)\underline{i}}^i\,,
\end{eqnarray}
and where $\hat e^i_{(0)\underline{i}}$ satisfies
\begin{equation}
\hat e^i_{(0)\underline{i}}\hat e_{(0)j}^{\underline{j}}\left(\delta^j_i-\left(1+\hat H_{(0)k}\hat N_{(0)}^k\right)^{-1}\hat H_{(0)i}\hat N_{(0)}^j\right)=\delta_{\underline{i}}^{\underline{j}}\,.
\end{equation}
The null bein $\hat H_{(0)}^{\hat a}$ is chosen such that it is given by the Killing vector $\partial_u$.

\subsection{Five-dimensional Ward identities}\label{subsec:Wardidentities}

It is of interest to know which boundary diffeomorphisms and conformal rescalings of the boundary metric preserve this foliation. Conformal rescalings of the boundary metric $\hat h_{(0)\hat a\hat b}$ and boundary diffeomorphisms are generated by Penrose--Brown--Henneaux (PBH) transformations \cite{Penrose:1986ca}, i.e. diffeomorphisms that preserve the gauge choice of the FG expansion. Infinitesimally these transformations act on the 5-dimensional fields as
\begin{eqnarray}
\delta \hat g_{\hat\mu\hat\nu} & = & \mathcal{L}_{\hat\xi}\hat g_{\hat\mu\hat\nu}\,,\\
\delta\hat\phi & = & \mathcal{L}_{\hat\xi}\hat\phi\,,\\
\delta\hat\chi & = & \mathcal{L}_{\hat\xi}\hat\chi\,,
\end{eqnarray}
such that $\mathcal{L}_{\hat\xi}\hat g_{rr}=\mathcal{L}_{\hat\xi}\hat g_{ra}=0$ so that the radial gauge of the 5-dimensional metric \eqref{eq: sol metric gauge} is preserved. The solution to these equations gives
\begin{eqnarray}
\hat\xi^r & = & r\hat\xi^r_{(0)}\,,\label{eq:xir}\\
\hat\xi^{\hat a} & = & \hat\xi^{\hat a}_{(0)}-\int\frac{dr}{r}\hat h^{\hat a\hat b}\partial_{\hat b}\hat\xi_{(0)}^r=\hat\xi^{\hat a}_{(0)}-\frac{1}{2}r^2\hat h_{(0)}^{\hat a\hat b}\partial_{\hat b}\hat\xi^r_{(0)}+O(r^4)\,,
\end{eqnarray}
where $\hat\xi^r_{(0)}$ and $\hat\xi^{\hat a}_{(0)}$ are independent of $r$. Acting with such diffeomorphisms assuming $\hat\xi_{(0)}^r\neq 0$ on the 5-dimensional solution leads to
\begin{eqnarray}
\delta \hat h_{\hat a\hat b} & = & \hat\xi^{\hat c}\partial_{\hat c}\hat h_{\hat a\hat b}+\hat h_{\hat c\hat b}\partial_{\hat a}\hat\xi^{\hat c}+\hat h_{\hat a\hat c}\partial_{\hat b}\hat\xi^{\hat c}+\hat\xi^r\partial_r\hat h_{\hat a\hat b}\,,\\
\delta\hat\phi & = & \hat\xi^{\hat a}\partial_{\hat a}\hat\phi+\hat\xi^r\partial_r\hat\phi\,,\\
\delta\hat\chi & = & \hat\xi^{\hat a}\partial_{\hat a}\hat\chi+\hat\xi^r\partial_r\hat\chi\,.
\end{eqnarray}
At leading order this leads to conformal rescalings and reparametrizations of the boundary metric $\hat h_{(0)\hat a\hat b}$ via
\begin{eqnarray}
\delta \hat h_{(0)\hat a\hat b} & = & \hat\xi^{\hat c}_{(0)}\partial_{\hat c}\hat h_{(0)\hat a\hat b}+\hat h_{(0)\hat c\hat b}\partial_{\hat a}\hat\xi^{\hat c}_{(0)}+\hat h_{(0)\hat a\hat c}\partial_{\hat b}\hat\xi^{\hat c}_{(0)}-2\hat\xi^r_{(0)}\hat h_{(0)\hat a\hat b}\,,\label{eq:trafoh0}\\
\delta\hat\phi_{(0)} & = & \hat\xi^{\hat a}_{(0)}\partial_{\hat a}\hat\phi_{(0)}\,,\label{eq:trafophi0}\\
\delta\hat\chi_{(0)} & = & \hat\xi^{\hat a}_{(0)}\partial_{\hat a}\hat\chi_{(0)}\,.\label{eq:trafochi0}
\end{eqnarray}

The relations \eqref{eq:tracet} and \eqref{eq:divt} are Ward identities for the local gauge transformations that preserve the radial gauge of the FG expansion. To derive the Ward identities we use the variation of the on-shell action, obtained by taking  \eqref{eq:deltaS} on-shell,
\begin{equation}\label{eq:variation-on-shell-action}
\delta S_{\text{ren}}^{\text{on-shell}}=-\frac{1}{2\kappa_5^2}\int d^4x\sqrt{-\hat h_{(0)}}\left(\hat t_{\hat a\hat b}\delta\hat h_{(0)}^{\hat a\hat b}+2\langle\hat{\mathcal{O}}_{\hat\phi} \rangle\delta\hat\phi_{(0)}+2\langle\hat{\mathcal{O}}_{\hat\chi} \rangle\delta\hat\chi_{(0)}-2\hat{\mathcal{A}}_{(0)}\frac{\delta r}{r}\right)\,,
\end{equation}
where the last term comes from the variation of $\log r$ in the counterterm action \eqref{eq:Sct1}. We next take for the variations equations \eqref{eq:trafoh0}--\eqref{eq:trafochi0} writing $\delta \hat h_{(0)}^{\hat a\hat b}$ as
\begin{equation}
\delta \hat h_{(0)}^{\hat a\hat b}=-\hat\nabla_{(0)}^{\hat a}\hat\xi_{(0)}^{\hat b}-\hat\nabla_{(0)}^{\hat b}\hat\xi_{(0)}^{\hat a}+2\hat\xi_{(0)}^r\hat h_{(0)}^{\hat a\hat b}
\end{equation}
as well as $\delta r=\hat\xi^r=r\hat\xi_{(0)}^r$ where we used \eqref{eq:xir}. The terms proportional to $\hat\xi_{(0)}^r$ give \eqref{eq:tracet} whereas the terms proportional to $\hat\xi_{(0)}^{\hat a}$ give \eqref{eq:divt}.

\section{Transformation to Radial Gauge}\label{app:radialgauge}

We address here how  to write the 4D Einstein frame metric in radial gauge, i.e. as
\begin{equation}\label{eq:coordinatetrafo}
ds^2=e^\Phi\frac{dr^2}{r^2}+h_{ab}dx^adx^b=l^2_{\text{Lif}}\left(\frac{dr'^2}{r'^2}+h'_{ab}dx'^adx'^b\right)\,,
\end{equation}
where $l^2_{\text{Lif}}$ is the Lifshitz radius. We will study this problem infinitesimally. To this end we write for the metric on the left hand side $\left(g'_{\mu\nu}+\delta g_{\mu\nu}\right)dx^\mu dx^\nu$ while we have on the right hand side $g'_{\mu\nu}dx'^\mu dx'^\nu$. We thus need to find an infinitesimal coordinate transformation such that
\begin{equation}
g_{\mu\nu}dx^\mu dx^\nu=\left(g'_{\mu\nu}+\delta g_{\mu\nu}\right)dx^\mu dx^\nu=g'_{\mu\nu}dx'^\mu dx'^\nu\,.
\end{equation}
For our purposes (see the discussion in section \ref{subsec:radialgaugeEframe}) it will prove convenient to work up to second order by which we mean that we expand the pure gauge perturbation $\delta g_{\mu\nu}$ as follows
\begin{equation}\label{eq:deltagmunu}
\delta g_{\mu\nu}=\epsilon\delta_{[1]}g_{\mu\nu}+\frac{1}{2}\epsilon^2\delta_{[2]}g_{\mu\nu}+O(\epsilon^3)\,.
\end{equation}
To achieve this we transform the left hand side using (see e.g. \cite{Bruni:1996im})
\begin{equation}
r=r'-\xi^r(r',x')+\frac{1}{2}\xi^\nu\partial'_\nu\xi^r+O(\epsilon^3)\,,\qquad x^a=x'^a-\xi^a(r',x')+\frac{1}{2}\xi^\nu\partial'_\nu\xi^a+O(\epsilon^3)\,,
\end{equation}
where we expand $\xi^\mu$ as
\begin{equation}\label{eq:expepsilon}
\xi^\mu=\epsilon\xi^\mu_{[1]}+\frac{1}{2}\epsilon^2\xi^\mu_{[2]}+O(\epsilon^3)\,.
\end{equation}

The metric in the primed coordinate system $g'_{\mu\nu}$ is related to the metric in the unprimed coordinate system $g_{\mu\nu}=g'_{\mu\nu}+\delta g_{\mu\nu}$ via
\begin{equation}\label{eq:gaugevariationmetric}
\delta g_{\mu\nu}=g_{\mu\nu}-g'_{\mu\nu}=\mathcal{L}_{\xi}g_{\mu\nu}-\frac{1}{2}\mathcal{L}_\xi\mathcal{L}_\xi g_{\mu\nu}+O(\epsilon^3)\,,
\end{equation}
where everything is a function of the primed coordinates and where $\mathcal{L}_\xi$ denotes the Lie derivative along $\xi^\mu$. To be more explicit about what we mean by $\delta g_{\mu\nu}$ we write
\begin{eqnarray}
\Phi & = & 2\log l_{\text{Lif}}+\delta\Phi\,,\label{eq:perturbationPhi}\\
h_{ab} & = & l^2_{\text{Lif}}h'_{ab}+\delta h_{ab}\,,\label{eq:gaugeperturbationhab}
\end{eqnarray}
where we use the following $\epsilon$ expansions
\begin{eqnarray}
\delta\Phi & = & \epsilon\delta_{[1]}\Phi+\frac{1}{2}\epsilon^2\delta_{[2]}\Phi+O(\epsilon^3)\,,\\
\delta h_{ab} & = & \epsilon\delta_{[1]}h_{ab}+\frac{1}{2}\epsilon^2\delta_{[2]}h_{ab}+O(\epsilon^3)\,,\label{eq:deltahab}
\end{eqnarray}
and expand the left hand side of \eqref{eq:coordinatetrafo} taking (dropping the prime on the coordinates)
\begin{eqnarray}
\delta g_{rr} & = & \frac{l^2_{\text{Lif}}}{r^2}\left(\delta\Phi+\frac{1}{2}(\delta\Phi)^2+O(\epsilon^3)\right)\,,\\
\delta g_{ar} & = & 0\,.
\end{eqnarray}
In other words we have
\begin{eqnarray}
\delta_{[1]}g_{rr} & = & \frac{l^2_{\text{Lif}}}{r^2}\delta_{[1]}\Phi\,,\label{eq:delta1grr}\\
\delta_{[1]}g_{ra} & = & 0\,,\label{eq:delta1gra}\\
\delta_{[2]}g_{rr} & = & \frac{l^2_{\text{Lif}}}{r^2}\left(\delta_{[2]}\Phi+(\delta_{[1]}\Phi)^2\right)\,,\label{eq:delta2grr}\\
\delta_{[2]}g_{ra} & = & 0\,.\label{eq:delta2gra}
\end{eqnarray}

Expanding the $rr$ component of \eqref{eq:gaugevariationmetric} up to second order in $\epsilon$ using \eqref{eq:expepsilon} as well as \eqref{eq:deltagmunu}, \eqref{eq:delta1grr} and \eqref{eq:delta2grr} we obtain
\begin{eqnarray}
\delta_{[1]}\Phi & = & 2\left(\partial_r\xi^r_{[1]}-\frac{1}{r}\xi^r_{[1]}\right)\,,\label{eq:delta1Phi}\\
\delta_{[2]}\Phi & = & 2\left(\partial_r\xi^r_{[2]}-\frac{1}{r}\xi^r_{[2]}\right)+2\xi_{[1]}^\mu\partial_\mu\left(\partial_r\xi^r_{[1]}-\frac{1}{r}\xi^r_{[1]}\right)\,.\label{eq:delta2Phi}
\end{eqnarray}
Doing the same for the $ar$ component of \eqref{eq:gaugevariationmetric} we find
\begin{equation}\label{eq:deltagar}
0=\delta g_{ar}=\epsilon\delta_{[1]}g_{ar}+\frac{1}{2}\epsilon^2\delta_{[2]}g_{ar}+O(\epsilon^3)\,,
\end{equation}
where
\begin{eqnarray}
\delta_{[1]}g_{ar} & = & \frac{l^2_{\text{Lif}}}{r^2}\left(\partial_a\xi^r_{[1]}+r^2h'_{ab}\partial_r\xi^b_{[1]}\right)\,,\\
\delta_{[2]}g_{ar} & = & \frac{l^2_{\text{Lif}}}{r^2}\left(\partial_a\left(\xi^r_{[2]}-\xi^c_{[1]}\partial_c\xi^r_{[1]}-\xi^r_{[1]}\partial_r\xi^r_{[1]}\right)+r^2h'_{ab}\partial_r\left(\xi^b_{[2]}-\xi^c_{[1]}\partial_c\xi_{[1]}^b-\xi^r_{[1]}\partial_r\xi_{[1]}^b\right)\right.\nonumber\\
&&\left.+2r^2h'_{ab}\partial_r\xi_{[1]}^c\partial_c\xi^b_{[1]}+2\partial_r\xi^r_{[1]}\partial_a\xi^r_{[1]}\right)\,.
\end{eqnarray}
A similar analysis for the $ab$ component of \eqref{eq:gaugevariationmetric} tells us that
\begin{equation}
h_{ab}=l^2_{\text{Lif}}h'_{ab}+\epsilon\delta_{[1]}h_{ab}+\frac{1}{2}\epsilon^2\delta_{[2]}h_{ab}+O(\epsilon^3)\,,
\end{equation}
where
\begin{eqnarray}
\delta_{[1]}h_{ab} & = & l^2_{\text{Lif}}\left(\xi^r_{[1]}\partial_rh'_{ab}+L_{\xi_{[1]}}h'_{ab}\right)\,,\\
\delta_{[2]}h_{ab} & = & l^2_{\text{Lif}}\left(\xi^r_{[2]}\partial_rh'_{ab}+L_{\xi_{[2]}}h'_{ab}+\xi^r_{[1]}\partial_r\left(\xi^r_{[1]}\partial_rh'_{ab}\right)+\xi^r_{[1]}\partial_r\left(L_{\xi_{[1]}}h'_{ab}\right)\right.\nonumber\\
&&\left.+L_{\xi_{[1]}}\left(\xi^r_{[1]}\partial_rh'_{ab}\right)+L_{\xi_{[1]}}L_{\xi_{[1]}}h'_{ab}\right)\,,
\end{eqnarray}
where $L_{\xi_{[1]}}$ and $L_{\xi_{[2]}}$ denote the Lie derivative along $\xi_{[1]}^a$ and $\xi_{[2]}^a$, respectively. We next invert the expression for $h_{ab}$ in terms of $h'_{ab}$ giving
\begin{eqnarray}
h'_{ab} & = & l^{-2}_{\text{Lif}}\left(h_{ab}-\epsilon\left(\xi^r_{[1]}\partial_r h_{ab}+L_{\xi_{[1]}}h_{ab}\right)-\frac{1}{2}\epsilon^2\left(\xi^r_{[2]}\partial_rh_{ab}+L_{\xi_{[2]}}h_{ab}-\xi^r_{[1]}\partial_r\left(\xi^r_{[1]}\partial_rh_{ab}\right)
\right.\right.\nonumber\\
&&\left.\left.-\xi_{[1]}^r\partial_r\left(L_{\xi_{[1]}}h_{ab}\right)-L_{\xi_{[1]}}\left(\xi^r_{[1]}\partial_r h_{ab}\right)-L_{\xi_{[1]}}L_{\xi_{[1]}}h_{ab}\right)+O(\epsilon^3)\right)\,.\label{eq:hprimeab}
\end{eqnarray}
Substituting this expression in \eqref{eq:deltagar} we find (after contraction with $\frac{l^2_{\text{Lif}}}{r^2}h^{ab}$) at first order in $\epsilon$
\begin{equation}\label{eq:delta1gar}
\frac{l^2_{\text{Lif}}}{r^2}h^{ab}\partial_a\xi^r_{[1]}+\partial_r\xi_{[1]}^b=0\,,
\end{equation}
and
\begin{eqnarray}
0 & = & \frac{l^2_{\text{Lif}}}{r^2}h^{ab}\left(\partial_a\left(\xi^r_{[2]}+\xi^c_{[1]}\partial_c\xi^r_{[1]}+\xi^r_{[1]}\partial_r\xi^r_{[1]}\right)-2\partial_r\xi^r_{[1]}\partial_a\xi^r_{[1]}+2\delta_{[1]}\Phi\partial_a\xi^r_{[1]}\right)\nonumber\\
&&+\partial_r\left(\xi_{[2]}^b+\xi^c_{[1]}\partial_c\xi^b_{[1]}+\xi^r_{[1]}\partial_r\xi^b_{[1]}\right)-2\partial_r\xi^c_{[1]}\partial_c\xi^b_{[1]}\,.\label{eq:delta2gar}
\end{eqnarray}
at second order in $\epsilon$.

The leading order behavior of $\delta_{[1]}\Phi$ and $\delta_{[2]}\Phi$ is given by order $r^0$ terms that we denote as $\delta_{[1]}\Phi_{(0)}$ and $\delta_{[2]}\Phi_{(0)}$, respectively. We further note that $h^{ab}$ starts at order $r^2$ as follows from the Kaluza--Klein reduction since $\hat h^{ab}=e^{-\Phi}h^{ab}$. From this we conclude that we have
\begin{eqnarray}
\xi^r_{[1]} & = & r\left(\log r\xi^r_{[1](0,1)}+\xi^r_{[1](0)}\right)+O(r^3\log^2 r)\,,\\
\xi^a_{[1]} & = & \xi^a_{[1](0)}+O(r^2\log r)\,,\\
\xi^r_{[2]} & = & r\left(\log r\xi^r_{[2](0,1)}+\xi^r_{[2](0)}\right)+O(r^3\log^2 r)\,,\\
\xi^a_{[2]} & = & \xi^a_{[2](0)}+O(r^2\log r)\,,
\end{eqnarray}
where
\begin{eqnarray}
\frac{1}{2}\delta_{[1]}\Phi_{(0)} & = & \xi^r_{[1](0,1)}\,,\\
\frac{1}{2}\delta_{[2]}\Phi_{(0)} & = & \xi^r_{[2](0,1)}+\xi^a_{[1](0)}\partial_a\xi^r_{[1](0,1)}\,.
\end{eqnarray}
These expansions for $\xi^\mu_{[1]}$ and $\xi^\mu_{[2]}$ solve equations \eqref{eq:delta1Phi}, \eqref{eq:delta2Phi}, \eqref{eq:delta1gar} and \eqref{eq:delta2gar} to leading order in $r${}\footnote{This result supersedes and corrects the result for a similar calculation performed in section 3.5 of \cite{Chemissany:2012du}.}.

\section{Reduction of the Anomaly Density}\label{app:reducedanomaly}

\noindent In this appendix we will express the anomaly
\eqref{eq:hatA0} in terms of the 4-dimensional sources by reducing
it. For simplicity we restrict ourselves to the case where
$\tau_{(0)a}$ is hypersurface orthogonal. Using the expressions
\eqref{eq:h0abin4dsources}--\eqref{eq:invh0uu} for the AlAdS$_5$
boundary metric in terms of the 4-dimensional sources we find for
the Christoffel connection of the AlAdS$_5$ boundary metric
\begin{eqnarray}
\hspace{-.5cm}\hat \Gamma^a_{(0)bc} & = & \Gamma^a_{(0)bc} +
\frac{1}{2}\sigma_{(0)}^a\left(A_{(0)b}\tau_{(0)c} + A_{(0)c}\tau_{(0)b}\right)\,,\\
\hspace{-.5cm}\hat \Gamma^u_{(0)bc} & = &
\tilde{K}_{(0)bc}-\frac{1}{2}\left(\tau_{(0)b}\partial_c\hat
h_{(0)}^{uu}+\tau_{(0)c}\partial_b\hat h_{(0)}^{uu}\right)\,,\\
\hspace{-.5cm}\hat \Gamma^a_{(0)bu} & = & \frac{1}{2}\sigma^a_{(0)}
\tau_{(0)b}\,,\\
\hspace{-.5cm}\hat \Gamma^u_{(0)ua} &=& -
\frac{1}{2}\left(\sigma_{(0)a} +
\tau_{(0)a}A_{(0)b}\sigma_{(0)}^b\right)\,,\\
\hspace{-.5cm}\hat \Gamma^a_{(0)uu} &=& \hat \Gamma^u_{(0)uu} = 0
\,,
\end{eqnarray}
where $\tilde{K}_{(0)bc}$ is the boost invariant extrinsic curvature
defined in \eqref{eq:boostinvextrinsiccurv}. The quantities
$\sigma_{(0)a}$ and $\sigma^a_{(0)}$ are defined in
\eqref{eq:defsigma0} and \eqref{eq:defsigmaup} respectively.

It follows that the curvature components of $\hat h_{(0)\hat a\hat
b}$ are given by
\begin{eqnarray}
\hat R_{(0)ab} & = & R_{(0)ab}+
\frac{1}{2}\nabla_{(0)b}\sigma_{(0)a} + \tau_{(0)(a}A_{(0)b)}
\mathcal{D}_{(0)c}\sigma_{(0)}^c + \frac{1}{2}\left( \sigma_{(0)}^c
\partial_c\hat{h}_{(0)}^{uu} - \left(A_{(0)c}\sigma_{(0)}^c\right)^2
\right)\tau_{(0)a} \tau_{(0)b}  \nonumber\\&& +
\sigma^c_{(0)}\left( \nabla_{(0)c}A_{(0)(b} - \tilde{K}_{(0)c(b} -
\frac{1}{2}A_{(0)c} \sigma_{(0)(b} -
\frac{3}{2}\sigma_{(0)c}A_{(0)(b}\right)\tau_{(0)a)} - \frac{1}{4}
\sigma_{(0)a} \sigma_{(0)b}  \,,\\
\hat R_{(0)au} & = & \frac{1}{2}\tau_{(0)a}\left(\mathcal{D}_{(0)b} \sigma_{(0)}^b -
\sigma_{(0)b}\sigma_{(0)}^b\right)\,,\\
\hat R_{(0)uu} & = & 0\,,\\
\hat R_{(0)} & = & \mathcal{R}_{(0)} + 2\mathcal{D}_{(0)a}
\sigma_{(0)}^a-\frac{3}{2}\sigma_{(0)a} \sigma_{(0)}^a\,,
\end{eqnarray}
where $\mathcal{D}_{(0)a}$ is the projected covariant derivative
defined in \eqref{eq: def projcovderiv}.
We repeat here the convention mentioned in section \ref{subsec:curvature} that indices of objects that are orthogonal to $v_{(0)}^a$ are raised with $\Pi_{(0)}^{ab}$ and likewise indices on tensors orthogonal to $\tau_{(0)a}$ are lowered with $\Pi_{(0)ab}$.
It then follows that
\begin{eqnarray}
\hat h_{(2)ab} &=&  - \frac{1}{2}\hat R_{(0)ab}  +
\frac{1}{4}\partial_a\phi_{(0)}\partial_b\phi_{(0)}
 + \frac{1}{4}e^{2\phi_{(0)}}
\partial_a\chi_{(0)}\partial_b\chi_{(0)}
\nonumber\\&& + \frac{1}{12}Q_{(0)}\left(A_{(0)a}\tau_{(0)b} +
A_{(0)b}\tau_{(0)a}+\Pi_{(0)ab}\right) \,,\\
\hat h_{(2)au} &=& \frac{1}{4}ke^{2\phi_{(0)}}\partial_a\chi_{(0)}
+\left( \frac{1}{12}Q_{(0)}  -
\frac{1}{4}\mathcal{D}_{(0)b}\sigma^b_{(0)}+
\frac{1}{4}\sigma_{(0)b}\sigma^b_{(0)}\right)\tau_{(0)a}\,,\\
\hat h_{(2)uu} &=& \frac{1}{4}k^2e^{2\phi_{(0)}} \,,\\
\hat \phi_{(2)} &=&
\frac{1}{4}\mathcal{D}_{(0)a}\left(\Pi^{ab}_{(0)}\partial_b\phi_{(0)}\right)
-\frac{1}{4}\sigma^a_{(0)}\partial_a\phi_{(0)} -
\frac{1}{2}k^2e^{2\phi_{(0)}}I_{(0)}\,, \\
\hat \chi_{(2)} &=&
\frac{1}{4}\mathcal{D}_{(0)a}\left(\Pi^{ab}_{(0)}\partial_b\chi_{(0)}\right)
-  \frac{1}{4}k\tilde{K}_{(0)ab}\Pi^{ab}_{(0)} -
\frac{1}{4}\sigma^a_{(0)}\partial_a\chi_{(0)}\nonumber\\&& -
\frac{1}{2}k\left(
v_{(0)}^a+\Pi^{ab}_{(0)}B_{(0)b}\right)\partial_a\phi_{(0)}\,,
\end{eqnarray}
where
\begin{eqnarray}
B_{(0)a} &=& A_{(0)a} - k^{-1}\partial_a\chi_{(0)}\,,\\
I_{(0)} &=&v_{(0)}^aB_{(0)a} + \frac{1}{2}\Pi^{ab}_{(0)}B_{(0)a}B_{(0)b}\,,\\
Q_{(0)} &= &\mathcal{R}_{(0)} + 2\mathcal{D}_{(0)a}\sigma^a_{(0)} -
\frac{3}{2}\sigma_{(0)a}\sigma^a_{(0)} - k^2e^{2\phi_{(0)}}I_{(0)} -
\frac{1}{2}\Pi^{ab}_{(0)}\partial_a\phi_{(0)}\partial_b\phi_{(0)}\,.
\end{eqnarray}
We find that the full anomaly is given by
\begin{eqnarray}
\hat{\mathcal{A}}_{(0)} &=&
\frac{1}{8}X_{(0)ab}X_{(0)cd}\left(\Pi^{ac}_{(0)}\Pi^{bd}_{(0)} -
\Pi^{ab}_{(0)}\Pi^{cd}_{(0)}\right) + \frac{1}{48}\left(3X_{(0)ab}\Pi_{(0)}^{ab}-
Q_{(0)}\right)^2 \nonumber\\
&& -
\frac{1}{4}k^2e^{2\phi_{(0)}}B_{(0)a}X_{(0)bc}\Pi_{(0)}^{ab}v_{(0)}^c
- \frac{1}{2}\hat h_{(2)uu}Y_{(0)ab} v_{(0)}^a v_{(0)}^b + \hat
\phi_{(2)}^2 + e^{2\phi_{(0)}}\hat \chi_{(2)}^2\,,\label{eq: full
hatA0}
\end{eqnarray}
with
\begin{eqnarray}
Y_{(0)ab} &= &R_{(0)ab}+ \frac{1}{2}\nabla_{(0)b}\sigma_{(0)a} -
\frac{1}{2}\partial_a\phi_{(0)}\partial_b\phi_{(0)} -
\frac{1}{2}k^2e^{2\phi_{(0)}}B_{(0)a}B_{(0)b} \,,\\
X_{(0)ab} &=& Y_{(0)ab} - \frac{1}{4}\sigma_{(0)a}\sigma_{(0)b} +
\frac{1}{2}\sigma_{(0)}^c\left(
\partial_c\hat h_{(0)}^{uu} - F_{(0)cd}v_{(0)}^d  -
A_{(0)c}A_{(0)d}\sigma_{(0)}^d\right) \tau_{(0)a} \tau_{(0)b}
\nonumber\\&& + \sigma^c_{(0)}\left(\nabla_{(0)c}A_{(0)(b} -
\tilde{K}_{(0)c(b} - \frac{1}{2} A_{(0)c}\sigma_{(0)(b} -
\frac{1}{2}\sigma_{(0)c}A_{(0)(b}\right)\tau_{(0)a)} \,.
\end{eqnarray}
This result for the reduced anomaly density is so far not yet a very
insightful expression. In section \ref{subsec:anomaly} we will
rewrite it using the natural curvature objects of torsional
Newton--Cartan as defined in section \ref{subsec:curvature} for the
case of hypersurface orthogonal $\tau_{(0)a}$, i.e. for TTNC
boundary geometry.

\section{Holographic Reconstruction}\label{app:reconstruction}
The relations between the 4- and 5-dimensional fields are given by
\begin{eqnarray}
  h_{ab}&=&e^{\Phi}\left(\hat{h}_{ab}-e^{-2\Phi}\hat{h}_{au}\hat{h}_{bu}\right) \label{5dmetric}\,, \\
  A_{a}&=& e^{-2\Phi}\hat{h}_{au}\,, \\
  \Phi &= &\frac{1}{2}\log \hat{h}_{uu}\,. \label{5dPhi}
\end{eqnarray}
Hence, for $\hat{h}_{(0)uu}=0$, we have:
\begin{eqnarray}
 \Phi &=& \Phi_{(0)}+r^2 \log r \Phi_{(2,1)} + r^2 \Phi_{(2)} + r^4 \log^2 r \Phi_{(4,2)} + r^4 \log r \Phi_{(4,1)} + r^4 \Phi_{(4)} \nonumber \\ 
 &&+ r^6  \log^3 r \Phi_{(6,3)}+ r^6 \log^2 r  \Phi_{(6,2)} + r^6 \log r \Phi_{(6,1)} + r^6 \Phi_{(6)} + \mathcal{O}\left(r^8 \log^4 r \right)\,, \label{phiexpansion} \\
 A_{a} &=& r^{-2}V_{(0)a} + \log r V_{(2,1)a} + V_{(2)a} + r^{2} \log^2 r V_{(4,2)a} + r^{2} \log r V_{(4,1)a} + r^2 V_{(4)a} \nonumber \\ 
 && + r^4  \log^3 r V_{(6,3)a} + r^4 \log^2 r V_{(6,2)a} + r^4 \log r V_{(6,1)a} + r^4 V_{(6)a} + \mathcal{O}\left(r^6 \log^4 r  \right)\,, \label{bexpansion} \\
 h_{ab} &= &r^{-4}\gamma_{(0)ab} + r^{-2} \log r \gamma_{(2,1)ab} + r^{-2} \gamma_{(2)ab} +  \log^2 r \gamma_{(4,2)ab} + \log r \gamma_{(4,1)ab} \nonumber \\ 
 && + \gamma_{(4)ab} + r^2 \log^3 r \gamma_{(6,3)ab} + r^2  \log^2 r \gamma_{(6,2)ab}+ r^2 \log r \gamma_{(6,1)ab} + r^2 \gamma_{(6)ab} \nonumber \\ 
 &&  + \mathcal{O}\left(r^4 \log^4 r \right)\,, \label{hexpansion} 
\end{eqnarray}
with the coefficients given by
\begin{eqnarray}
 e^{2\Phi_{(0)}}&=&\hat{h}_{(2)uu}=-\frac{1}{4}e^{3\Phi_{(0)}}\left(\varepsilon_{(0)}^{abc}e^{\underline{t}}_{(0)a}\partial_b e^{\underline{t}}_{(0)c}\right)^2+\frac{k^2}{4}e^{2\phi_{(0)}}\,, \label{eq:constraintsources}\\
 \Phi_{(2,1)}&=&\frac{1}{2}e^{-2\Phi_{(0)}}\hat{h}_{(4,1)uu}\,, \\
 \Phi_{(2)}&=&\frac{1}{2}e^{-2\Phi_{(0)}}\left(-\frac{1}{2}e^{\Phi_{(0)}}T_{(0)}^{\underline{t}}+\hat X_{uu}\right)\,, \\
 \Phi_{(4,2)}&=& -\Phi_{(2,1)}^{2}\,, \\
 \Phi_{(4,1)}&=&\frac{1}{2}e^{-2\Phi_{(0)}}\hat{h}_{(6,1)uu} - 2\Phi_{(2)}\Phi_{(2,1)}\,, \\
 \Phi_{(4)}&=& \frac{1}{2}e^{-2\Phi_{(0)}}\hat{h}_{(6)uu} - \Phi_{(2)}^{2}\,, \\
 \Phi_{(6,3)}&=&\frac{4}{3}\Phi_{(2,1)}^{3}\,, \\
 \Phi_{(6,2)}&=&\frac{1}{2}e^{-2\Phi_{(0)}}\hat{h}_{(8,2)uu}-2\Phi_{(4,1)}\Phi_{(2,1)}\,, \\
 \Phi_{(6,1)}&=&\frac{1}{2}e^{-2\Phi_{(0)}}\hat{h}_{(8,1)uu} - 2\Phi_{(2,1)}\Phi_{(2)}^{2} - 2\Phi_{(4)}\Phi_{(2,1)} - 2 \Phi_{(4,1)}\Phi_{(2)}\,, \\
 \Phi_{(6)}&=&\frac{1}{2}e^{-2\Phi_{(0)}}\hat{h}_{(8)uu}-\frac{2}{3}\Phi_{(2)}^{3}-2\Phi_{(4)}\Phi_{(2)}\,,\\
 V_{(0)a}&=&e^{-3\Phi_{(0)}/2}e^{\underline{t}}_{(0)a}\,, \\
 V_{(2,1)a}&=&-2\Phi_{(2,1)}V_{(0)a}\,, \\
 V_{(2)a}&=&e^{-2\Phi_{(0)}}\hat{h}_{(2)au}-2\Phi_{(2)}V_{(0)a}\,,  \\
 V_{(4,2)a}&=& 4\Phi_{(2,1)}^{2}V_{(0)a}\,, \\
 V_{(4,1)a} &=&e^{-2\Phi_{(0)}}\hat{h}_{(4,1)au} - 2\Phi_{(2,1)}V_{(2)a} - 2\Phi_{(4,1)}V_{(0)a}\,, \\
 V_{(4)a} &=&e^{-2\Phi_{(0)}}\left(\frac{1}{2}e^{\Phi_{(0)}}S_{(0)a}^{\underline{t}}+\hat X_{au}\right) - 2\Phi_{(2)}^{2}V_{(0)a}-2\Phi_{(2)}V_{(2)a}-2\Phi_{(4)}V_{(0)a}\,, \\
 V_{(6,3)a}&=&-8\Phi_{(2,1)}^{3}V_{(0)a}\,, \\
 V_{(6,2)a} &=&4\Phi_{(2,1)}\Phi_{(4,1)}V_{(0)a} - 2\Phi_{(2,1)}V_{(4,1)a}\,, \\
 V_{(6,1)a} &=&e^{-2\Phi_{(0)}}\hat{h}_{(6,1)au} -2 \Phi_{(2)}V_{(4,1)a} - 2\Phi_{(2,1)}V_{(4)a} - 4\Phi_{(2,1)}\Phi_{(2)}V_{(2)a} \nonumber \\ 
 & &- 2\Phi_{(4,1)}V_{(2)a}  - 4\Phi_{(4,1)}\Phi_{(2)}V_{(0)a} - 2\Phi_{(6,1)}V_{(0)a}\,,  \\
 V_{(6)a} &=&e^{-2\Phi_{(0)}}\hat{h}_{(6)au} - 2\Phi_{(2)}V_{(4)a} - 2\Phi_{(2)}^{2}V_{(2)a} - 2\Phi_{(4)}V_{(2)a} - \frac{4}{3}\Phi_{(2)}^{3}V_{(0)a} \nonumber \\ 
 && - 4 \Phi_{(4)}\Phi_{(2)}V_{(0)a} - 2\Phi_{(6)}V_{(0)a}\,,\\
\gamma_{(0)ab} & = & -e^{\underline{t}}_{(0)a}e^{\underline{t}}_{(0)b}\,,\\
\gamma_{(2,1)ab} & = & -\Phi_{(2,1)}\gamma_{(0)ab}\,,\\
\gamma_{(2)ab} & = & e^{3\Phi_{(0)}/2}\left(e^{\underline{t}}_{(0)a}A_{(0)b}+e^{\underline{t}}_{(0)b}A_{(0)a}\right)+\delta_{\underline{i}\underline{j}}e^{\underline{i}}_{(0)a}e^{\underline{j}}_{(0)b}+3\Phi_{(2)}\gamma_{(0)ab}\nonumber\\
&& - e^{3\Phi_{(0)}}\left(V_{(0)a}V_{(2)b} + V_{(0)b}V_{(2)a} \right)\,,\\
\gamma_{(4,2)ab} & = & \left(\Phi_{(4,2)}-\frac{3}{2}\Phi_{(2,1)}^2\right)\gamma_{(0)ab}\,,\\
\gamma_{(4,1)ab} & = &  -e^{3\Phi_{(0)}}\left( V_{(0)a}V_{(4,1)b} + V_{(0)b}V_{(4,1)a} \right)+\Phi_{(2,1)}\gamma_{(2)ab}\nonumber\\
&&+3\left(\Phi_{(4,1)}-\Phi_{(2,1)}\Phi_{(2)}\right)\gamma_{(0)ab}\,,\\
\gamma_{(4)ab} & = & e^{\Phi_{(0)}}\hat{h}_{(2)ab} - e^{3\Phi_{(0)}}\left(V_{(0)a}V_{(4)b} + V_{(0)b}V_{(4)a} \right)-e^{3\Phi_{(0)}}V_{(2)a}V_{(2)b}\nonumber\\
&& - 2e^{3\Phi_{(0)}}\Phi_{(2)}\left( V_{(0)a}V_{(2)b} + V_{(0)b}V_{(2)a} \right) +\Phi_{(2)}\gamma_{(2)ab}\nonumber\\
&&+\left(3\Phi_{(4)}-\frac{5}{2}\Phi_{(2)}^2\right)\gamma_{(0)ab}\,,\\
\gamma_{(6,3)ab} & = & \left(\Phi_{(6,3)}-\Phi_{(2,1)}\Phi_{(4,2)}-\frac{29}{6}\Phi_{(2,1)}^3\right)\gamma_{(0)ab}\,,\\
\gamma_{(6,2)ab} & = & -\Phi_{(2,1)}\gamma_{(4,1)ab}+\left(\Phi_{(4,2)}+\frac{3}{2}\Phi_{(2,1)}^2\right)\gamma_{(2)ab}+\left(\Phi_{(6,2)}+4\Phi_{(4,1)}\Phi_{(2,1)}\right.\nonumber\\
&&\left.+3\Phi_{(2,1)}^2\Phi_{(2)}\right)\gamma_{(0)ab}\,,\\
\gamma_{(6,1)ab} & = & e^{\Phi_{(0)}}\hat h_{(4,1)ab}-e^{3\Phi_{(0)}}\left( V_{(0)a}V_{(6,1)b} + V_{(0)b}V_{(6,1)a} \right)-2e^{3\Phi_{(0)}}\Phi_{(2,1)}V_{(2)a}V_{(2)b}   \nonumber \\ 
&& -2e^{3\Phi_{(0)}}\Phi_{(4,1)}\left( V_{(0)a}V_{(2)b} + V_{(0)b}V_{(2)a} \right) - e^{3\Phi_{(0)}}\left( V_{(2)a}V_{(4,1)b} + V_{(2)a}V_{(4,1)b} \right) \nonumber \\ 
&& - 2 e^{3\Phi_{(0)}}\Phi_{(2)}\left( V_{(0)a}V_{(4,1)b} + V_{(0)b}V_{(4,1)a} \right) +\Phi_{(2,1)}\gamma_{(4)ab}+\Phi_{(2)}\gamma_{(4,1)ab}\nonumber \\ 
&&+\left(\Phi_{(4,1)}-\Phi_{(2,1)}\Phi_{(2)}\right)\gamma_{(2)ab}+\left(\Phi_{(4)}-\frac{1}{2}\Phi_{(2)}^2\right)\gamma_{(2,1)ab}\nonumber\\
&&+\left(3\Phi_{(6,1)}+3\Phi_{(4,1)}\Phi_{(2)}-5\Phi_{(4)}\Phi_{(2,1)}-\frac{7}{2}\Phi_{(2,1)}\Phi_{(2)}^2\right)\gamma_{(0)ab}\,,\\
\gamma_{(6)ab} & = & e^{\Phi_{(0)}}\left(\frac{1}{2}\hat t_{ab}+\hat X_{ab}\right) - e^{3\Phi_{(0)}}\left(V_{(0)a}V_{(6)b} +  V_{(0)b}V_{(6)a} \right) - 2e^{3\Phi_{0}}\Phi_{(2)}V_{(2)a}V_{(2)b} \nonumber \\ 
&& - e^{3\Phi_{(0)}}\left( V_{(4)a}V_{(2)b} + V_{(4)b}V_{(2)a} \right) - 2e^{3\Phi_{(0)}}\Phi_{(2)}\left( V_{(4)a}V_{(0)b} + V_{(4)b}V_{(0)a} \right) \nonumber \\ 
&& - 2 e^{3\Phi_{(0)}}\Phi_{(2)}^{2}\left( V_{(2)a}V_{(0)b} + V_{(2)b}V_{(0)a} \right) - 2e^{3\Phi_{(0)}}\Phi_{(4)}\left( V_{(2)a}V_{(0)b} + V_{(2)b}V_{(0)a} \right) \nonumber \\ 
&& +\Phi_{(2)}\gamma_{(4)ab}+\left(\Phi_{(4)}-\frac{1}{2}\Phi_{(2)}^2\right)\gamma_{(2)ab}+ 3\left(\Phi_{(6)}+\Phi_{(4)}\Phi_{(2)} +\frac{1}{2} \Phi_{(2)}^3\right)\gamma_{(0)ab}\,,\nonumber\\
\end{eqnarray}
where $\hat t_{ab}$ is given in equation \eqref{eq:hattab} and $\hat X_{ab}$ is given in \eqref{eq: Xab}.

\section{A Hyperscaling $\theta=-1$ and $z=1$ UV Completion}\label{subsec:h0uuneq0}

In this appendix we discuss the consequences of having $\hat h_{(0)uu}>0$ in
the Fefferman-Graham expansion, i.e. performing the reduction with a spacelike circle on the boundary. Consider the following 5-dimensional solution
\begin{eqnarray}
d\hat s^2 & = & \frac{1}{r^2}\left(2dtdu+dx^2+dy^2\right)+\frac{dr^2}{r^2}+\left(\frac{1}{r^2}+\frac{k^2}{4}g_s^2\right)du^2\,,\\
\hat\chi & = & ku\,,\\
\hat\phi & = & \log g_s\,.
\end{eqnarray}
This solution can be obtained from \eqref{eq:Schz=0} by applying to it the following diffeomorphism $t\rightarrow t-u/2$. However this diffeomorphism does not correspond to a local symmetry of the reduced theory so upon performing a reduction we obtain a solution that is not related to a $z=2$ Lifshitz space-time by some local symmetry. The 4-dimensional solution reads
\begin{eqnarray}
ds^2 & = & \frac{1}{r}\left(1+\frac{k^2}{4}g_s^2r^2\right)^{1/2}\left[-\frac{1}{r^2}\left(1+\frac{k^2}{4}g_s^2r^2\right)^{-1}dt^2+\frac{1}{r^2}\left(dx^2+dy^2+dr^2\right)\right]\,,\label{eq:hyperscalingUV1}\\
\Phi & = & -\log r+\frac{1}{2}\log\left(1+\frac{k^2}{4}g_s^2r^2\right)\,,\label{eq:hyperscalingUV2}\\
A & = & \left(1+\frac{k^2}{4}g_s^2r^2\right)^{-1}dt\,,\label{eq:hyperscalingUV3}
\end{eqnarray}
with the 4-dimensional axion-dilaton equal to a constant. If we put $k=0$ the solution is a $\theta=-1$ and $z=1$ hyperscaling violating space-time where $\theta$ is defined as in \cite{Huijse:2011ef}. For $k\neq 0$ it is asymptotically a $\theta=-1$ and $z=1$ hyperscaling violating space-time (see \cite{Kanitscheider:2008kd} for holography for space-times that are conformally AdS).

In general, from the reduction ansatz
\eqref{eq:KKansatzmetric} and the expansion \eqref{eq: sol metric} we find that the 4-dimensional expansions are given by
\begin{eqnarray}
 \Phi & = & -\log r + \Phi_{(0)} + r^2 \Phi_{(2)} + r^4 \log r
\Phi_{(4,1)} + r^4 \Phi_{(4)} + O\left(r^6 \log r \right)\,,\label{eq:UVexpansion1}\\
A_{a} & = & V_{(0)a}+r^2 V_{(2)a} + r^4 \log r V_{(4,1)a} + r^4 V_{(4)a} +
O\left(r^6 \log r \right)\,,\\
 h_{ab} & = & r^{-3}\gamma_{(0)ab} + r^{-1}\gamma_{(2)ab}+ r \log r \gamma_{(4,1)ab} + r
\gamma_{(4)ab} + O\left(r^3 \log r \right)\,,
\end{eqnarray}
where the coefficients are given by
\begin{eqnarray}
 \Phi_{(0)}&=&\frac{1}{2}\log \hat{h}_{(0)uu}\,, \\
 \Phi_{(2)}&=&\frac{1}{2}e^{-2\Phi_{(0)}}\hat{h}_{(2)uu}\,, \\
 \Phi_{(4,1)}&=&\frac{1}{2}e^{-2\Phi_{(0)}}\hat{h}_{(4,1)uu}\,, \\
 \Phi_{(4)}&=&\frac{1}{2}e^{-2\Phi_{(0)}}\left( \frac{1}{2}\hat
t_{uu}+\hat{X}_{uu}\right)-\Phi_{(2)}^{2}\,,\\
V_{(0)a}&=& e^{-2\Phi_{(0)}}\hat{h}_{(0)au}\,, \\
 V_{(2)a}&=&e^{-2\Phi_{(0)}}\hat{h}_{(2)au} - 2\Phi_{(2)}V_{(0)a}\,, \\
 V_{(4,1)a}&=& e^{-2\Phi_{(0)}}\hat{h}_{(4,1)au} -2 \Phi_{(4,1)}V_{(0)a}\,,\\
 V_{(4)a} &=&e^{-2\Phi_{(0)}}\left( \frac{1}{2}\hat
t_{au}+\hat{X}_{au}\right)- 2\Phi_{(4)}V_{(0)a} -
2\Phi_{(2)}\left(\Phi_{(2)}V_{(0)a}+ V_{(2)a} \right)\,,\\
\gamma_{(0)ab} &=&
e^{\Phi_{(0)}}\hat{h}_{(0)ab}-e^{3\Phi_{(0)}}V_{(0)a}V_{(0)b}\,, \\
 \gamma_{(2)ab} &=& e^{\Phi_{(0)}}\hat{h}_{(2)ab} + \Phi_{(2)}\gamma_{(0)ab} - 2
e^{3\Phi_{(0)}}\Phi_{(2)}V_{(0)a}V_{(0)b} \nonumber\\
 && -e^{3\Phi_{(0)}}\left(V_{(0)a}V_{(2)b} + V_{(0)b}V_{(2)a} \right) \,, \\
 \gamma_{(4,1)ab} &=& e^{\Phi_{(0)}}\hat{h}_{(4,1)ab} - 2
e^{3\Phi_{(0)}}\Phi_{(4,1)}V_{(0)a}V_{(0)b} +
\Phi_{(4,1)}\gamma_{(0)ab}\nonumber\\
 && -e^{3\Phi_{(0)}}\left(V_{(0)a}V_{(4,1)b} + V_{(0)b}V_{(4,1)a}
\right)\,, \\
 \gamma_{(4)ab} &=&e^{\Phi_{(0)}}\left(\frac{1}{2}\hat
t_{ab}+\hat{X}_{ab}\right) - e^{3\Phi_{(0)}}V_{(2)a}V_{(2)b} -
2e^{3\Phi_{(0)}}\Phi_{(2)}\left(V_{(0)a}V_{(2)b} + V_{(0)b}V_{(2)a} \right)
\nonumber \\
 && -e^{3\Phi_{(0)}}\left( V_{(0)a}V_{(4)b} + V_{(0)b}V_{(4)a} \right) - 2
e^{3\Phi_{(0)}}\left(\Phi_{(4)}+\Phi_{(2)}^{2}\right)V_{(0)a}V_{(0)b} \nonumber \\
 &&  +\Phi_{(2)}\gamma_{(2)ab}
+\left(\Phi_{(4)}- \frac{1}{2}\Phi_{(2)}^2\right) \gamma_{(0)ab}  \,.\label{eq:UVexpansionlast}
\end{eqnarray}

So far we have focused on the UV near $r=0$. The solution \eqref{eq:hyperscalingUV1}--\eqref{eq:hyperscalingUV3} can also be written as follows
\begin{eqnarray}
ds^2 & = & \left(\frac{k^2g_s^2}{4}+\frac{1}{r^2}\right)^{1/2}\left[-\frac{1}{r^4}\left(\frac{k^2g_s^2}{4}+\frac{1}{r^2}\right)dt^2+\frac{1}{r^2}\left(dx^2+dy^2+dr^2\right)\right]\,,\label{eq:IRperspective1}\\
e^{2\Phi} & = & \frac{k^2g_s^2}{4}+\frac{1}{r^2}\,,\label{eq:IRperspective2}\\
A & = & \frac{1}{r^2}\left(\frac{k^2g_s^2}{4}+\frac{1}{r^2}\right)^{-1}dt\,.\label{eq:IRperspective3}
\end{eqnarray}
Writing it like this makes it manifest that in the IR for large $r$ the solution asymptotes to a $z=2$ Lifshitz space-time. We have thus found an interpolating solution from a $\theta=-1$ and $z=1$ UV to a $z=2$ Lifshitz IR. We conclude that the two classes of solutions obtained by dimensional reduction with $h_{(0)uu}=0$ and $h_{(0)uu}>0$ have very different UV behavior but agree in the IR.

\addcontentsline{toc}{section}{References}
\small

\providecommand{\href}[2]{#2}\begingroup\raggedright\endgroup


\end{document}